\documentclass[a4paper,11pt]{article}
\pdfoutput=1
\usepackage{jheppub}
\usepackage[utf8]{inputenc}
\usepackage{url}

\usepackage[vcentermath]{youngtab}
\usepackage{pdflscape}
\usepackage{tikz}
\usetikzlibrary{positioning}
\usetikzlibrary{arrows}
\usetikzlibrary{decorations.pathreplacing}
\usetikzlibrary{shapes.geometric}
\usetikzlibrary{calc}
\usepackage{refcount}

\usepackage{enumitem}

\usepackage{pdflscape}

\usepackage{MnSymbol}

\usepackage{subcaption}

\tikzset{hasse/.style={circle, fill,inner sep=2pt}}
\tikzset{gauge/.style={inner sep=1mm,draw=none,fill=white,minimum size=2mm,circle, draw}}
\tikzset{flavour/.style={draw=none,minimum size=0.3mm,fill=white, regular polygon,regular polygon sides=4,draw}}
\tikzset{boxb/.style={draw=none,minimum size=2mm, regular polygon,regular polygon sides=4,draw}}
\tikzset{boxp/.style={draw=none,color=purple,minimum size=2mm, regular polygon,regular polygon sides=4,draw}}
\tikzset{boxg/.style={draw=none,color=black!30!green,minimum size=2mm, regular polygon,regular polygon sides=4,draw}}
\tikzset{boxo/.style={draw=none,color=orange,minimum size=2mm, regular polygon,regular polygon sides=4,draw}}

\usepackage{amsthm}
\newtheorem{ex}{Example}
\newtheorem{defn}{Definition}
\newtheorem{obs}{Observation}

\newcommand{\HC}{\mathfrak{H}_C}
\newcommand{\HH}{\mathfrak{H}_H}

\preprint{Imperial/TP/20/AH/02}
\title{Hasse Diagrams for $\mathbf{3d}$ $\mathbf{\mathcal{N}=4}$ Quiver Gauge Theories -- Inversion and the full Moduli Space}

\author{Julius F. Grimminger, }
\author{Amihay Hanany}
\affiliation{Theoretical Physics Group, The Blackett Laboratory, Imperial College London, Prince Consort Road
London, SW7 2AZ, UK}
\emailAdd{julius.grimminger17@imperial.ac.uk}
\emailAdd{a.hanany@imperial.ac.uk}

\abstract{We study Hasse diagrams of moduli spaces of $\mathrm{3d}$ $\mathcal{N}=4$ quiver gauge theories. The goal of this work is twofold: 1) We introduce the notion of inverting a Hasse diagram and conjecture that the Coulomb branch and Higgs branch Hasse diagrams of certain theories are related through this operation. 2) We introduce a Hasse diagram to map out the entire moduli space of the theory, including the Coulomb, Higgs and mixed branches. For theories whose Higgs and Coulomb branch Hasse diagrams are related by inversion it is straight forward to generate the Hasse diagram of the entire moduli space. We apply inversion of the Higgs branch Hasse diagram in order to obtain the Coulomb branch Hasse diagram for bad theories and obtain results consistent with the literature. For theories whose Higgs and Coulomb branch Hasse diagrams are not related by inversion it is nevertheless possible to produce the Hasse diagram of the full moduli space using different methods. We give examples for Hasse diagrams of the entire moduli space of theories with \emph{enhanced} Coulomb branches.}

\begin{document}

\maketitle

\section{Introduction}
The analysis of Coulomb and Higgs branches of $3d$ $\mathcal{N}=4$ quiver gauge theories has been of interest to both the physics community and the mathematics community since the advent of supersymmetry and lead to many interesting discoveries. While (classical) Higgs branches are hyper-K\"ahler quotients \cite{Hitchin:1986ea,Argyres:1996eh,Antoniadis:1996ra}, Coulomb branches require a mathematical construction in their own right \cite{Nakajima:2015txa,Nakajima2015,Braverman:2016wma}. The Coulomb branch receives quantum corrections, and a formula to compute its exact non-perturbative Hilbert series was given in \cite{Cremonesi:2013lqa}, restricting to quivers which are called good or ugly in \cite{Gaiotto:2008ak}. An explicit construction of the Coulomb branch based on an abelianisation approach was given in \cite{Bullimore:2015lsa}.

The dynamics of a Lagrangian gauge theory is governed by the Higgs mechanism \cite{englert1964broken,higgs1964broken,guralnik1964global,kibble1967symmetry}, where the gauge group is broken spontaneously by the vacuum expectation value of a scalar in the theory, i.e.\ a choice of vacuum. The moduli space of a gauge theory is the space of all of its gauge-equivalent vacua. For most of the cases we consider in this note, both the Coulomb and Higgs branch are so called \emph{symplectic singularities}\footnote{As a slight abuse of notation, we refer to a variety as a symplectic singularity, if it's Hasse diagram has only one lowest and one highest leaf. If there are multiple highest leaves we call it a union of symplectic singularities, if there are multiple lowest leaves, we call it a variety with symplectic singularities.}, which were introduced in \cite{beauville2000symplectic}, and as such admit a foliation (stratification) into a finite number of \emph{symplectic leaves} \cite{kaledin2006symplectic}. The leaves are partially ordered by inclusion of their closures and to each leaf there is a \emph{transverse slice} in the closure of any higher\footnote{Here by \emph{higher} leaf $\mathcal{L}_h$ we mean a leaf in the symplectic singularity whose closure $\bar{\mathcal{L}}_h$ contains the \emph{lower} leaf $\mathcal{L}_l$; i.e. $\mathcal{L}_l\subset\bar{\mathcal{L}}_h$} leaf. For two neighbouring leaves we call the transverse slice an \emph{elementary slice}\footnote{In this case the lower leaf is also called a \emph{minimal degeneration}.}. The closure of a symplectic leaf is the union with all of its lower leaves. This structure can be summarised in a \emph{Hasse diagram} which depicts the partial order induced by inclusion of closures. In the following for a given quiver $\mathsf{Q}$ we use the notation $\HC(\mathsf{Q})$ for the Hasse diagram of the Coulomb branch of $\mathsf{Q}$ and $\HH(\mathsf{Q})$ for the Hasse diagram of the Higgs branch of $\mathsf{Q}$. It should be noted, that two different symplectic singularities may have the same Hasse diagram. An example of this phenomenon is given in the Appendix \ref{app:add}. A nice review of the mathematics of symplectic singularities is \cite{2005math.....10346F}.

The different leaves that make up the moduli space correspond to different sets of massless states including gauge fields in the adjoint of the different subgroups that the gauge group may be broken to dynamically and matter fields in nontrivial representations. Transverse slices correspond to the moduli one has to tune in order to move from one phase of the theory to another, corresponding to moving from one leaf in the moduli space to another. The transverse slices can be seen as moduli spaces in their own right, and correspond the moduli spaces of the interacting part of theories that are reached by Higgsing, after massive states are integrated out. For the Higgs branch this was analysed in detail in \cite{Bourget:2019aer}. In the math literature it is known that singular hyper-K\"ahler quotients have a stratification into symplectic leaves, which was proven in \cite{nakajima1994instantons,dancer1997geometry} building on the work of \cite{sjamaar1991stratified} \footnote{J.F.G. thanks Andrew Dancer for patient explanations.}.

Kraft and Procesi \cite{kraft1980minimal,Kraft1982} used Hasse diagrams to describe the geometry of closures of nilpotent orbits, a result reproduced from brane physics in \cite{Cabrera:2016vvv,Cabrera:2017njm}. As shown in \cite{Bourget:2019aer} one can determine the Hasse diagram of a Coulomb branch of a theory using an operation called \emph{quiver subtraction}, which was first introduced in \cite{Cabrera:2018ann}. One can compute the Hasse diagram of a classical Higgs branch through the partial Higgs mechanism, or if a magnetic quiver is known, through quiver subtraction on the magnetic quiver \cite{Bourget:2019aer}. Hasse diagrams for singular hyper-K\"ahler quotients were studied in \cite{mayrand2018stratified}. For unitary quivers a procedure to produce the Higgs branch (quiver variety) Hasse diagram is given in \cite{2016arXiv160200164B} \footnote{The authors thank Antoine Bourget for a wonderful journal club talk on \cite{2016arXiv160200164B}, and Travis Schedler for helpful comments.}. Outside the realm of symplectic singularities, Hasse diagrams were introduced for Coulomb branches of $4d$ $\mathcal{N}=3$ theories, so called triply special K\"ahler spaces, in \cite{Argyres:2019yyb}.

\begin{figure}
    \centering
    \begin{tikzpicture}
        \draw (0,0) ellipse (3cm and 2cm);
        \draw (-1.1,-1) circle (1.5cm);
        \draw (1.1,-1) circle (1.5cm);
        \node at (0,0.95) {symplectic singularities};
        \node at (-1.3,-0.9) {HK};
        \node at (1.3,-0.9) {CB};
        \node[rotate=90] at (0,-1) {3d MS};
        \node at (6.4,0.3) {HK - hyper-K\"ahler Quotient};
        \node at (5.95,-0.3) {CB - Coulomb Branch};
        \node at (6,-0.9) {3d MS - 3d Mirror Symmetry};
    \end{tikzpicture}
    \caption{One has to make a distinction between the various geometric spaces dealt with in the realm 3d $\mathcal{N}=4$ gauge theories. Examples of symplectic singularities are given through both the hyper-K\"ahler quotient and the Coulomb branch construction. The space of symplectic singularites which can be constructed as a Coulomb branch and as a hyper-K\"ahler quotient, is the realm of the famous 3d mirror symmetry. However, there are Coulomb branches for which no hyper-K\"ahler quotient construction is known and vice versa. Furthermore there are hyper-K\"ahler quotients and Coulomb branches, which are not symplectic singularities. If the hyper-K\"ahler quotient is a union of symplectic singularities, the individual cones may be described as the Coulomb branches of a set of magnetic quivers. The hyper-K\"ahler quotient need not accurately describe the Higgs branch of the quantum moduli space of a theory, an example is given in Section \ref{Bad}.}
    \label{fig:overview}
\end{figure}
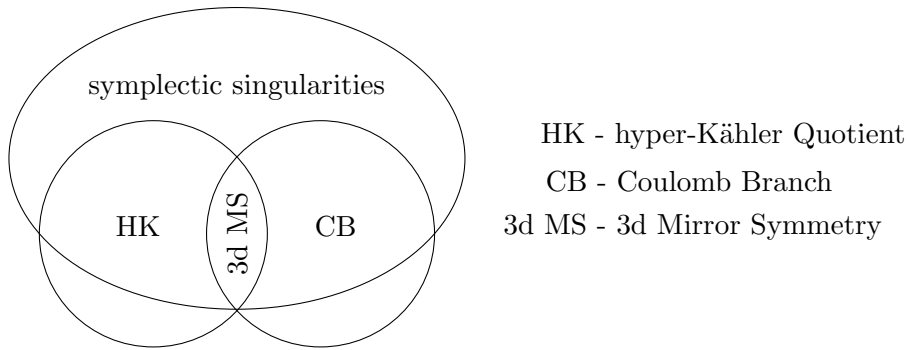
In Figure \ref{fig:overview} an overview of how the notions of hyper-K\"ahler quotients, Coulomb branches and symplectic singularities interplay is given. There is a large class of examples of symplectic singularities, which can be constructed as the Higgs branch of one theory and the Coulomb branch of another theory. This is the playground for the 3d mirror symmetry of \cite{Intriligator:1996ex} \footnote{Two theories do not have to be 3d mirror duals just because the Coulomb branch of one is the Higgs branch of the other, a simple counter example is O(2) with 2 fundamental hypermultiplets and the affine $\hat{D}_4$ Dynkin quiver. While the Coulomb branch of the O(2) theory is the Higgs branch of the affine quiver, the Higgs branch of the O(2) theory is not the Coulomb branch of the affine quiver and they are not 3d mirror duals.}. There are several examples of hyper-K\"ahler quotients, which are symplectic singularites, where no description as a Coulomb branch is known; an example are the Kleinian singularities associated to exceptional groups. Furthermore there are hyper-K\"ahler quotients which are a union of several symplectic singularities, which may individually be described as Coulomb branches of magnetic quivers, however can not be constructed as a single Coulomb branch. Also, there are Coulomb branches, which are symplectic singularities, for which no description as a hyper-K\"ahler quotient is known; for example the minimal nilpotent orbits of the E-type exceptional groups. Neither Higgs branches nor Coulomb branches need to be symplectic singularities. A simple example is the theory of SU(2) with 2 fundamental hypermultiplets, which is discussed in detail at the beginning of section \ref{Bad}, neither the Higgs nor Coulomb branch are symplectic singularities in this case. The example illustrates a further complication in the study of moduli spaces: The classical Higgs branch, computed through the hyper-K\"ahler quotient, may not appear in this form in the quantum moduli space of the theory. In the case of SU(2) with 2 fundamental hypermultiplets the classical Higgs branch consists of two cones intersecting at the origin. In the quantum theory the two cones are separated along the Coulomb branch \cite{Seiberg:1996nz,Assel:2018exy}. In section \ref{Bad} we use Hasse diagrams to describe the moduli space of a theory, even when it is not a (union of) symplectic singularities.

One of our central tools are brane set-ups of \cite{Hanany:1996ie} (and their relatives). They are configurations of NS5, D3 and D5 branes in Type IIB String Theory, such that on the worldvolume of the D3 branes $1/4$ of the 32 supercharges remain unbroken. The effective theory on the D3 branes can be described as a $3d$ $\mathcal{N}=4$ quiver gauge theory with $SO(1,2)$ Lorentz-symmetry and $SU(2)_H\times SU(2)_C$ R-symmetry.\\
\begin{table}[h]
\begin{center}
\begin{tabular}{c|c|c|c|c|c|c|c|c|c|c}
& $x^0$ & $x^1$ & $x^2$ & $x^3$ & $x^4$ & $x^5$ & $x^6$ & $x^7$ & $x^8$ & $x^9$\\
\hline
NS5 & x & x & x & x & x & x & & & &  \\
\hline
D3 & x & x & x & & & & x & & & \\
\hline
D5 & x & x & x & & & & & x & x & x\\
\multicolumn{1}{c}{} & \multicolumn{3}{c}{\upbracefill} & \multicolumn{3}{c}{\upbracefill}& \multicolumn{1}{c}{}& \multicolumn{3}{c}{\upbracefill}\\
\multicolumn{1}{c}{} & \multicolumn{3}{c}{$SO(1,2)$} & \multicolumn{3}{c}{$SU(2)_C$} & \multicolumn{1}{c}{}& \multicolumn{3}{c}{$SU(2)_H$}
\end{tabular}
\caption{The 'x' mark the spacetime directions spanned by the various branes. The groups $SO(1,2)$, $SU(2)_H$ and $SU(2)_C$ can be inferred.}
\label{setup}
\end{center}
\end{table}

When all the D3 branes are suspended between NS5 branes the effective field theory on them contains massless vector multiplets and one moves on the Coulomb branch of the quiver gauge theory. When all (or a maximal number)\footnote{When all the D3 branes can be suspended between D5 branes one speaks of complete Higgsing, in this case the gauge group can be completely broken. Otherwise one speaks of incomplete Higgsing.} of D3 branes are suspended between D5 branes the effective field theory on them contains massless hyper multiplets and one moves on the Higgs branch of the quiver gauge theory. When there are D3 branes between NS5 branes and D5 branes respectively one moves on a mixed branch. The $SU(2)_H$ acts on the Higgs branch but not on the Coulomb branch, the $SU(2)_C$ acts on the Coulomb branch but not on the Higgs branch, and both act on the mixed branches.\\

All theories studied in the following are $3d$ $\mathcal{N}=4$ quiver gauge theories which can be completely Higgsed and have unitary gauge nodes or unitary magnetic quivers\footnote{The concept of magnetic quivers, introduced in \cite{Hanany:1996ie}, was already used in \cite{DelZotto:2014kka,Cremonesi:2015lsa,Ferlito:2017xdq,Hanany:2018vph,Cabrera:2018jxt} and had its name revived in \cite{Cabrera:2019izd} and \cite{Cabrera:2019dob}. For good or ugly 3 dimensional theories with Higgs branches that consist of only one cone the magnetic quiver is the well known 3d mirror dual \cite{Intriligator:1996ex}.}. All masses and Fayet-Iliopoulus parameters are set to zero.\\

Consider SQED with N flavours. The corresponding quiver is
\begin{equation}
    \begin{tikzpicture}
        \node[gauge] (1) at (0,0) {};
        \node[flavour] (2) at (0,1) {};
        \draw (1)--(2);
        \node at (-0.3,0) {$1$};
        \node at (-0.4,1) {$N$};
    \end{tikzpicture}\,,
    \label{sqed}
\end{equation}
which we write in text form as $(1)-[N]$. The corresponding brane system to \eqref{sqed} is (red = NS5, blue = D5, black = D3)
\begin{equation}
    \begin{tikzpicture}
        \draw[red] (0,0)--(0,4) (4,0)--(4,4);
        \draw[blue] (0.5,1)--(1.5,3) (2.5,1)--(3.5,3);
        \draw[dashed] (-1,2)--(5,2);
        \node at (-1.5,2) {origin};
        \draw (0,2)--(4,2);
        \begin{scope}[shift={(-2,1)}]
        \draw[->] (-3,1)--(-2,1);
        \node at (-1.5,1) {$(x^6)$};
        \draw[->] (-3,1)--(-3,2);
        \node at (-3,2.3) {$(x^3,x^4,x^5)$};
        \draw[->] (-3,1)--(-3.5,0);
        \node at (-4.5,0) {$(x^7,x^8,x^9)$};
        \end{scope}
        \draw [decorate,decoration={brace,amplitude=5pt}] (1.3,3.1)--(3.7,3.1);
        \node at (2.5,3.5) {N};
        \node at (1.7,1.4) {\dots};
    \end{tikzpicture}\,,
\end{equation}
The D5 branes are drawn to point out of the page. The Coulomb branch and the Higgs branch are realised in the brane system as: a) = Coulomb, b) = Higgs
\begin{equation}
    \begin{tikzpicture}
        \draw[red] (0,0)--(0,4) (4,0)--(4,4);
        \draw[blue] (0.5,1)--(1.5,3) (2.5,1)--(3.5,3);
        \draw[dashed] (-1,2)--(5,2);
        \draw (0,0.5)--(4,0.5);
        \node at (1.7,1.4) {\dots};
        \begin{scope}[shift={(8,0)}]
        \draw[red] (0,0)--(0,4) (4,0)--(4,4);
        \draw[blue] (0.5,1)--(1.5,3) (2.5,1)--(3.5,3);
        \draw[dashed] (-1,2)--(5,2);
        \draw (0,2)--(1,2) (3,2)--(4,2) (0.75,1.5)--(1.25,1.5);
        \node at (1.7,1.4) {\dots};
        \end{scope}
        \begin{scope}[shift={(8-0.2,-0.4)}]
        \draw (2.75,1.5)--(2.25,1.5);
        \end{scope}
        \node at (-1,3) {a)};
        \node at (7,3) {b)};
    \end{tikzpicture}\,.
\end{equation}
The Coulomb branch consists of two symplectic leaves, the origin, of dimension $0$, and the origin-less Kleinian singularity $A_{N-1}-\{0\}=\mathbb{C}^2/\mathbb{Z}_N-\{0\}$, of dimension $1$\footnote{All dimensions are quaternionic.}. The Higgs branch also consists of two leaves, the origin and the minimal nilpotent orbit of $sl(N,\mathbb{C})$ denoted $a_{N-1}$, of dimension $N-1$. The respective Hasse diagrams are
\begin{equation}
    \HC=\quad\begin{tikzpicture}
        \node[hasse] (1) at (0,0) {};
        \node[hasse] (2) at (0,2) {};
        \draw (1)--(2);
        \node at (0.5,1) {$A_{N-1}$};
        \node at (-0.3,0) {$0$};
        \node at (-0.3,2) {$1$};
    \end{tikzpicture}\qquad
    \HH=\quad\begin{tikzpicture}
        \node[hasse] (1) at (0,0) {};
        \node[hasse] (2) at (0,2) {};
        \draw (1)--(2);
        \node at (0.5,1) {$a_{N-1}$};
        \node at (-0.3,0) {$0$};
        \node at (-0.7,2) {$N-1$};
    \end{tikzpicture}.
    \label{eq:Klein_Hasse}
\end{equation}
The information displayed in the Hasse diagram is:
\begin{enumerate}
    \item Black dots \begin{tikzpicture}
        \node[hasse] at (0,0) {};
    \end{tikzpicture} with a number $n$ next to it: denote a leaf of quaternionic dimension $n$.
    \item A line $|$ with a label next to it, between two black dots: denotes the elementary slice between two neighbouring leaves.
\end{enumerate}
\begin{figure}[t]
    \makebox[\textwidth][c]{
    \begin{tikzpicture}
        % A-type
        \node[gauge] (1) at (0,0) {};
        \node[gauge] (2) at (2,0) {};
        \node[gauge] (3) at (1,1) {};
        \draw (0.5,0)--(1)--(3)--(2)--(1.5,0);
        \node at (1,0) {\dots};
        \node at (0.5,1) {$1$};
        \node at (0,-0.35) {$1$};
        \node at (2,-0.35) {$1$};
        \draw [decorate,decoration={brace,amplitude=5pt}] (2.3,-0.5)--(-0.3,-0.5);
        \node at (1,-1) {$n$};
        
        % D-type
        \begin{scope}[shift={(5,0.5)}]
        \node[gauge] (4) at (0,0.5) {};
        \node[gauge] (5) at (0,-0.5) {};
        \node[gauge] (6) at (0.5,0) {};
        \node[gauge] (7) at (3,0.5) {};
        \node[gauge] (8) at (3,-0.5) {};
        \node[gauge] (9) at (2.5,0) {};
        \draw (4)--(6)--(5) (6)--(1,0) (2,0)--(9) (7)--(9)--(8);
        \node at (1.5,0) {\dots};
        \node at (-0.3,0.5) {$1$};
        \node at (-0.3,-0.5) {$1$};
        \node at (3.3,0.5) {$1$};
        \node at (3.3,-0.5) {$1$};
        \node at (0.5,-0.35) {$2$};
        \node at (2.5,-0.35) {$2$};
        \draw [decorate,decoration={brace,amplitude=5pt}] (2.7,-0.5)--(0.3,-0.5);
        \node at (1.5,-1) {$n-3$};
        \end{scope}
        
        % E6
        \begin{scope}[shift={(11,0)}]
        \node[gauge] (10) at (0,0) {};
        \node[gauge] (11) at (1,0) {};
        \node[gauge] (12) at (2,0) {};
        \node[gauge] (13) at (2,1) {};
        \node[gauge] (14) at (2,2) {};
        \node[gauge] (15) at (3,0) {};
        \node[gauge] (16) at (4,0) {};
        \draw (10)--(11)--(12)--(13)--(14) (12)--(15)--(16);
        \node at (0,-0.35) {$1$};
        \node at (1,-0.35) {$2$};
        \node at (2,-0.35) {$3$};
        \node at (3,-0.35) {$2$};
        \node at (4,-0.35) {$1$};
        \node at (1.7,1) {$2$};
        \node at (1.7,2) {$1$};
        \end{scope}
        
        % E7
        \begin{scope}[shift={(0,-3)}]
        \node[gauge] (17) at (0,0) {};
        \node[gauge] (18) at (1,0) {};
        \node[gauge] (19) at (2,0) {};
        \node[gauge] (20) at (3,0) {};
        \node[gauge] (21) at (4,0) {};
        \node[gauge] (22) at (5,0) {};
        \node[gauge] (23) at (6,0) {};
        \node[gauge] (24) at (3,1) {};
        \draw (17)--(18)--(19)--(20)--(21)--(22)--(23) (24)--(20);
        \node at (0,-0.35) {$1$};
        \node at (1,-0.35) {$2$};
        \node at (2,-0.35) {$3$};
        \node at (3,-0.35) {$4$};
        \node at (4,-0.35) {$3$};
        \node at (5,-0.35) {$2$};
        \node at (6,-0.35) {$1$};
        \node at (2.7,1) {$2$};
        \end{scope}
        
        % E8
        \begin{scope}[shift={(8,-3)}]
        \node[gauge] (25) at (0,0) {};
        \node[gauge] (26) at (1,0) {};
        \node[gauge] (27) at (2,0) {};
        \node[gauge] (28) at (3,0) {};
        \node[gauge] (29) at (4,0) {};
        \node[gauge] (30) at (5,0) {};
        \node[gauge] (31) at (6,0) {};
        \node[gauge] (32) at (7,0) {};
        \node[gauge] (33) at (5,1) {};
        \draw (25)--(26)--(27)--(28)--(29)--(30)--(31)--(32) (33)--(30);
        \node at (0,-0.35) {$1$};
        \node at (1,-0.35) {$2$};
        \node at (2,-0.35) {$3$};
        \node at (3,-0.35) {$4$};
        \node at (4,-0.35) {$5$};
        \node at (5,-0.35) {$6$};
        \node at (6,-0.35) {$4$};
        \node at (7,-0.35) {$2$};
        \node at (4.7,1) {$3$};
        \end{scope}
        \node at (-1,1) {a)};
        \node at (4,1) {b)};
        \node at (10,1) {c)};
        \node at (-1,-2) {d)};
        \node at (7,-2) {e)};
        \node at (0,-4) {};
    \end{tikzpicture}}
    \caption{Affine ADE Dynkin quivers. a) $\hat{A}_n$ b) $\hat{D}_n$ c) $\hat{E}_6$ d) $\hat{E}_7$ e) $\hat{E}_8$. Their Coulomb branches are the minimal nilpotent orbit closure of the corresponding algebra, written $a_n$, $d_n$, and $e_n$ respectively. Their Higgs branches are the Kleinian singularities corresponding to the algebra, written $A_n$, $D_n$, and $E_n$ respectively. It should be clear from context, when a capital letter refers to the Kleinian singularity rather than a Dynkin diagram or an algebra. The Hasse diagrams for both the Coulomb and Higgs branches are given in \eqref{eq:aff_Hasse}.}
    \label{fig:aff_quivers}
\end{figure}
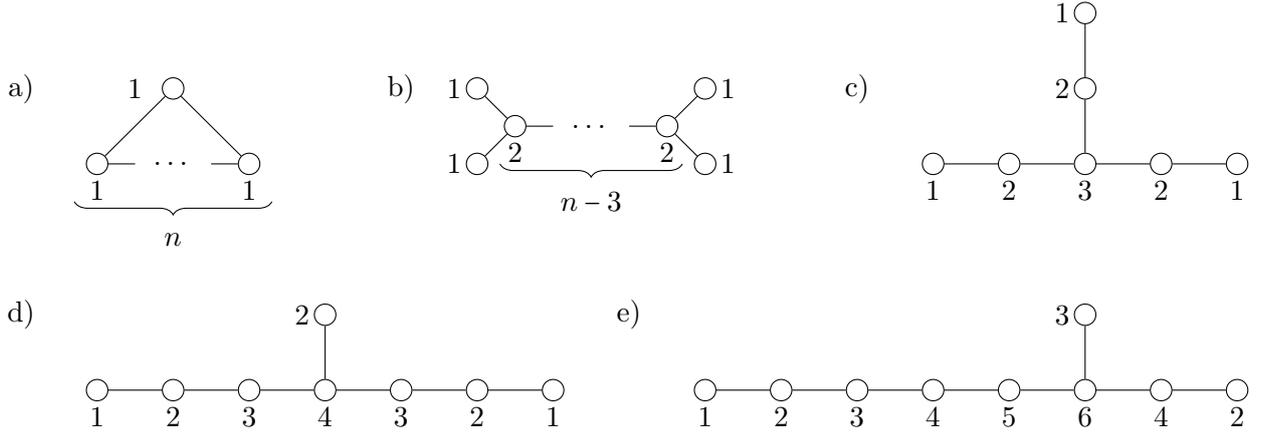

Now consider the affine Dynkin quivers of ADE (Figure \ref{fig:aff_quivers}). Their Coulomb branches are minimal nilpotent orbit closures, while their Higgs branches are Kleinian singularities. Both their Higgs and Coulomb branches consist of two symplectic leaves and the transverse slice is the the branch itself. The respective Hasse diagrams are
\begin{equation}
    \HC=\quad\begin{tikzpicture}
        \node[hasse] (1) at (0,0) {};
        \node[hasse] (2) at (0,2) {};
        \draw (1)--(2);
        \node at (1.5,1) {$a_n$, $d_n$ or $e_n$};
        \node at (-0.3,0) {$0$};
        \node at (-0.3,2) {$x$};
    \end{tikzpicture}\qquad
    \HH=\quad\begin{tikzpicture}
        \node[hasse] (1) at (0,0) {};
        \node[hasse] (2) at (0,2) {};
        \draw (1)--(2);
        \node at (1.5,1) {$A_N$, $D_n$ or $E_n$};
        \node at (-0.3,0) {$0$};
        \node at (-0.3,2) {$1$};
    \end{tikzpicture}.
    \label{eq:aff_Hasse}
\end{equation}
Where closures of minimal nilpotent orbits are denoted with a lower case and Kleinian singularities with an upper case letter, note that $A_1=a_1$, and
\begin{equation}
    x=\left\{\begin{array}{l l}
         n & \textnormal{for }a_n\\
         2n-3 & \textnormal{for }d_n\\
         11 & \textnormal{for }e_6\\
         17 & \textnormal{for }e_7\\
         29 & \textnormal{for }e_8
    \end{array}\right.
\end{equation}
is the the dimension of the Coulomb Branch of the associated quiver\footnote{It is the dual coxeter number $-1$ of the associated algebra.}. The Hasse diagrams $\HC$ and $\HH$ of the examples discussed are related by exchanging minimal nilpotent orbit closures with Kleinian singularities associated to the same group. This is an example of \emph{inversion} of a Hasse diagram, when there is only one transverse slice.\\

In Section \ref{Inversion} we introduce the concept of inversion for a more general set up, and give examples of theories where inversion does or does not relate the Coulomb and Higgs branch Hasse diagrams. In Section \ref{Moduli} we discuss how this concept enables us to produce a Hasse diagram for the entire moduli space of the theory, i.e. the Coulomb, Higgs and mixed branches, starting from the Hasse diagram of one branch. We comment on the physical interpretation of the leaves and transverse slices that make up the moduli space, and the physical meaning of the quivers involved in the quiver subtraction procedure. In Section \ref{Bad} we propose an application of inversion to bad theories when there is complete Higgsing. In Section \ref{noninv} we analise the Hasse diagram of entire moduli spaces for theories where inversion does not relate the Higgs and Coulomb branch Hasse diagrams using arguments from brane constructions explained in Appendix \ref{app:O}. In Section \ref{exotic} we use the lessons from Section \ref{Moduli} and \ref{noninv} in order to obtain Hasse diagrams of the entire moduli space in cases where there is no brane construction and inversion has to be used in a less straight forward manner. In Section \ref{Outlook} we give an overview of some open questions.

\section{Inversion of a Hasse diagram}
\label{Inversion}
The aim of this section is to explore the relationship between the Hasse diagram of the Coulomb branch and Higgs branch of a theory. In the following a Hasse diagram $\mathfrak{H}$ describes the stratification of symplectic singularities into symplectic leaves and the elementary transverse slices between neighbouring leaves, as introduced in the last section.
\begin{defn}
When all elementary slices are minimal nilpotent orbit closures or Kleinian singularities of type ADE the Hasse diagram is called \emph{invertible}.
\end{defn}
\begin{defn}
The operation of \emph{inversion} $\mathfrak{I}$ of an invertible Hasse diagram $\mathfrak{H}$ is defined by inverting the partial ordering (turning the Hasse diagram upside down) and exchanging minimal nilpotent orbits and Kleinian singularities with each other (exchanging lower case and upper case letters).
\end{defn}
\begin{ex}\normalfont As an example of inverting a Hasse diagram we pick a simple Hasse diagram $\mathfrak{H}$ consisting of 3 leaves and 2 elementary slices which are (from the top) $a_1$ and $a_3$. Next to it we give the inversion of the Hasse diagram denoted $\mathfrak{I}(\mathfrak{H})$ which consits of the same number of leaves and elementary slices which are (from the top) $A_3$ and $A_1$
\begin{equation}
    \mathfrak{H}=\quad\begin{tikzpicture}
        \node[hasse] (1) at (0,0) {};
        \node[hasse] (2) at (0,2) {};
        \node[hasse] (3) at (0,4) {};
        
        \draw (1)--(2)--(3);
        
        \node at (0.5,1) {$a_3$};
        \node at (0.5,3) {$a_1$};
        \node at (-0.3,0) {$0$};
        \node at (-0.3,2) {$3$};
        \node at (-0.3,4) {$4$};
    \end{tikzpicture}\qquad
    \mathfrak{I}(\mathfrak{H})=\quad\begin{tikzpicture}
        \node[hasse] (1) at (0,0) {};
        \node[hasse] (2) at (0,2) {};
        \node[hasse] (3) at (0,4) {};
        
        \draw (1)--(2)--(3);
        
        \node at (0.5,1) {$A_1$};
        \node at (0.5,3) {$A_3$};
        \node at (-0.3,0) {$0$};
        \node at (-0.3,2) {$1$};
        \node at (-0.3,4) {$2$};
    \end{tikzpicture}\;.
    \label{eq:example}
\end{equation}
\end{ex}
We conjecture that the concept of inversion may be applied to the Hasse diagrams of Coulomb and Higgs branches of a theory.
\begin{obs}
There exists a set of $3d$ $\mathcal{N}=4$ quiver gauge theories for which the Hasse diagrams of the Higgs and Coulomb branch, called $\HH$ and $\HC$ respectively, are related by inversion:
\begin{equation}
    \mathfrak{I}(\HH)=\HC\quad\text{and}\quad\mathfrak{I}(\HC)=\HH\,.
\end{equation}
\end{obs}
The Hasse diagrams in the example \eqref{eq:example} above are those associated to the quiver\footnote{The association is not unique! The quiver [4]-(2)-(1)-[1] and its 3d Mirror dual [2]-(1)-(1)-(1)-(1)-[1] also have the Hasse diagrams \eqref{eq:example} associated to them; just as \eqref{eq:3leafexample} and its 3d Mirror dual \eqref{eq:3leafexampledual}. See the Appendix \ref{app:add} for details.}
\begin{equation}
    \begin{tikzpicture}
        \node[gauge] (1) at (0,0) {};
        \node[gauge] (2) at (-1,0) {};
        \node[gauge] (3) at (1,0) {};
        \node[flavour] (4) at (0,1) {};
        \draw(2)--(1)--(3) (1)--(4);
        
        \node at (-1,-0.5) {1};
        \node at (0,-0.5) {2};
        \node at (1,-0.5) {1};
        \node at (0.5,1) {2};
    \end{tikzpicture}\hspace{2cm} (\HH=\mathfrak{H}\textnormal{ and }\HC=\mathfrak{I}(\mathfrak{H}))\,,
    \label{eq:3leafexample}
\end{equation}
and of course to its 3d Mirror dual,
\begin{equation}
    \begin{tikzpicture}
        \node[gauge] (1) at (0,0) {};
        \node[flavour] (4) at (0,1) {};
        \draw(1)--(4);
        
        \node at (0.5,0) {2};
        \node at (0.5,1) {4};
    \end{tikzpicture}\hspace{3cm} (\HC=\mathfrak{H}\textnormal{ and }\HH=\mathfrak{I}(\mathfrak{H}))\,,
    \label{eq:3leafexampledual}
\end{equation}
whose Higgs and Coulomb branches are related by inversion. In the case where one branch is a minimal nilpotent orbit of ADE, the inversion is shown to hold in the introduction. For linear unitary quivers which are good or ugly we can derive the inversion by applying quiver subtraction (see \cite{Cabrera:2018ann} and Appendix A in \cite{Bourget:2019aer}) to obtain $\HC$ and reading the Higgs branch of each subtracted quiver to build $\HH$ from the bottom up, which was already noted in \cite{Cabrera:2016vvv} and \cite{Rogers:2018dez,Rogers:2019pqe}. Since quiver subtraction is nothing but a Kraft-Procesi transition \cite{Cabrera:2016vvv} in the brane system, the inversion property can also be seen from moving between different phases in the brane system.

As a non-trivial example involving D-type transitions one can study the Higgs and Coulomb branch of
\begin{equation}
    \begin{tikzpicture}
    \node[gauge] (1) at (0,0) {};
    \node[flavour] (2) at (0,1) {};
    \draw (1)--(2);
    \node at (0.7,0) {$SU(3)$};
    \node at (0.5,1) {$6$};
    \end{tikzpicture}\,.
    \label{eq:SUexample}
\end{equation}
One can perform quiver subtraction on the quiver \eqref{eq:SUexample} to obtain the Hasse diagram of the Coulomb branch
\begin{equation}
    \HC\eqref{eq:SUexample}=\quad\begin{tikzpicture}
        \node[hasse] (1) at (0,0) {};
        \node[hasse] (2) at (0,2) {};
        \node[hasse] (3) at (0,4) {};
        
        \draw (1)--(2)--(3);
        
        \node at (0.5,1) {$D_4$};
        \node at (0.5,3) {$A_5$};
        \node at (-0.3,0) {$0$};
        \node at (-0.3,2) {$1$};
        \node at (-0.3,4) {$2$};
    \end{tikzpicture}\,.
\end{equation}
\eqref{eq:SUexample} has the 3d Mirror dual
\begin{equation}
    \begin{tikzpicture}
    \node[gauge] (1) at (-2,0) {};
    \node[gauge] (2) at (-1,0) {};
    \node[gauge] (3) at (0,0) {};
    \node[gauge] (4) at (1,0) {};
    \node[gauge] (5) at (2,0) {};
    \node[gauge] (6) at (-0.6,0.8) {};
    \node[gauge] (7) at (0.6,0.8) {};
    \draw (1)--(2)--(3)--(4)--(5) (6)--(3)--(7);
    \node at (-2,-0.5) {$1$};
    \node at (-1,-0.5) {$2$};
    \node at (0,-0.5) {$3$};
    \node at (1,-0.5) {$2$};
    \node at (2,-0.5) {$1$};
    \node at (-1,0.8) {$1$};
    \node at (1,0.8) {$1$};
    \end{tikzpicture}\,.
    \label{eq:SUexamplemirror}
\end{equation}
The Hasse diagram of the Higgs branch of \eqref{eq:SUexample} (Coulomb branch of the mirror \eqref{eq:SUexamplemirror}) is straight forward to obtain from quiver subtraction on the mirror quiver \eqref{eq:SUexamplemirror}
\begin{equation}
    \HH\eqref{eq:SUexample}=\HC\eqref{eq:SUexamplemirror}=\quad\begin{tikzpicture}
        \node[hasse] (1) at (0,0) {};
        \node[hasse] (2) at (0,2) {};
        \node[hasse] (3) at (0,4) {};
        
        \draw (1)--(2)--(3);
        
        \node at (0.5,1) {$a_5$};
        \node at (0.5,3) {$d_4$};
        \node at (-0.3,0) {$0$};
        \node at (-0.3,2) {$5$};
        \node at (-0.3,4) {$10$};
    \end{tikzpicture}\quad=\mathfrak{I}(\HC\eqref{eq:SUexample})\,.
\end{equation}
Both Hasse diagrams \eqref{eq:SUexample} and \eqref{eq:SUexamplemirror} are related by inversion.\\

The inversion procedure can be used to make predictions for the Higgs branch Hasse diagrams of good or ugly quivers with no known Lagrangian 3d Mirror dual. An example of such a quiver is
\begin{equation}
    \begin{tikzpicture}
    \node[gauge,label=below:{1}] (1) at (-1,0) {};
    \node[gauge,label=below:{2}] (2) at (0,0) {};
    \node (3) at (1,0) {$\cdots$};
    \node[gauge,label=below:{9}] (4) at (2,0) {};
    \node[gauge,label=below:{10}] (5) at (3,0) {};
    \node[gauge,label=below:{6}] (6) at (4,0) {};
    \node[gauge,label=below:{2}] (7) at (5,0) {};
    \node[gauge,label=right:{5}] (55) at (3,1) {};
    \draw (1)--(2)--(3)--(4)--(5)--(6)--(7) (5)--(55);
    \end{tikzpicture}
    \label{eq:classSmirror}
\end{equation}
The Coulomb branch Hasse diagram of \ref{eq:classSmirror} is readily computed from quiver subtraction
\begin{equation}
    \HC\eqref{eq:classSmirror}=\quad
    \begin{tikzpicture}
    \node[hasse] (1) at (0,3) {};
    \node[hasse] (2) at (0,2) {};
    \node[hasse] (3) at (0,1) {};
    \node[hasse] (4) at (0,0) {};
    \draw (1)--(2)--(3)--(4);
    \node at (0.4,2.5) {$e_8$};
    \node at (0.4,1.5) {$d_{10}$};
    \node at (0.4,0.5) {$d_{12}$};
    \end{tikzpicture}\;.
\end{equation}
its inversion is
\begin{equation}
    \mathcal{I}(\HC\eqref{eq:classSmirror})=\quad
    \begin{tikzpicture}
    \node[hasse] (1) at (0,3) {};
    \node[hasse] (2) at (0,2) {};
    \node[hasse] (3) at (0,1) {};
    \node[hasse] (4) at (0,0) {};
    \draw (1)--(2)--(3)--(4);
    \node at (0.4,2.5) {$D_{12}$};
    \node at (0.4,1.5) {$D_{10}$};
    \node at (0.4,0.5) {$E_{8}$};
    \end{tikzpicture}
    =\HH\eqref{eq:classSmirror}\;,
\end{equation}
where the equality to the right of the Hasse diagram is conjectured. The quiver \eqref{eq:classSmirror} can be obtained form a brane system \cite{Cabrera:2019izd} with an NS5 brane on an $O8^-$ plane. For this brane system there is a good control over the Kraft-Procesi transitions with simply-laced magnetic quivers, whose Higgs branch can be computed in contrast to non-simply laced quivers, hence the inversion property is to be expected. This result remains to be checked with the methods of \cite{2016arXiv160200164B}. In Section \ref{Bad} we see yet another application of inversion, in the case of a bad quiver with a known Higgs branch Hasse diagram.\\

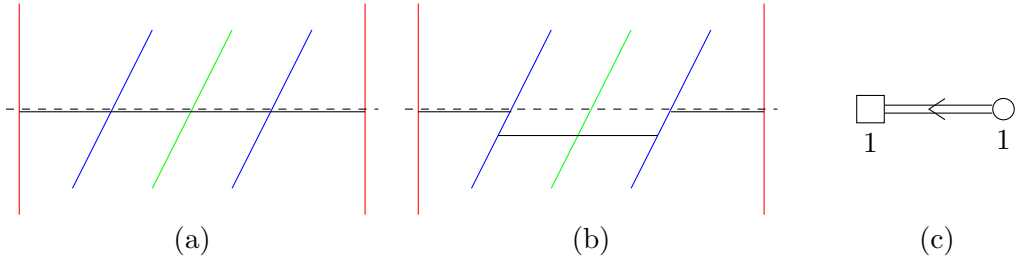
\begin{figure}[h]
    \centering
    \begin{tikzpicture}[scale=0.35]
    \draw[red] (0,-4)--(0,4) (13,-4)--(13,4);
    \draw[blue] (2,-3)--(5,3) (8,-3)--(11,3);
    \draw[green] (5,-3)--(8,3);
    \draw[dashed] (-0.5,0)--(13.5,0);
    \draw (0,-0.1)--(13,-0.1);
    \begin{scope}[shift={(15,0)}]
    \draw[red] (0,-4)--(0,4) (13,-4)--(13,4);
    \draw[blue] (2,-3)--(5,3) (8,-3)--(11,3);
    \draw[green] (5,-3)--(8,3);
    \draw[dashed] (-0.5,0)--(13.5,0);
    \draw (0,-0.1)--(3.5,-0.1) (3,-1)--(9,-1) (9.5,-0.1)--(13,-0.1); 
    \end{scope}
    \begin{scope}[shift={(32,0)}]
    \node[flavour,label=below:{$1$}] (1) at (0,0) {};
    \node[gauge,label=below:{$1$}] (2) at (5,0) {};
    \draw (2.8,0.4)--(2.2,0)--(2.8,-0.4);
    \draw[transform canvas={yshift=-1.5pt}] (1)--(2);
    \draw[transform canvas={yshift=1.5pt}] (1)--(2);
    \end{scope}
    \node at (6.5,-5) {(a)};
    \node at (21.5,-5) {(b)};
    \node at (34.5,-5) {(c)};
    \end{tikzpicture}
    \caption{Brane set up for $O(1)$ with 1 flavour. The green line represents an $O5^+$ plane. The moduli space of $O(1)$ with 1 flavour consists of only its Higgs branch. The Higgs branch consists of two leaves depicted in the brane constructions (a) and (b). In (c) the magnetic quiver is presented, its Coulomb branch is $\mathbb{C}^2/\mathbb{Z}_2$}
    \label{fig:O1N1}
\end{figure}

Finally we can give an example for which the Coulomb branch and Higgs branch Hasse diagram are not related by inversion. Take
\begin{equation}
    \begin{tikzpicture}
    \node[gauge,label=right:{$O(1)$}] (1) at (0,0) {};
    \node[flavour,label=right:{$1$}] (2) at (0,1) {};
    \draw (1)--(2);
    \end{tikzpicture}
    \label{O1N1quiv}
\end{equation}
the Higgs branch of this theory is $A_1=\mathbb{C}^2/\mathbb{Z}_2$\footnote{This space should really be called $c_1$, as the Higgs branch of $O(1)$ with N flavours is $c_N$.}, the Coulomb branch of this theory is trivial (Figure \ref{fig:O1N1}). The relevant Hasse diagrams are:

\begin{equation}
    \HH(\eqref{O1N1quiv})=\quad\begin{tikzpicture}
    \node[hasse] (1) at (0,0) {};
    \node[hasse] (2) at (0,1) {};
    \draw (1)--(2);
    \node at (0.4,0.5) {$A_1$};
    \end{tikzpicture}\neq\mathcal{I}(\HC(\eqref{O1N1quiv}))\;,\qquad\HC(\eqref{O1N1quiv})=\quad\begin{tikzpicture}
    \node[hasse] (1) at (0,0) {};
    \end{tikzpicture}\quad\neq\mathcal{I}(\HH(\eqref{O1N1quiv}))
\end{equation}

Inversion does not relate the Higgs and Coulomb branch Hasse diagrams of just any theory and a precise set of admissible quivers has to be found.

\section{The Hasse diagram of the full moduli space -- invertible}
\label{Moduli}
In this section we use Hasse diagrams to explore not only the Coulomb or Higgs branch of a theory, but its entire moduli space $\mathcal{M}$, including mixed branches. This enables us to understand all different phases of spontaneous symmetry breaking and the massless spectra of the theory at given points in the moduli space. Mixed branches as algebraic varieties have already been studied for $U(n)$ and $USp(2n)$ SQCD theories with fundamental matter in \cite{Assel:2017jgo} and \cite{Assel:2018exy} respectively. For an analysis of the Hilbert series of mixed branches of linear unitary quiver gauge theories, see \cite{Carta:2016fjb}.\\

\begin{figure}[h]
    \makebox[\textwidth][c]{
    \begin{tikzpicture}[scale=0.5]
        \draw[red] (-2,0)--(-2,4) (0,0)--(0,4) (4,0)--(4,4) (6,0)--(6,4);
        \draw[blue] (0.5,1)--(1.5,3) (2.5,1)--(3.5,3);
        \draw[dashed] (-3,2)--(7,2);
        \draw (-2,2)--(0,2) (0,2.1)--(4,2.1) (0,1.9)--(4,1.9) (4,2)--(6,2);
        \begin{scope}[shift={(-6,8)}]
        \draw[red] (-2,0)--(-2,4) (0,0)--(0,4) (4,0)--(4,4) (6,0)--(6,4);
        \draw[blue] (0.5,1)--(1.5,3) (2.5,1)--(3.5,3);
        \draw[dashed] (-3,2)--(7,2);
        \draw (0,2)--(4,2) (-2,0.5)--(0,0.5) (0,0.3)--(4,0.3) (4,0.5)--(6,0.5);
        \end{scope}
        \begin{scope}[shift={(6,8)}]
        \draw[red] (-2,0)--(-2,4) (0,0)--(0,4) (4,0)--(4,4) (6,0)--(6,4);
        \draw[blue] (0.5,1)--(1.5,3) (2.5,1)--(3.5,3);
        \draw[dashed] (-3,2)--(7,2);
        \draw (0,2.1)--(1,2.1) (3,2.1)--(4,2.1) (0.75,1.5)--(2.75,1.5) (-2,1.9)--(6,1.9);
        \end{scope}
        \begin{scope}[shift={(-12,16)}]
        \draw[red] (-2,0)--(-2,4) (0,0)--(0,4) (4,0)--(4,4) (6,0)--(6,4);
        \draw[blue] (0.5,1)--(1.5,3) (2.5,1)--(3.5,3);
        \draw[dashed] (-3,2)--(7,2);
        \draw (0,0.7)--(4,0.7) (-2,0.5)--(0,0.5) (0,0.3)--(4,0.3) (4,0.5)--(6,0.5);
        \end{scope}
        \begin{scope}[shift={(0,16)}]
        \draw[red] (-2,0)--(-2,4) (0,0)--(0,4) (4,0)--(4,4) (6,0)--(6,4);
        \draw[blue] (0.5,1)--(1.5,3) (2.5,1)--(3.5,3);
        \draw[dashed] (-3,2)--(7,2);
        \draw (0,2)--(1,2) (3,2)--(4,2) (0.75,1.5)--(2.75,1.5) (-2,0.5)--(0,0.5) (0,0.3)--(4,0.3) (4,0.5)--(6,0.5);
        \end{scope}
        \begin{scope}[shift={(12,16)}]
        \draw[red] (-2,0)--(-2,4) (0,0)--(0,4) (4,0)--(4,4) (6,0)--(6,4);
        \draw[blue] (0.5,1)--(1.5,3) (2.5,1)--(3.5,3);
        \draw[dashed] (-3,2)--(7,2);
        \draw (-2,1.9)--(1,1.9) (3,1.9)--(6,1.9) (0,2.1)--(1,2.1) (3,2.1)--(4,2.1) (0.75,1.5)--(2.75,1.5) (0.75-0.2,1.5-0.4)--(2.75-0.2,1.5-0.4);
        \end{scope}
        \draw (-1,5)--(-3,7) (5,5)--(7,7) (-7,13)--(-9,15) (-1,13)--(1,15) (5,13)--(3,15) (9,13)--(11,15);
        \node at (-3,6) {\color{red}$a_3$};
        \node at (7,6) {\color{blue}$A_1$};
        \node at (-9,14) {\color{red}$a_1$};
        \node at (-1,14) {\color{blue}$A_1$};
        \node at (5,14) {\color{red}$a_3$};
        \node at (11,14) {\color{blue}$A_3$};
    \end{tikzpicture}}
    \caption{Brane configurations corresponding to the leaves in \eqref{eq:nminA3hasse}. In \cite{Cabrera:2016vvv} Hasse diagrams were computed moving along the top row. The commutativity of red and blue lines in the full Hasse diagram allows to construct the Hasse diagram of a single branch in this way. See the appendix \ref{app:O} for a discussion of how to get moduli spaces and Hasse diagrams from brane constructions}
    \label{fig:nminA3branes}
\end{figure}
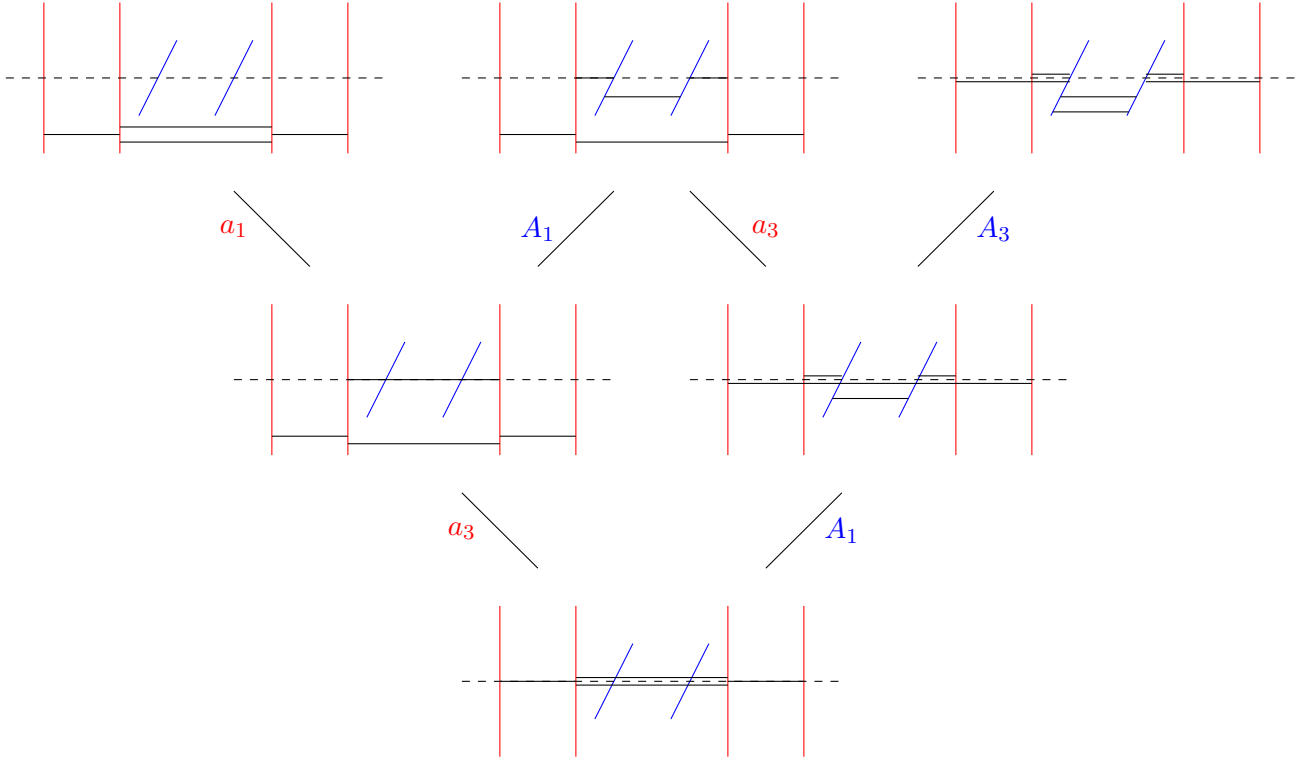

Moving from one leaf to a lower leaf, i.e. to a degeneration, on any branch physically implies the appearance of extra massless states. Generically these extra massless states open up transverse directions\footnote{This may not always be the case, see for example the discussion on $O(k)$ gauge theories.} into another branch. For example moving downwards in the $\HC$ more hypermultiplets become massless and their scalars are able to obtain vevs parametrising a transverse slice to the appropriate mixed branch. Reaching the origin all the hypers are massless and their possible vevs parametrise the full Higgs branch. Equivalently, moving downwards in $\HH$, more vector multiplets become massless providing Coulomb branch directions. We can draw the Hasse diagram of the full moduli space and use {\color{red}red} for transverse slices which are {\color{red}Coulomb branch directions} and {\color{blue}blue} for those that are {\color{blue}Higgs branch directions}, i.e. specifying what part of the R-symmetry acts on the transverse slices. For example the moduli space of SQED with N flavours has the Hasse diagram:
\begin{equation}
    \begin{tikzpicture}
        \node[hasse] (1) at (0,0) {};
        \node[hasse] (2) at (-1,2) {};
        \node[hasse] (3) at (1,2) {};
        
        \draw[red] (2)--(1);
        \draw[blue] (1)--(3);
        
        \node at (-1,1) {$A_{N-1}$};
        \node at (1,1) {$a_{N-1}$};
        \node at (-0.3,0) {\color{red}$0$};
        \node at (+0.3,0) {\color{blue}$0$};
        \node at (-1-0.3,2) {\color{red}$1$};
        \node at (-1+0.3,2) {\color{blue}$0$};
        \node at (1+0.7,2) {\color{blue}$N-1$};
        \node at (1-0.3,2) {\color{red}$0$};
    \end{tikzpicture}\quad\begin{tikzpicture}
    \node at (-1,0) {for};
        \node[gauge] (1) at (0,0) {};
        \node[flavour] (4) at (0,1) {};
        \draw(1)--(4);
        
        \node at (0.5,0) {1};
        \node at (0.5,1) {N};
    \end{tikzpicture}\,.
\end{equation}
The dimension of a Coulomb branch part of a leaf is given in red, while blue is used for the Higgs branch. The Coulomb branch dimension is the number of abelian vector multiplets that do not couple to any other massless fields, the Higgs branch dimension is the number of free hypermultiplets respectively. The Hasse diagram for \eqref{eq:3leafexample} is
\begin{equation}
\begin{tikzpicture}
    \node[hasse] (1) at (0,0) {};
    \node[hasse] (2) at (-1,2) {};
    \node[hasse] (3) at (-2,4) {};
    \node[hasse] (4) at (1,2) {};
    \node[hasse] (5) at (2,4) {};
    \node[hasse] (6) at (0,4) {};
    
    \draw[red] (1)--(2)--(3) (4)--(6);
    \draw[blue] (1)--(4)--(5) (2)--(6);
    
    \node at (-1,1) {$a_3$};
    \node at (-2,3) {$a_1$};
    \node at (1,1) {$A_1$};
    \node at (2,3) {$A_3$};
    \node at (-1,3) {$A_1$};
    \node at (1,3) {$a_3$};
    \node at (-0.3,0) {\color{red}$0$};
    \node at (0.3,0) {\color{blue}$0$};
    \node at (-1-0.3,2) {\color{red}$3$};
    \node at (-1+0.3,2) {\color{blue}$0$};
    \node at (1-0.3,2) {\color{red}$0$};
    \node at (1+0.3,2) {\color{blue}$1$};
    \node at (2-0.3,4) {\color{red}$0$};
    \node at (2+0.3,4) {\color{blue}$2$};
    \node at (0-0.3,4) {\color{red}$3$};
    \node at (0+0.3,4) {\color{blue}$1$};
    \node at (-2+0.3,4) {\color{blue}$0$};
    \node at (-2-0.3,4) {\color{red}$4$};
\end{tikzpicture}\quad\begin{tikzpicture}
    \node at (-2,0) {for};
    \node at (0,-2) {};
        \node[gauge] (1) at (0,0) {};
        \node[gauge] (2) at (-1,0) {};
        \node[gauge] (3) at (1,0) {};
        \node[flavour] (4) at (0,1) {};
        \draw(2)--(1)--(3) (1)--(4);
        
        \node at (-1,-0.5) {1};
        \node at (0,-0.5) {2};
        \node at (1,-0.5) {1};
        \node at (0.5,1) {2};
    \end{tikzpicture}\,.
\label{eq:nminA3hasse}
\end{equation}
The dimension of the mixed branch is ${\color{red}3}+{\color{blue}1}=4$. The brane configuration for each leaf in \eqref{eq:nminA3hasse} are given in Figure \ref{fig:nminA3branes}. We can identify the Hasse diagrams of the various branches (the closures of the three biggest leaves), and also see how they intersect:
\begin{equation}
\begin{tikzpicture}
    \node[hasse] (1) at (0,0) {};
    \node[hasse] (2) at (-1,2) {};
    \node[hasse] (3) at (-2,4) {};
    \node[hasse] (4) at (1,2) {};
    \node[hasse] (5) at (2,4) {};
    \node[hasse] (6) at (0,4) {};
    
    \draw[red] (1)--(2)--(3) (4)--(6);
    \draw[blue] (1)--(4)--(5) (2)--(6);
    
    \node at (-1,1) {$a_3$};
    \node at (-2,3) {$a_1$};
    \node at (1,1) {$A_1$};
    \node at (2,3) {$A_3$};
    \node at (-1,3) {$A_1$};
    \node at (1,3) {$a_3$};
    \draw[thick, rounded corners] (-0.5,-0.5) -- (-3.5, 4.5) -- (-1, 4.5) -- (0.5, -0.5) --  cycle;
    \node at (-2,4.7) {$\mathcal{C}$};
    \draw[thick, rounded corners] (0.5,-0.7) -- (3.5, 4.5) -- (1, 4.5) -- (-0.7, -0.7) --  cycle;
    \node at (2,4.7) {$\mathcal{H}$};
    \draw[thick, rounded corners] (0,4.5) -- (2.5,2) -- (1, -1) -- (-1, -1) -- (-2.5, 2) --  cycle;
    \node at (0,4.7) {$\mathrm{mixed}$};
\end{tikzpicture}\quad\begin{tikzpicture}
    \node at (0,-2) {};
    \node at (1,1) {$\mathcal{M}=\mathcal{C}\cup\mathrm{mixed}\cup\mathcal{H}$};
    \end{tikzpicture}
\end{equation}
Following the reasoning of \cite{Gaiotto:2008ak} the moduli space of this theory can be expressed in the form
\begin{equation}
    \mathcal{M}\eqref{eq:3leafexample}=\bigcup_{\alpha=0}^2\mathcal{C}_{2-\alpha}\times\mathcal{H}_{\alpha}\,,
\end{equation}
where $\mathcal{C}_2=\mathcal{C}\eqref{eq:3leafexample}$, $\mathcal{C}_1=a_3$, $\mathcal{C}_0=\{0\}$ and $\mathcal{H}_2=\mathcal{H}\eqref{eq:3leafexample}$, $\mathcal{H}_1=A_1$, $\mathcal{H}_0=\{0\}$. The associated Hasse diagrams are
\begin{equation}
    \begin{tikzpicture}
        \node at (0,0) {$\mathfrak{h}_{\mathcal{C}_2}=$};
        \node[hasse] (1) at (1,-1) {};
        \node[hasse] (2) at (1,0) {};
        \node[hasse] (3) at (1,1) {};
        \draw[red] (1)--(2)--(3); \node at (1.5,-0.5) {$a_3$}; \node at (1.5,0.5) {$a_1$};
        \node at (3,0) {$\mathfrak{h}_{\mathcal{C}_1}=$};
        \node[hasse] (7) at (4,-0.5) {};
        \node[hasse] (8) at (4,0.5) {};
        \draw[red] (7)--(8); \node at (4.5,0) {$a_3$};
        \node at (6,0) {$\mathfrak{h}_{\mathcal{C}_0}=$};
        \node[hasse] at (7,0) {};
        \node at (0,-2) {$\mathfrak{h}_{\mathcal{H}_0}=$};
        \node[hasse] at (1,-2) {};
        \node at (3,-2) {$\mathfrak{h}_{\mathcal{H}_1}=$};
        \node[hasse] (9) at (4,-2.5) {};
        \node[hasse] (10) at (4,-1.5) {};
        \draw[blue] (9)--(10); \node at (4.5,-2) {$A_1$};
        \node at (6,-2) {$\mathfrak{h}_{\mathcal{H}_2}=$};
        \node[hasse] (4) at (7,-3) {};
        \node[hasse] (5) at (7,-2) {};
        \node[hasse] (6) at (7,-1) {};
        \draw[blue] (4)--(5)--(6); \node at (7.5,-2.5) {$A_1$}; \node at (7.5,-1.5) {$A_3$};
    \end{tikzpicture}\;.
\end{equation}
The entire moduli space of \eqref{eq:3leafexample} consists of: its Coulomb branch; a mixed branch which is the product of the Coulomb branch of $[1]-(1)-(1)-(1)-[1]$ and the Higgs branch of $(1)-[2]$; and its Higgs branch. In particular every operator on the mixed branch is a product of gauge invariant operators from the Higgs branch and gauge invariant operators from the Coulomb branch.\\

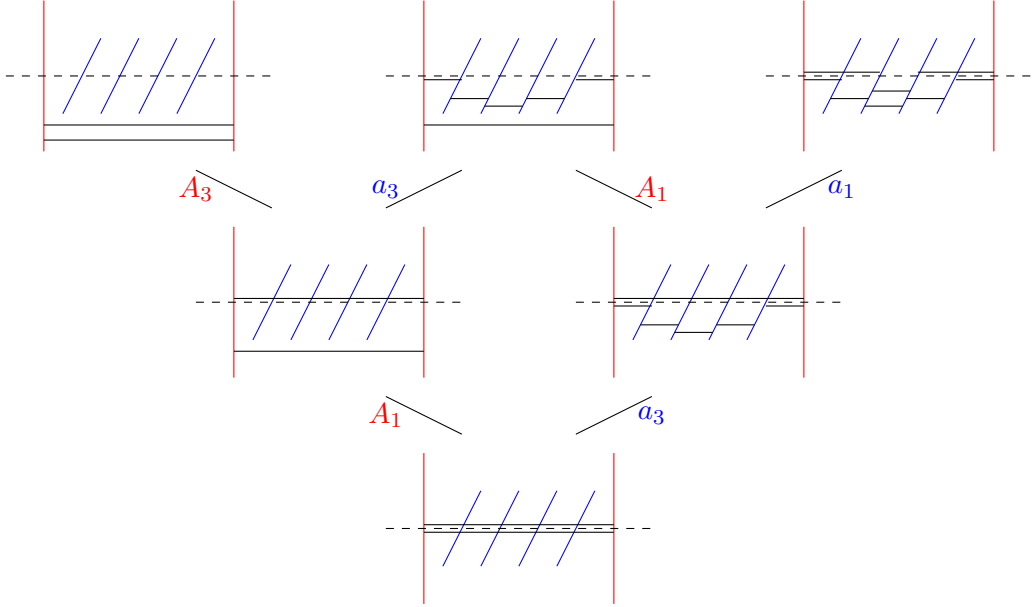
\begin{figure}[h]
    \makebox[\textwidth][c]{
    \begin{tikzpicture}[scale=0.5]
    \draw[red] (0,0)--(0,4) (5,0)--(5,4);
    \draw[blue] (0.5,1)--(1.5,3) (1.5,1)--(2.5,3) (2.5,1)--(3.5,3) (3.5,1)--(4.5,3);
    \draw[dashed] (-1,2)--(6,2);
    \draw (0,1.9)--(5,1.9) (0,2.1)--(5,2.1);
    \begin{scope}[shift={(-5,6)}]
    \draw[red] (0,0)--(0,4) (5,0)--(5,4);
    \draw[blue] (0.5,1)--(1.5,3) (1.5,1)--(2.5,3) (2.5,1)--(3.5,3) (3.5,1)--(4.5,3);
    \draw[dashed] (-1,2)--(6,2);
    \draw (0,0.7)--(5,0.7) (0,2.1)--(5,2.1);
    \end{scope}
    \begin{scope}[shift={(-10,12)}]
    \draw[red] (0,0)--(0,4) (5,0)--(5,4);
    \draw[blue] (0.5,1)--(1.5,3) (1.5,1)--(2.5,3) (2.5,1)--(3.5,3) (3.5,1)--(4.5,3);
    \draw[dashed] (-1,2)--(6,2);
    \draw (0,0.3)--(5,0.3) (0,0.7)--(5,0.7);
    \end{scope}
    \begin{scope}[shift={(0,12)}]
    \draw[red] (0,0)--(0,4) (5,0)--(5,4);
    \draw[blue] (0.5,1)--(1.5,3) (1.5,1)--(2.5,3) (2.5,1)--(3.5,3) (3.5,1)--(4.5,3);
    \draw[dashed] (-1,2)--(6,2);
    \draw (0,1.9)--(1,1.9) (4,1.9)--(5,1.9) (0,0.7)--(5,0.7) (0.7,1.4)--(1.7,1.4) (1.6,1.2)--(2.6,1.2) (2.7,1.4)--(3.7,1.4);
    \end{scope}
    \begin{scope}[shift={(5,6)}]
    \draw[red] (0,0)--(0,4) (5,0)--(5,4);
    \draw[blue] (0.5,1)--(1.5,3) (1.5,1)--(2.5,3) (2.5,1)--(3.5,3) (3.5,1)--(4.5,3);
    \draw[dashed] (-1,2)--(6,2);
    \draw (0,1.9)--(1,1.9) (4,1.9)--(5,1.9) (0,2.1)--(5,2.1) (0.7,1.4)--(1.7,1.4) (1.6,1.2)--(2.6,1.2) (2.7,1.4)--(3.7,1.4);
    \end{scope}
    \begin{scope}[shift={(10,12)}]
    \draw[red] (0,0)--(0,4) (5,0)--(5,4);
    \draw[blue] (0.5,1)--(1.5,3) (1.5,1)--(2.5,3) (2.5,1)--(3.5,3) (3.5,1)--(4.5,3);
    \draw[dashed] (-1,2)--(6,2);
    \draw (0,1.9)--(1,1.9) (4,1.9)--(5,1.9) (0,2.1)--(2,2.1) (3,2.1)--(5,2.1) (0.7,1.4)--(1.7,1.4) (1.6,1.2)--(2.6,1.2) (2.7,1.4)--(3.7,1.4) (1.8,1.6)--(2.8,1.6);
    \end{scope}
    \draw (1,4.5)--(-1,5.5) (4,4.5)--(6,5.5) (-4,10.5)--(-6,11.5) (-1,10.5)--(1,11.5) (4,11.5)--(6,10.5) (9,10.5)--(11,11.5);
    \node at (-1,5) {\color{red}$A_1$};
    \node at (-6,11) {\color{red}$A_3$};
    \node at (6,11) {\color{red}$A_1$};
    \node at (6,5) {\color{blue}$a_3$};
    \node at (-1,11) {\color{blue}$a_3$};
    \node at (11,11) {\color{blue}$a_1$};
    \end{tikzpicture}}
    \caption{Brane configurations corresponding to the leaves in \eqref{eq:nminA3dualhasse}. See Appendix \ref{app:O} for more information.}
    \label{fig:nminA3dualbranes}
\end{figure}

Note that 3d Mirror symmetry exchanges the red and blue colours in \eqref{eq:3leafexample}, hence the Hasse diagram of the full moduli space of \eqref{eq:3leafexampledual} is (the diagram is reflected along a vertical axis to keep blue on the right and red on the left.):
\begin{equation}
    \begin{tikzpicture}
    \node[hasse] (1) at (0,0) {};
    \node[hasse] (2) at (-1,2) {};
    \node[hasse] (3) at (-2,4) {};
    \node[hasse] (4) at (1,2) {};
    \node[hasse] (5) at (2,4) {};
    \node[hasse] (6) at (0,4) {};
    
    \draw[red] (1)--(2)--(3) (4)--(6);
    \draw[blue] (1)--(4)--(5) (2)--(6);
    
    \node at (-1,1) {$A_1$};
    \node at (-2,3) {$A_3$};
    \node at (1,1) {$a_3$};
    \node at (2,3) {$a_1$};
    \node at (-1,3) {$a_3$};
    \node at (1,3) {$A_1$};
    \node at (-0.3,0) {\color{red}$0$};
    \node at (0.3,0) {\color{blue}$0$};
    \node at (-1-0.3,2) {\color{red}$1$};
    \node at (-1+0.3,2) {\color{blue}$0$};
    \node at (1-0.3,2) {\color{red}$0$};
    \node at (1+0.3,2) {\color{blue}$3$};
    \node at (2-0.3,4) {\color{red}$0$};
    \node at (2+0.3,4) {\color{blue}$4$};
    \node at (0-0.3,4) {\color{red}$1$};
    \node at (0+0.3,4) {\color{blue}$3$};
    \node at (-2+0.3,4) {\color{blue}$0$};
    \node at (-2-0.3,4) {\color{red}$2$};
\end{tikzpicture}\quad\begin{tikzpicture}
    \node at (-1,0) {for};
    \node at (0,-2) {};
        \node[gauge] (1) at (0,0) {};
        \node[flavour] (4) at (0,1) {};
        \draw(1)--(4);
        
        \node at (0.5,0) {2};
        \node at (0.5,1) {4};
    \end{tikzpicture}\quad\begin{tikzpicture}
    \node at (-2.5,0) {3d MS};
    \draw[<->] (-3.2,-0.3)--(-1.7,-0.3);
    \node at (0,-2) {};
        \node[gauge] (1) at (0,0) {};
        \node[gauge] (2) at (-1,0) {};
        \node[gauge] (3) at (1,0) {};
        \node[flavour] (4) at (0,1) {};
        \draw(2)--(1)--(3) (1)--(4);
        
        \node at (-1,-0.5) {1};
        \node at (0,-0.5) {2};
        \node at (1,-0.5) {1};
        \node at (0.5,1) {2};
    \end{tikzpicture}\,.
\label{eq:nminA3dualhasse}
\end{equation}
this agrees with the described structure of mixed branches in \cite{Assel:2017jgo}. The brane configurations for all leaves in \eqref{eq:nminA3dualhasse} are depicted in Figure \ref{fig:nminA3dualbranes}.

As a second example consider the quiver
\begin{equation}
    \begin{tikzpicture}
        \node[gauge] (1) at (0,0) {};
        \node[gauge] (2) at (1,0) {};
        \node[gauge] (3) at (2,0) {};
        \node[flavour] (4) at (0,1) {};
        \node[flavour] (5) at (2,1) {};
        
        \draw (1)--(2)--(3) (1)--(4) (3)--(5);
        
        \node at (0,-0.5) {2};
        \node at (1,-0.5) {2};
        \node at (2,-0.5) {2};
        \node at (-0.5,1) {2};
        \node at (2.5,1) {3};
    \end{tikzpicture}\,.
    \label{eq:secondquiver}
\end{equation}
The Hasse diagram of the Coulomb branch and Higgs branch of \eqref{eq:secondquiver} are
\begin{equation}
    \HC=\begin{tikzpicture}
    \node[hasse] (0) at (0,-1) {};
    \node[hasse] (1) at (0,0) {};
    \node[hasse] (2) at (0,1) {};
    \node[hasse] (3) at (0,2) {};
    \node[hasse] (4) at (-0.5,3) {};
    \node[hasse] (5) at (0.5,4) {};
    \node[hasse] (6) at (0,5) {};
    
    \draw[red] (0)--(1)--(2)--(3)--(4)--(6) (3)--(5)--(6);
    
    \node at (-0.5,-0.5) {$a_2$};
    \node at (-0.5,0.5) {$A_1$};
    \node at (-0.5,1.5) {$A_1$};
    \node at (-1,2.5) {$A_1$};
    \node at (1,3) {$A_2$};
    \node at (-1,4) {$A_2$};
    \node at (1,4.5) {$A_1$};
    \node at (-0.3,-1) {\color{red}$0$};
    \node at (-0.3,0) {\color{red}$2$};
    \node at (-0.3,1) {\color{red}$3$};
    \node at (-0.3,2) {\color{red}$4$};
    \node at (-0.5-0.3,3) {\color{red}$5$};
    \node at (0.5-0.3,4) {\color{red}$5$};
    \node at (-0.3,5) {\color{red}$6$};
    \end{tikzpicture}\qquad , \qquad
    \HH=\mathfrak{I}(\HC)=\begin{tikzpicture}
    \node[hasse] (7) at (0,0) {};
    \node[hasse] (8) at (0.5,1) {};
    \node[hasse] (9) at (-0.5,2) {};
    \node[hasse] (10) at (0,3) {};
    \node[hasse] (11) at (0,4) {};
    \node[hasse] (12) at (0,5) {};
    \node[hasse] (13) at (0,6) {};
    
    \draw[blue] (7)--(8)--(10) (7)--(9)--(10)--(11)--(12)--(13);
    
    \node at (1,0.5) {$a_1$};
    \node at (-1,1) {$a_2$};
    \node at (1,2) {$a_2$};
    \node at (-1,2.5) {$a_1$};
    \node at (0.5,3.5) {$a_1$};
    \node at (0.5,4.5) {$a_1$};
    \node at (0.5,5.5) {$A_2$};
    \node at (0.3,0) {\color{blue}$0$};
    \node at (0.5+0.3,1) {\color{blue}$1$};
    \node at (-0.5+0.3,2) {\color{blue}$2$};
    \node at (0.3,3) {\color{blue}$3$};
    \node at (0.3,4) {\color{blue}$4$};
    \node at (0.3,5) {\color{blue}$5$};
    \node at (0.3,6) {\color{blue}$6$};
    \end{tikzpicture}\,,
    \label{eq:second_quiver_hasses}
\end{equation}
and the Hasse diagram for the full moduli space is
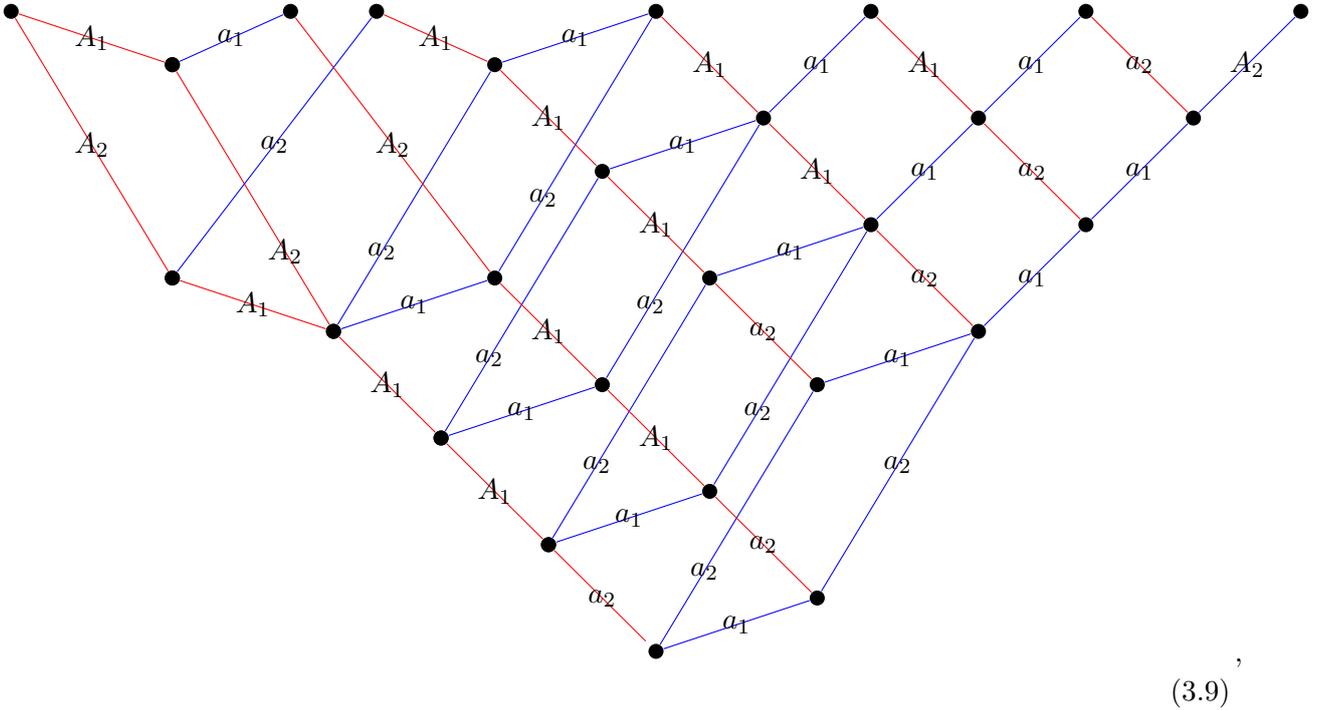
\begin{equation}
    \makebox[\textwidth][c]{\begin{tikzpicture}[scale=2]
% nodes to anchor
\begin{scope}[rotate=45]
\node (100) at (0,1) {};
\node (0) at (0,0) {};
\end{scope}
\begin{scope}[shift={(100)},rotate=45]
    \node[hasse] (1) at (0,0) {};
    \node[hasse] (2) at (0,1) {};
    \node[hasse] (3) at (0,2) {};
    \node[hasse] (4) at (-0.5,3) {};
    \node[hasse] (5) at (0.5,4) {};
    \node[hasse] (6) at (0,5) {};
    
    \draw[red] (0)--(1)--(2)--(3)--(4)--(6) (3)--(5)--(6);
    
    \node at ($(0)!0.5!(1)$) {$a_2$};
    \node at ($(1)!0.5!(2)$) {$A_1$};
    \node at ($(2)!0.5!(3)$) {$A_1$};
    \node at ($(3)!0.5!(4)$) {$A_1$};
    \node at ($(3)!0.3!(5)$) {$A_2$};
    \node at ($(4)!0.5!(6)$) {$A_2$};
    \node at ($(5)!0.5!(6)$) {$A_1$};
\end{scope}
\begin{scope}[rotate=-45]
    \node[hasse] (101) at (0,0) {};
    \node[hasse] (102) at (0.5,1) {};
    \node[hasse] (103) at (-0.5,2) {};
    \node[hasse] (104) at (0,3) {};
    \node[hasse] (105) at (0,4) {};
    \node[hasse] (106) at (0,5) {};
    \node[hasse] (107) at (0,6) {};
    
    \draw[blue] (101)--(102)--(104) (101)--(103)--(104)--(105)--(106)--(107);
    
    \node at ($(101)!0.5!(102)$) {$a_1$};
    \node at ($(101)!0.3!(103)$) {$a_2$};
    \node at ($(102)!0.5!(104)$) {$a_2$};
    \node at ($(103)!0.5!(104)$) {$a_1$};
    \node at ($(104)!0.5!(105)$) {$a_1$};
    \node at ($(105)!0.5!(106)$) {$a_1$};
    \node at ($(106)!0.5!(107)$) {$A_2$};
\end{scope}
\begin{scope}[shift={(100)},rotate=-45]
    \node[hasse] (7) at (0,0) {};
    \node[hasse] (8) at (0.5,1) {};
    \node[hasse] (9) at (-0.5,2) {};
    \node[hasse] (10) at (0,3) {};
    \node[hasse] (11) at (0,4) {};
    \node[hasse] (12) at (0,5) {};
    
    \draw[blue] (7)--(8)--(10) (7)--(9)--(10)--(11)--(12);
    
    \node at ($(7)!0.5!(8)$) {$a_1$};
    \node at ($(7)!0.3!(9)$) {$a_2$};
    \node at ($(8)!0.3!(10)$) {$a_2$};
    \node at ($(9)!0.5!(10)$) {$a_1$};
    \node at ($(10)!0.5!(11)$) {$a_1$};
    \node at ($(11)!0.5!(12)$) {$a_1$};
\end{scope}
\begin{scope}[shift={(2)},rotate=-45]
    \node[hasse] (13) at (0,0) {};
    \node[hasse] (14) at (0.5,1) {};
    \node[hasse] (15) at (-0.5,2) {};
    \node[hasse] (16) at (0,3) {};
    \node[hasse] (17) at (0,4) {};
    
    \draw[blue] (13)--(14)--(16) (13)--(15)--(16)--(17);
    
    \node at ($(13)!0.5!(14)$) {$a_1$};
    \node at ($(13)!0.3!(15)$) {$a_2$};
    \node at ($(14)!0.3!(16)$) {$a_2$};
    \node at ($(15)!0.5!(16)$) {$a_1$};
    \node at ($(16)!0.5!(17)$) {$a_1$};
\end{scope}
\begin{scope}[shift={(3)},rotate=-45]
    \node[hasse] (18) at (0,0) {};
    \node[hasse] (19) at (0.5,1) {};
    \node[hasse] (20) at (-0.5,2) {};
    \node[hasse] (21) at (0,3) {};
    
    \draw[blue] (18)--(19)--(21) (18)--(20)--(21);
    
    \node at ($(18)!0.5!(19)$) {$a_1$};
    \node at ($(18)!0.3!(20)$) {$a_2$};
    \node at ($(19)!0.3!(21)$) {$a_2$};
    \node at ($(20)!0.5!(21)$) {$a_1$};
\end{scope}
\begin{scope}[shift={(4)},rotate=-45]
    \node[hasse] (22) at (0,0) {};
    \node[hasse] (23) at (-0.3,2.2) {};
    
    \draw[blue] (22)--(23);
    
    \node at ($(22)!0.5!(23)$) {$a_2$};
\end{scope}
\begin{scope}[shift={(5)},rotate=-45]
    \node[hasse] (24) at (0,0) {};
    \node[hasse] (25) at (0.3,0.8) {};
    
    \draw[blue] (24)--(25);
    
    \node at ($(24)!0.5!(25)$) {$a_1$};
\end{scope}

\draw[red] (102)--(8)--(14)--(19)--(25) (103)--(9)--(15)--(20)--(23) (104)--(10)--(16)--(21) (105)--(11)--(17) (106)--(12);

\node at ($(102)!0.5!(8)$) {$a_2$}; %
\node at ($(8)!0.5!(14)$) {$A_1$};
\node at ($(14)!0.5!(19)$) {$A_1$};
\node at ($(19)!0.5!(25)$) {$A_2$};

\node at ($(103)!0.5!(9)$) {$a_2$}; %
\node at ($(9)!0.5!(15)$) {$A_1$};
\node at ($(15)!0.5!(20)$) {$A_1$};
\node at ($(20)!0.5!(23)$) {$A_1$};

\node at ($(104)!0.5!(10)$) {$a_2$}; %
\node at ($(10)!0.5!(16)$) {$A_1$};
\node at ($(16)!0.5!(21)$) {$A_1$};

\node at ($(105)!0.5!(11)$) {$a_2$}; %
\node at ($(11)!0.5!(17)$) {$A_1$};

\node at ($(106)!0.5!(12)$) {$a_2$}; %
    \end{tikzpicture}}\,,
    \label{eq:second_quiver_full_hasse}
\end{equation}
where the dimensions of the leaves are suppressed for aesthetic reasons. They can be obtained by adding the dimensions of the elementary slices. The algorithm for computing the Hasse diagram for the full moduli space from the Hasse diagram of the Coulomb branch (or Higgs branch) is as follows:
\begin{enumerate}
    \item Draw in red: The Hasse diagram of $\HC$.
    \item Draw in blue: For every node in $\HC$ add the inversion of the associated sub-diagram going from this node to the top. For the lowest node this gives $\HH$.
    \item Draw in red: For every node in $\HH$ add the inversion of the associated sub-diagram going to the top and link it up with the diagrams added in 2. such that Coulomb and Higgs (red and blue) directions commute.
\end{enumerate}
The same rules apply when exchanging Coulomb and red with Higgs and blue respectively in the above.\\

A lot of physical information is contained in the transverse slices in the Moduli space of the theory. An analysis of this is the content of the next section. 

\subsection{Symplectic Leaves, Transverse Slices and Transverse Space}
In the following we make a distinction between two different types of subspaces of the moduli space:
\begin{enumerate}
    \item The \emph{Transverse slice} $\mathsf{T}$ between two leaves $\mathcal{L}_1$ and $\mathcal{L}_2$ (where $\mathcal{L}_1\subset\bar{\mathcal{L}}_2$): $\mathsf{T}(\mathcal{L}_1,\mathcal{L}_2)$ This is the transverse slice to a leaf $\mathcal{L}_1$ inside the closure of a bigger leaf $\mathcal{L}_2$.
    \item The \emph{Transverse space} $\mathcal{T}$ of a leaf $\mathcal{L}$: $\mathcal{T}(\mathcal{L})$ is the transverse space to a single leaf $\mathcal{L}$ inside the entire moduli space $\mathcal{M}$. A transverse slice can be obtained from taking the intersection $\mathsf{T}(\mathcal{L}_1,\mathcal{L}_2)=\mathcal{T}(\mathcal{L}_1)\bigcap\bar{\mathcal{L}}_2$.
\end{enumerate}
Geometrically, we could allow $\mathcal{L}_2$ to be a union of leaves, however in the following we reserve the name transverse slice for a slice between two single leaves.\\

From the point of view of field theory, the two spaces have distinct meanings:
\begin{enumerate}
    \item Transverse slice $\mathsf{T}(\{0\},\mathcal{L})$ = moduli we have tuned
    \item Transverse space $\mathcal{T}(\mathcal{L})$ = moduli we have left to tune = moduli space of the Higgsed theory
    \item Transverse slice $\mathsf{T}(\mathcal{L}_1,\mathcal{L}_2)$ = moduli we need to tune in order to move from phase ``1" to phase ``2" of the theory. How to compute this from a brane construction is shown in the appendix \ref{app:O}.
\end{enumerate}

\begin{figure}[h]
\centering
\begin{tikzpicture}
    \begin{scope}[shift={(0,-6)}]
    \node[hasse] (1) at (0,0) {};
    \node[hasse] (2) at (-1,2) {};
    \node[hasse] (3) at (-2,4) {};
    \node[hasse] (4) at (1,2) {};
    \node[hasse] (5) at (2,4) {};
    \node[hasse] (6) at (0,4) {};
    
    \draw[red] (1)--(2)--(3) (4)--(6);
    \draw[blue] (1)--(4)--(5) (2)--(6);
    
    \node at (-1,1) {$a_3$};
    \node at (-2,3) {$a_1$};
    \node at (1,1) {$A_1$};
    \node at (2,3) {$A_3$};
    \node at (-1,3) {$A_1$};
    \node at (1,3) {$a_3$};
    \node[boxb] at (1) {};
    \node[boxg] at (4) {};
    \node[boxp] at (5) {};
    \node[boxo] at (2) {};
    \node[boxp] at (3) {};
    \node[boxp] at (6) {};
    \node at (-0.3,0) {\color{red}$0$};
    \node at (0.3,0) {\color{blue}$0$};
    \node at (-1-0.3,2) {\color{red}$3$};
    \node at (-1+0.3,2) {\color{blue}$0$};
    \node at (1-0.3,2) {\color{red}$0$};
    \node at (1+0.3,2) {\color{blue}$1$};
    \node at (2-0.3,4) {\color{red}$0$};
    \node at (2+0.3,4) {\color{blue}$2$};
    \node at (0-0.3,4) {\color{red}$3$};
    \node at (0+0.3,4) {\color{blue}$1$};
    \node at (-2+0.3,4) {\color{blue}$0$};
    \node at (-2-0.3,4) {\color{red}$4$};
    \end{scope}
    \begin{scope}[shift={(7,0)}]
        \node[hasse] (7) at (0,0) {};
        \node[hasse] (8) at (0,2) {};
        \node[hasse] (9) at (0,4) {};
        
        \draw[blue] (7)--(8)--(9);
        
        \node at (-0.5,1) {$A_1$};
        \node at (-0.5,3) {$A_3$};
        
        \draw[orange] (0.3,0)--(0.5,0)--(0.5,1.9)--(0.3,1.9);
        \draw[black!30!green] (0.3,2.1)--(0.5,2.1)--(0.5,4)--(0.3,4);
        \draw (-1.3,0)--(-1.5,0)--(-1.5,4)--(-1.3,4);
        \draw[purple] (-0.3,4.1)--(-0.5,4.1)--(-0.5,3.9)--(-0.3,3.9) (-0.3,-0.1)--(-0.5,-0.1)--(-0.5,0.1)--(-0.3,0.1) (-0.3,1.9)--(-0.5,1.9)--(-0.5,2.1)--(-0.3,2.1);
		\node at (-0.8,4) {$\{1\}$};
		\node at (-0.8,2) {$\{1\}$};
		\node at (-0.8,0) {$\{1\}$};
     \end{scope}
     \begin{scope}[shift={(9,0.5)}]
		\node[gauge] (10) at (0,0) {};
		\node[flavour] (11) at (0,1) {};
		\draw (10)--(11);
		\node at (-0.4,0) {$1$};
		\node at (-0.4,1) {$2$};
		\end{scope}
     \begin{scope}[shift={(10,2.75)}]
		\node[gauge] (12) at (0,0) {};
		\node[flavour] (13) at (0,1) {};
		\node[gauge] (14) at (-1,0) {};
		\node[gauge] (15) at (-2,0) {};
		\node[flavour] (16) at (-2,1) {};
		\draw (13)--(12)--(14)--(15)--(16);
		\node at (0,-0.4) {$1$};
		\node at (-1,-0.4) {$1$};
		\node at (-2,-0.4) {$1$};
		\node at (0.4,1) {$1$};
		\node at (-2.4,1) {$1$};
		\end{scope}
		\begin{scope}[shift={(1,0)}]
        \node[hasse] (27) at (0,0) {};
        \node[hasse] (28) at (0,2) {};
        \node[hasse] (29) at (0,4) {};
        
        \draw[red] (27)--(28)--(29);
        
        \node at (0.5,1) {$a_3$};
        \node at (0.5,3) {$a_1$};
        
        \draw[black!30!green] (-0.3,0)--(-0.5,0)--(-0.5,1.9)--(-0.3,1.9);
        \draw[orange] (-0.3,2.1)--(-0.5,2.1)--(-0.5,4)--(-0.3,4);
        \draw (1.3,0)--(1.5,0)--(1.5,4)--(1.3,4);
        \draw[purple] (0.3,4.1)--(0.5,4.1)--(0.5,3.9)--(0.3,3.9) (0.3,-0.1)--(0.5,-0.1)--(0.5,0.1)--(0.3,0.1) (0.3,1.9)--(0.5,1.9)--(0.5,2.1)--(0.3,2.1);
		\node at (0.8,4) {$\{1\}$};
		\node at (0.8,2) {$\{1\}$};
		\node at (0.8,0) {$\{1\}$};
     \end{scope}
		\begin{scope}[shift={(-1,2.75)}]
		\node[gauge] (110) at (0,0) {};
		\node[flavour] (111) at (0,1) {};
		\draw (110)--(111);
		\node at (-0.4,0) {$1$};
		\node at (-0.4,1) {$2$};
		\end{scope}
     \begin{scope}[shift={(0,0.5)}]
		\node[gauge] (112) at (0,0) {};
		\node[flavour] (113) at (0,1) {};
		\node[gauge] (114) at (-1,0) {};
		\node[gauge] (115) at (-2,0) {};
		\node[flavour] (116) at (-2,1) {};
		\draw (113)--(112)--(114)--(115)--(116);
		\node at (0,-0.4) {$1$};
		\node at (-1,-0.4) {$1$};
		\node at (-2,-0.4) {$1$};
		\node at (0.4,1) {$1$};
		\node at (-2.4,1) {$1$};
		\end{scope}
		\begin{scope}[shift={(5,1.5)}]
		\node[gauge] (117) at (0,0) {};
		\node[gauge] (118) at (-1,0) {};
		\node[gauge] (119) at (-2,0) {};
		\node[flavour] (120) at (-1,1) {};
		\draw (117)--(118)--(119) (118)--(120);
		\node at (0,-0.4) {$1$};
		\node at (-1,-0.4) {$2$};
		\node at (-2,-0.4) {$1$};
		\node at (-1.4,1) {$2$};
		\end{scope}
		\node at (-3,4) {a)};
		\node at (1,-0.5) {$\HC$};
		\node at (7,-0.5) {$\HH$};
		\node at (-3,-5) {b)};
		\node at (6,-5) {c)};
		\begin{scope}[shift={(8,-5)}]
		    \node[hasse] (200) at (0,0) {};
		    \node[hasse] (201) at (-1,2) {};
		    \node[hasse] (202) at (1,2) {};
		    
		    \draw[red] (200)--(201);
		    \draw[blue] (200)--(202);
		    \node at (-1,1) {$a_1$};
		    \node at (1,1) {$A_1$};
		    \node at (-0.3,0) {\color{red}0};
		    \node at (-1-0.3,2) {\color{red}1};
		    \node at (1-0.3,2) {\color{red}0};
		    \node at (0.3,0) {\color{blue}0};
		    \node at (-1+0.3,2) {\color{blue}0};
		    \node at (1+0.3,2) {\color{blue}1};
            \node[boxo] at (0,0) {};
		\end{scope}
    \end{tikzpicture}
    \caption{a) (left) The Hasse diagram of the Coulomb branch, which in principle contains all of the information, encoded in the transverse slices. a) (right) The Hasse diagram of the Higgs branch, which in principle contains all of the information, encoded in the transverse slices. All the trivial transverse slices (purple) correspond to points in the highest leaves of the entire moduli space. All the elementary slices (green, orange) correspond to the transverse spaces of minimal degenerations of the highest leaves in the Hasse diagram of the full moduli space. b) The Hasse diagram of the moduli space of the theory \eqref{eq:3leafexample}. The coloured boxes around the nodes are in bijection with the transverse slices in the Higgs branch as well as the Coulomb branch. All effective theories for this example can be read off the brane diagrams in Figure \ref{fig:nminA3branes}. c) As an example: The transverse space to the orange leaf inside the entire moduli space, the moduli space of U(1) with 2 Flavours.}
    \label{fig:slices}
\end{figure}
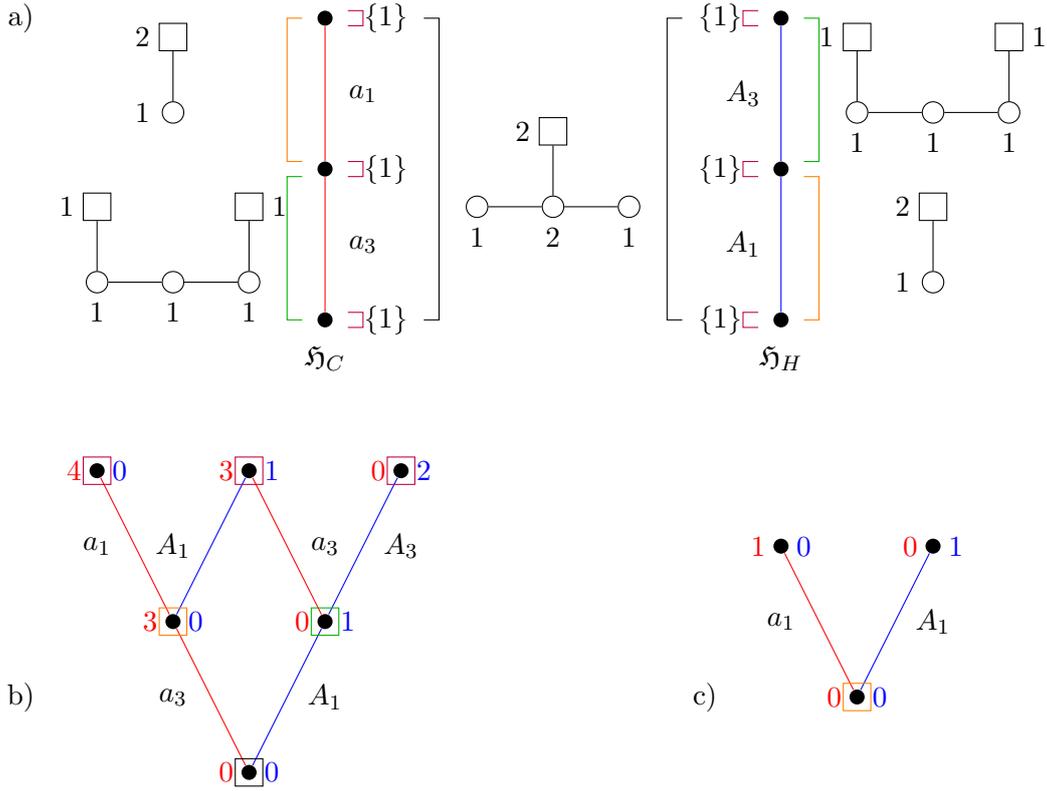

To each leaf in the moduli space of a theory (point in the Hasse diagram) there is a transverse space in the full moduli space (all lines and points higher in the Hasse diagram emanating from the selected point). This transverse space is the moduli space of a different theory, specifically a theory obtained from partial Higgsing along Coulomb, Higgs or mixed directions. To a leaf itself we can not associate a new theory, it only carries information about the massless spectrum of the original theory at this point in its moduli space, such as abelian vector multiplets that do not couple to other massless fields (Coulomb dimension of the leaf) and free neutral hypermultiplets (Higgs dimension of the leaf), the unbroken gauge group and the matter charged under it, on a general point on the leaf. The quiver representing the unbroken gauge group and the matter charged under it has as a moduli space the transverse space to the specified leaf. An example is given in Figure \ref{fig:slices}.

For the theories we study, i.e. with invertible associated Hasse diagrams, all this information is contained in the Coulomb/Higgs branch alone. In Figure \ref{fig:slices} each transverse slice in the Coulomb and Higgs branch respectively is associated to a Higgsable theory and its corresponding moduli space is identified in the Hasse diagram of the full moduli space of original theory. In general we can identify the following theories:\\

\noindent\underline{Transverse slices in the Coulomb branch:}
\begin{itemize}
    \item Slice from any point to the top: Coulomb branch of a new theory obtained from Higgsing along the Coulomb branch of the original theory.
    \item Slice from a point that is not the origin to a leaf that is not the top: Coulomb branch of a new theory obtained from Higgsing along a mixed branch of the original theory.
    \item Slice from the origin to a leaf that is not the top: Coulomb branch of a new theory obtained from Higgsing along the Higgs branch of the original theory.
\end{itemize}
\noindent\underline{Transverse slices in the Higgs branch:}
\begin{itemize}
    \item Slice from any point to the top: Higgs branch of a new theory obtained from Higgsing along the Higgs branch of the original theory.
    \item Slice from a point that is not the origin to a leaf that is not the top: Higgs branch of a new theory obtained from Higgsing along a mixed branch of the original theory.
    \item Slice from the origin to a leaf that is not the top: Higgs branch of a new theory obtained from Higgsing along the Coulomb branch of the original theory.
\end{itemize}

An analysis of effective theories was performed in the realm of nilpotent orbits and Slodowy slices in \cite{Rogers:2018dez,Hanany:2019tji}, their `descendants' or `Slodowy intersection' figures contain the information of the effective massless interacting theory on each leaf in the full moduli space, or in other words the theories associated to all transverse spaces inside the full moduli space. It should be noted, that our analysis is not restricted to nilpotent orbits, see for example \eqref{eq:secondquiver}-\eqref{eq:second_quiver_full_hasse}.\\

Since the Hasse diagram of Coulomb branch is computed using quiver subtraction, we discuss this operation in the following section.

\subsection{Quiver Subtraction}
We can now turn to the question what quivers we obtain from quiver subtraction. Quiver subtraction is an operation that lets one compute the Hasse diagram of the Coulomb branch of a good or ugly $3d$ $\mathcal{N}=4$ quiver gauge theory, see \cite{Cabrera:2018ann} and appendix A.2 of \cite{Bourget:2019aer}. From now on we shall call this quiver the electric quiver. To each electric quiver we can associate a finite amount of magnetic quivers whose union of Coulomb branches make up the Higgs branch\footnote{One magnetic quiver per cone in the Higgs branch.} of the electric quiver. One can now perform quiver subtraction on the magnetic quivers to obtain the Hasse diagram of the classical Higgs branch. We consider the case where there is only one magnetic quiver, also called a 3d mirror dual, as in the case of all examples above.\\

\noindent\underline{Quiver subtraction on the electric quiver:} We obtain 
\begin{itemize}
    \item Hasse diagram of the Coulomb branch
    \item Electric quivers of the theories obtained from partial Higgsing along the Higgs branch
\end{itemize}
\noindent\underline{Quiver subtraction on the magnetic quiver:} We obtain
\begin{itemize}
    \item Hasse diagram of the Higgs branch
    \item Magnetic quivers of the theories obtained from partial Higgsing along the Coulomb branch
\end{itemize}

Starting with the electric quiver, one can in principle compute the magnetic quiver at every step of the quiver subtraction. As well as for the magnetic quivers, one can compute the electric quiver at every step of the quiver subtraction. In this way one is able to obtain all theories that can be obtained by partially Higgsing the theory along any direction in the moduli space. However this is not an easy task for several reasons, some of which are:
\begin{itemize}
    \item The rules for quivers subtraction are only partially understood.
    \item 3d mirror duals have only been found for certain `nice' quivers and 3d mirror symmetry can be difficult to implement.
\end{itemize}

\subsection{ADE singularities and minimal nilpotent orbits}
Turning again to the simplest theories in Table \ref{tab:ADEguys}, those which have Higgs branches/Coulomb branches that are either ADE singularities or ADE minimal nilpotent orbits. For these theories the entire moduli space is very simple, it is the union of the Coulomb branch and the Higgs branch intersecting at the origin. \\

Note that there are (unitary-)orthosymplectic realisations of the quivers in Table \ref{tab:ADEguys} which also follow inversion. They are discussed in \cite{Bourget:2020xdz}.

\begin{landscape}
\begin{table}[h]
    \makebox[\textwidth][l]{
\begin{tabular}{c|c|c|c|c}
        \begin{tikzpicture}
    \node[gauge,label=right:{\small$U(1)$}] (1) at (0,0) {};
    \node[flavour,label=right:{\small$N$}] (2) at (0,1) {};
    \draw (1)--(2);
\end{tikzpicture} & \begin{tikzpicture}
    \node[gauge,label=right:{\small$SU(2)$}] (1) at (0,0) {};
    \node[flavour,label=right:{\small$D_{2N}$}] (2) at (0,1) {};
    \draw (1)--(2);
\end{tikzpicture} & \begin{tikzpicture}
    \node at (0,0) {};
    \node at (0,1) {};
    \node at (0,0.7) {no known quiver};
\end{tikzpicture} & \begin{tikzpicture}
    \node at (0,0) {};
    \node at (0,1) {};
    \node at (0,0.7) {no known quiver};
\end{tikzpicture} & \begin{tikzpicture}
    \node at (0,0) {};
    \node at (0,1) {};
    \node at (0,0.7) {no known quiver};
\end{tikzpicture}\\
%%%%%%%%%%%%%%%%%%%%%%%%%%%%%%%%%%%%%%%%%%%%%%%%%%%%%%%%%%%%%%%%%%%%%%%%
\begin{tikzpicture}
    \node[hasse] (1) at (0,0) {};
    \node[hasse] (2) at (-0.5,1) {};
    \node[hasse] (3) at (0.5,1) {};
    \draw[red] (1)--(2);
    \draw[blue] (1)--(3);
    \node at (-0.7,0.5) {$A_n$};
    \node at (0.7,0.5) {$a_n$};
\end{tikzpicture} & \begin{tikzpicture}
    \node[hasse] (1) at (0,0) {};
    \node[hasse] (2) at (-0.5,1) {};
    \node[hasse] (3) at (0.5,1) {};
    \draw[red] (1)--(2);
    \draw[blue] (1)--(3);
    \node at (-0.7,0.5) {$D_n$};
    \node at (0.7,0.5) {$d_n$};
\end{tikzpicture} & \begin{tikzpicture}
    \node[hasse] (1) at (0,0) {};
    \node[hasse] (2) at (-0.5,1) {};
    \node[hasse] (3) at (0.5,1) {};
    \draw[red] (1)--(2);
    \draw[blue] (1)--(3);
    \node at (-0.7,0.5) {$E_6$};
    \node at (0.7,0.5) {$e_6$};
\end{tikzpicture} & \begin{tikzpicture}
    \node[hasse] (1) at (0,0) {};
    \node[hasse] (2) at (-0.5,1) {};
    \node[hasse] (3) at (0.5,1) {};
    \draw[red] (1)--(2);
    \draw[blue] (1)--(3);
    \node at (-0.7,0.5) {$E_7$};
    \node at (0.7,0.5) {$e_7$};
\end{tikzpicture} & \begin{tikzpicture}
    \node[hasse] (1) at (0,0) {};
    \node[hasse] (2) at (-0.5,1) {};
    \node[hasse] (3) at (0.5,1) {};
    \draw[red] (1)--(2);
    \draw[blue] (1)--(3);
    \node at (-0.7,0.5) {$E_8$};
    \node at (0.7,0.5) {$e_8$};
\end{tikzpicture}\\
%%%%%%%%%%%%%%%%%%%%%%%%%%%%%%%%%%%%%%%%%%%%%%%%%%%%%%%%%%%%%%%%%%%%%%%%
\begin{tikzpicture}
    \node (1) at (0,0) {};
    \node (2) at (0,1.5) {};
    \draw[<->] (1)--(2);
    \node at (0.7,0.75) {3d MS};
\end{tikzpicture} & \begin{tikzpicture}
    \node (1) at (0,0) {};
    \node (2) at (0,1.5) {};
    \draw[<->] (1)--(2);
    \node at (0.7,0.75) {3d MS};
\end{tikzpicture} & \begin{tikzpicture}
    \node (1) at (0,0) {};
    \node (2) at (0,1.5) {};
    \draw[<->] (1)--(2);
    \node at (0.7,0.75) {3d MS};
\end{tikzpicture} & \begin{tikzpicture}
    \node (1) at (0,0) {};
    \node (2) at (0,1.5) {};
    \draw[<->] (1)--(2);
    \node at (0.7,0.75) {3d MS};
\end{tikzpicture} & \begin{tikzpicture}
    \node (1) at (0,0) {};
    \node (2) at (0,1.5) {};
    \draw[<->] (1)--(2);
    \node at (0.7,0.75) {3d MS};
\end{tikzpicture} \\
%%%%%%%%%%%%%%%%%%%%%%%%%%%%%%%%%%%%%%%%%%%%%%%%%%%%%%%%%%%%%%%%%%%%%%%%
\begin{tikzpicture}
    \node[hasse] (1) at (0,0) {};
    \node[hasse] (2) at (-0.5,1) {};
    \node[hasse] (3) at (0.5,1) {};
    \draw[red] (1)--(2);
    \draw[blue] (1)--(3);
    \node at (-0.7,0.5) {$a_n$};
    \node at (0.7,0.5) {$A_n$};
\end{tikzpicture} & \begin{tikzpicture}
    \node[hasse] (1) at (0,0) {};
    \node[hasse] (2) at (-0.5,1) {};
    \node[hasse] (3) at (0.5,1) {};
    \draw[red] (1)--(2);
    \draw[blue] (1)--(3);
    \node at (-0.7,0.5) {$d_n$};
    \node at (0.7,0.5) {$D_n$};
\end{tikzpicture} & \begin{tikzpicture}
    \node[hasse] (1) at (0,0) {};
    \node[hasse] (2) at (-0.5,1) {};
    \node[hasse] (3) at (0.5,1) {};
    \draw[red] (1)--(2);
    \draw[blue] (1)--(3);
    \node at (-0.7,0.5) {$e_6$};
    \node at (0.7,0.5) {$E_6$};
\end{tikzpicture} & \begin{tikzpicture}
    \node[hasse] (1) at (0,0) {};
    \node[hasse] (2) at (-0.5,1) {};
    \node[hasse] (3) at (0.5,1) {};
    \draw[red] (1)--(2);
    \draw[blue] (1)--(3);
    \node at (-0.7,0.5) {$e_7$};
    \node at (0.7,0.5) {$E_7$};
\end{tikzpicture} & \begin{tikzpicture}
    \node[hasse] (1) at (0,0) {};
    \node[hasse] (2) at (-0.5,1) {};
    \node[hasse] (3) at (0.5,1) {};
    \draw[red] (1)--(2);
    \draw[blue] (1)--(3);
    \node at (-0.7,0.5) {$e_8$};
    \node at (0.7,0.5) {$E_8$};
\end{tikzpicture} \\
%%%%%%%%%%%%%%%%%%%%%%%%%%%%%%%%%%%%%%%%%%%%%%%%%%%%%%%%%%%%%%%%%%%%%%%%
         \begin{tikzpicture}[scale=0.8]
        \node[gauge] (1) at (0,0) {};
        \node[gauge] (2) at (2,0) {};
        \node[gauge] (3) at (1,1) {};
        \draw (0.5,0)--(1)--(3)--(2)--(1.5,0);
        \node at (1,0) {\dots};
        \node at (0.5,1) {\tiny$1$};
        \node at (0,-0.35) {\tiny$1$};
        \node at (2,-0.35) {\tiny$1$};
        \draw [decorate,decoration={brace,amplitude=5pt}] (2.3,-0.5)--(-0.3,-0.5);
        \node at (1,-1) {$n$};
    \end{tikzpicture}&    \begin{tikzpicture}[scale=0.8]
        \node[gauge] (4) at (0,0.5) {};
        \node[gauge] (5) at (0,-0.5) {};
        \node[gauge] (6) at (0.5,0) {};
        \node[gauge] (7) at (3,0.5) {};
        \node[gauge] (8) at (3,-0.5) {};
        \node[gauge] (9) at (2.5,0) {};
        \draw (4)--(6)--(5) (6)--(1,0) (2,0)--(9) (7)--(9)--(8);
        \node at (1.5,0) {\dots};
        \node at (-0.3,0.5) {\tiny$1$};
        \node at (-0.3,-0.5) {\tiny$1$};
        \node at (3.3,0.5) {\tiny$1$};
        \node at (3.3,-0.5) {\tiny$1$};
        \node at (0.5,-0.35) {\tiny$2$};
        \node at (2.5,-0.35) {\tiny$2$};
        \draw [decorate,decoration={brace,amplitude=5pt}] (2.7,-0.5)--(0.3,-0.5);
        \node at (1.5,-1) {$n-3$};
    \end{tikzpicture}&    \begin{tikzpicture}[scale=0.8]
        \node[gauge] (10) at (0,0) {};
        \node[gauge] (11) at (1,0) {};
        \node[gauge] (12) at (2,0) {};
        \node[gauge] (13) at (2,1) {};
        \node[gauge] (14) at (2,2) {};
        \node[gauge] (15) at (3,0) {};
        \node[gauge] (16) at (4,0) {};
        \draw (10)--(11)--(12)--(13)--(14) (12)--(15)--(16);
        \node at (0,-0.35) {\tiny$1$};
        \node at (1,-0.35) {\tiny$2$};
        \node at (2,-0.35) {\tiny$3$};
        \node at (3,-0.35) {\tiny$2$};
        \node at (4,-0.35) {\tiny$1$};
        \node at (1.7,1) {\tiny$2$};
        \node at (1.7,2) {\tiny$1$};
    \end{tikzpicture}&    \begin{tikzpicture}[scale=0.8]
        \node[gauge] (17) at (0,0) {};
        \node[gauge] (18) at (1,0) {};
        \node[gauge] (19) at (2,0) {};
        \node[gauge] (20) at (3,0) {};
        \node[gauge] (21) at (4,0) {};
        \node[gauge] (22) at (5,0) {};
        \node[gauge] (23) at (6,0) {};
        \node[gauge] (24) at (3,1) {};
        \draw (17)--(18)--(19)--(20)--(21)--(22)--(23) (24)--(20);
        \node at (0,-0.35) {\tiny$1$};
        \node at (1,-0.35) {\tiny$2$};
        \node at (2,-0.35) {\tiny$3$};
        \node at (3,-0.35) {\tiny$4$};
        \node at (4,-0.35) {\tiny$3$};
        \node at (5,-0.35) {\tiny$2$};
        \node at (6,-0.35) {\tiny$1$};
        \node at (2.7,1) {\tiny$2$};
    \end{tikzpicture}&    \begin{tikzpicture}[scale=0.8]
        \node[gauge] (25) at (0,0) {};
        \node[gauge] (26) at (1,0) {};
        \node[gauge] (27) at (2,0) {};
        \node[gauge] (28) at (3,0) {};
        \node[gauge] (29) at (4,0) {};
        \node[gauge] (30) at (5,0) {};
        \node[gauge] (31) at (6,0) {};
        \node[gauge] (32) at (7,0) {};
        \node[gauge] (33) at (5,1) {};
        \draw (25)--(26)--(27)--(28)--(29)--(30)--(31)--(32) (33)--(30);
        \node at (0,-0.35) {\tiny$1$};
        \node at (1,-0.35) {\tiny$2$};
        \node at (2,-0.35) {\tiny$3$};
        \node at (3,-0.35) {\tiny$4$};
        \node at (4,-0.35) {\tiny$5$};
        \node at (5,-0.35) {\tiny$6$};
        \node at (6,-0.35) {\tiny$4$};
        \node at (7,-0.35) {\tiny$2$};
        \node at (4.7,1) {\tiny$3$};
    \end{tikzpicture}
\end{tabular}}
    \caption{The full Hasse diagram of theories with ADE type Kleinian singularities or minimal nilpotent orbit closures as Coulomb / Higgs branches.}
    \label{tab:ADEguys}
\end{table}

\end{landscape}

\subsection{Single gauge group}
In the following the Hasse diagram of the entire moduli space of several good theories with a single gauge node are given, in the case where the Higgs and Coulomb Hasse diagrams are related by inversion.

\subsubsection{U(k)-[N],$N\geq2k$}
For a single unitary gauge node and enough fundamental flavour for the theory to be good, we obtain as the Hasse diagram of the entire moduli space:

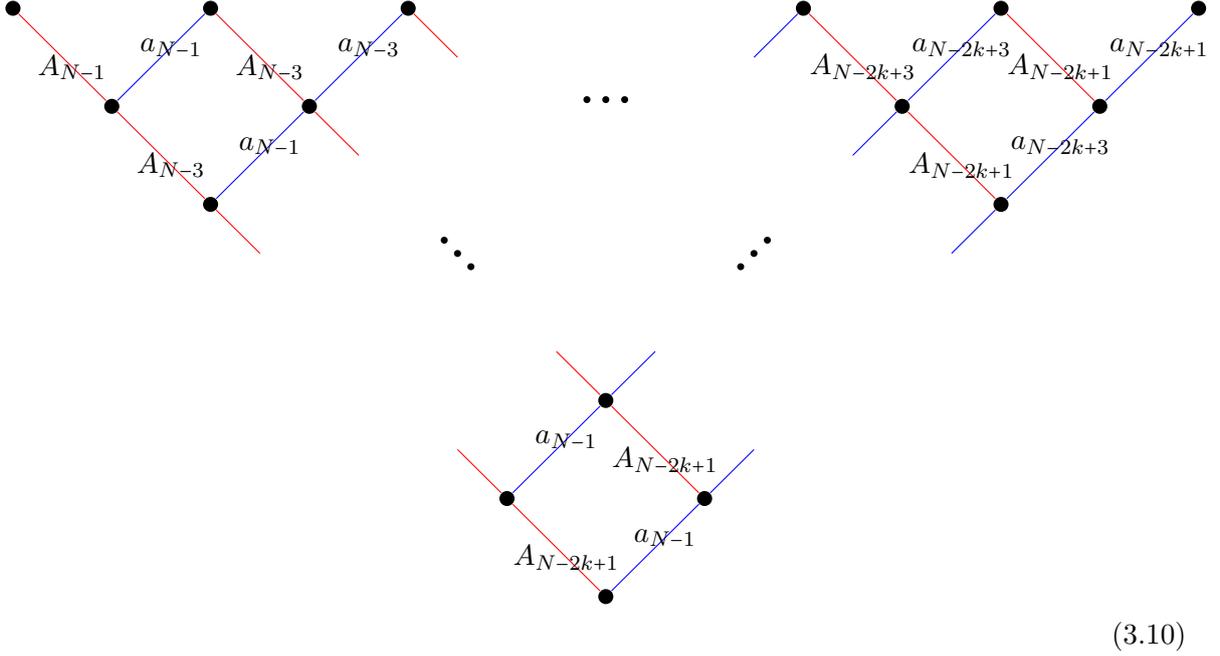
\begin{equation}
    \makebox[\textwidth][c]{\begin{tikzpicture}[scale=1.3]
    \node[hasse] (a1) at (0,0) {};
    \node[hasse] (a2) at (1,1) {};
    \node[hasse] (a3) at (4,4) {};
    \node[hasse] (a4) at (5,5) {};
    \node[hasse] (a5) at (6,6) {};
    \node[hasse] (b1) at (-1,1) {};
    \node[hasse] (b2) at (0,2) {};
    \node[hasse] (b3) at (3,5) {};
    \node[hasse] (b4) at (4,6) {};
    \node[hasse] (c) at (2,6) {};
    \node[hasse] (d1) at (-4,4) {};
    \node[hasse] (d2) at (-3,5) {};
    \node[hasse] (d3) at (-2,6) {};
    \node[hasse] (e1) at (-5,5) {};
    \node[hasse] (e2) at (-4,6) {};
    \node[hasse] (f1) at (-6,6) {};
    \draw[blue] (a1)--(a2)--(1.5,1.5) (3.5,3.5)--(a3)--(a4)--(a5) (b1)--(b2)--(0.5,2.5) (2.5,4.5)--(b3)--(b4) (1.5,5.5)--(c) (d1)--(d2)--(d3) (e1)--(e2);
    \draw[red] (a1)--(b1)--(-1.5,1.5) (-3.5,3.5)--(d1)--(e1)--(f1) (a2)--(b2)--(-0.5,2.5) (-2.5,4.5)--(d2)--(e2) (-1.5,5.5)--(d3) (a3)--(b3)--(c) (a4)--(b4);
    \node at (1.5,3.5) {\Huge$\udots$};
    \node at (-1.5,3.5) {\Huge$\ddots$};
    \node at (0,5) {\Huge$\cdots$};
    \node at ($(a1)!0.6!(a2)$) {$a_{N-1}$};
    \node at ($(a3)!0.6!(a4)$) {$a_{N-2k+3}$};
    \node at ($(a4)!0.6!(a5)$) {$a_{N-2k+1}$};
    \node at ($(b1)!0.6!(b2)$) {$a_{N-1}$};
    \node at ($(b3)!0.6!(b4)$) {$a_{N-2k+3}$};
    \node at ($(d1)!0.6!(d2)$) {$a_{N-1}$};
    \node at ($(d2)!0.6!(d3)$) {$a_{N-3}$};
    \node at ($(e1)!0.6!(e2)$) {$a_{N-1}$};
    \node at ($(a1)!0.4!(b1)$) {$A_{N-2k+1}$};
    \node at ($(d1)!0.4!(e1)$) {$A_{N-3}$};
    \node at ($(e1)!0.4!(f1)$) {$A_{N-1}$};
    \node at ($(a2)!0.4!(b2)$) {$A_{N-2k+1}$};
    \node at ($(d2)!0.4!(e2)$) {$A_{N-3}$};
    \node at ($(a3)!0.4!(b3)$) {$A_{N-2k+1}$};
    \node at ($(b3)!0.4!(c)$) {$A_{N-2k+3}$};
    \node at ($(a4)!0.4!(b4)$) {$A_{N-2k+1}$};
    \end{tikzpicture}}
\end{equation}
The moduli space $\mathcal{M}$ of $U(k)-[N]$ consists of $k+1$ branches: its Coulomb branch, $k-1$ mixed branches, and its Higgs branch:

\begin{equation}
    \begin{split}
        \mathcal{M}(U(k)-[N])\quad=\quad&\mathcal{C}(U(k)-[N])\cup\\
        &\bigcup_{l=1}^{k-1}\mathcal{C}(U(k-l)-[N-2l])\times\mathcal{H}(U(l)-[N])\\
        &\cup\mathcal{H}(U(k)-[N])\;.
    \end{split}
\end{equation}

\pagebreak

\subsubsection{SU(k)-[N], $N\geq2k$}
For a single special unitary gauge node and enough fundamental flavour for the theory to be good, we obtain as the Hasse diagram of the entire moduli space:

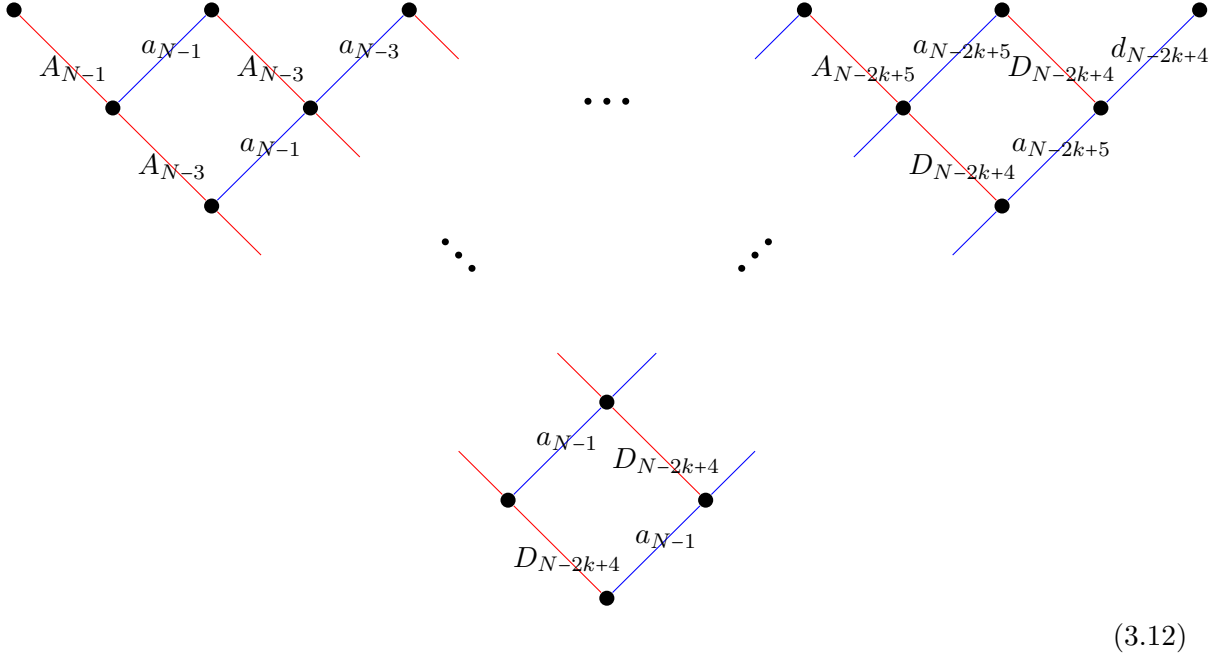
\begin{equation}
    \makebox[\textwidth][c]{\begin{tikzpicture}[scale=1.3]
    \node[hasse] (a1) at (0,0) {};
    \node[hasse] (a2) at (1,1) {};
    \node[hasse] (a3) at (4,4) {};
    \node[hasse] (a4) at (5,5) {};
    \node[hasse] (a5) at (6,6) {};
    \node[hasse] (b1) at (-1,1) {};
    \node[hasse] (b2) at (0,2) {};
    \node[hasse] (b3) at (3,5) {};
    \node[hasse] (b4) at (4,6) {};
    \node[hasse] (c) at (2,6) {};
    \node[hasse] (d1) at (-4,4) {};
    \node[hasse] (d2) at (-3,5) {};
    \node[hasse] (d3) at (-2,6) {};
    \node[hasse] (e1) at (-5,5) {};
    \node[hasse] (e2) at (-4,6) {};
    \node[hasse] (f1) at (-6,6) {};
    \draw[blue] (a1)--(a2)--(1.5,1.5) (3.5,3.5)--(a3)--(a4)--(a5) (b1)--(b2)--(0.5,2.5) (2.5,4.5)--(b3)--(b4) (1.5,5.5)--(c) (d1)--(d2)--(d3) (e1)--(e2);
    \draw[red] (a1)--(b1)--(-1.5,1.5) (-3.5,3.5)--(d1)--(e1)--(f1) (a2)--(b2)--(-0.5,2.5) (-2.5,4.5)--(d2)--(e2) (-1.5,5.5)--(d3) (a3)--(b3)--(c) (a4)--(b4);
    \node at (1.5,3.5) {\Huge$\udots$};
    \node at (-1.5,3.5) {\Huge$\ddots$};
    \node at (0,5) {\Huge$\cdots$};
    \node at ($(a1)!0.6!(a2)$) {$a_{N-1}$};
    \node at ($(a3)!0.6!(a4)$) {$a_{N-2k+5}$};
    \node at ($(a4)!0.6!(a5)$) {$d_{N-2k+4}$};
    \node at ($(b1)!0.6!(b2)$) {$a_{N-1}$};
    \node at ($(b3)!0.6!(b4)$) {$a_{N-2k+5}$};
    \node at ($(d1)!0.6!(d2)$) {$a_{N-1}$};
    \node at ($(d2)!0.6!(d3)$) {$a_{N-3}$};
    \node at ($(e1)!0.6!(e2)$) {$a_{N-1}$};
    \node at ($(a1)!0.4!(b1)$) {$D_{N-2k+4}$};
    \node at ($(d1)!0.4!(e1)$) {$A_{N-3}$};
    \node at ($(e1)!0.4!(f1)$) {$A_{N-1}$};
    \node at ($(a2)!0.4!(b2)$) {$D_{N-2k+4}$};
    \node at ($(d2)!0.4!(e2)$) {$A_{N-3}$};
    \node at ($(a3)!0.4!(b3)$) {$D_{N-2k+4}$};
    \node at ($(b3)!0.4!(c)$) {$A_{N-2k+5}$};
    \node at ($(a4)!0.4!(b4)$) {$D_{N-2k+4}$};
    \end{tikzpicture}}
\end{equation}
The moduli space $\mathcal{M}$ of $SU(k)-[N]$ consists of $k$ branches: its Coulomb branch, $k-2$ mixed branches, and its Higgs branch:

\begin{equation}
    \begin{split}
        \mathcal{M}(SU(k)-[N])\quad=\quad&\mathcal{C}(SU(k)-[N])\cup\\
        &\bigcup_{l=1}^{k-2}\mathcal{C}(SU(k-l)-[N-2l])\times\mathcal{H}(U(l)-[N])\\
        &\cup\mathcal{H}(SU(k)-[N])\;.
    \end{split}
\end{equation}

\pagebreak

\subsubsection{Sp(k)-[D${}_N$], $N>2k$}
For a single symplectic gauge node and enough fundamental flavour for the theory to be good, we obtain as the Hasse diagram of the entire moduli space:

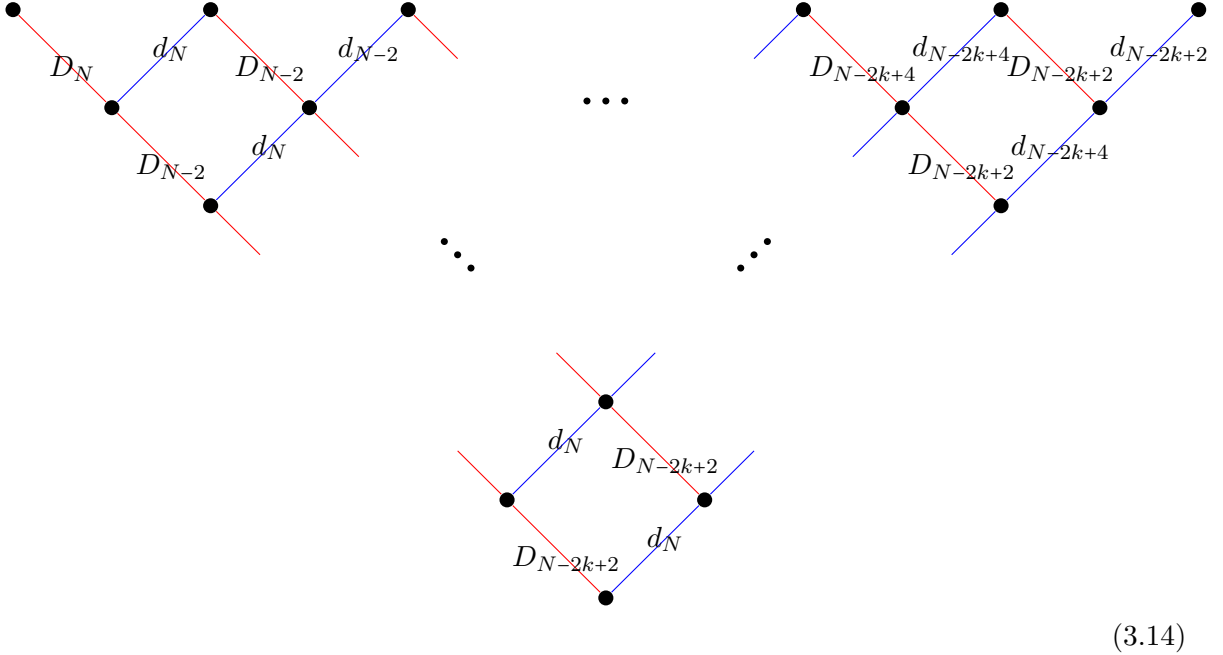
\begin{equation}
    \makebox[\textwidth][c]{\begin{tikzpicture}[scale=1.3]
    \node[hasse] (a1) at (0,0) {};
    \node[hasse] (a2) at (1,1) {};
    \node[hasse] (a3) at (4,4) {};
    \node[hasse] (a4) at (5,5) {};
    \node[hasse] (a5) at (6,6) {};
    \node[hasse] (b1) at (-1,1) {};
    \node[hasse] (b2) at (0,2) {};
    \node[hasse] (b3) at (3,5) {};
    \node[hasse] (b4) at (4,6) {};
    \node[hasse] (c) at (2,6) {};
    \node[hasse] (d1) at (-4,4) {};
    \node[hasse] (d2) at (-3,5) {};
    \node[hasse] (d3) at (-2,6) {};
    \node[hasse] (e1) at (-5,5) {};
    \node[hasse] (e2) at (-4,6) {};
    \node[hasse] (f1) at (-6,6) {};
    \draw[blue] (a1)--(a2)--(1.5,1.5) (3.5,3.5)--(a3)--(a4)--(a5) (b1)--(b2)--(0.5,2.5) (2.5,4.5)--(b3)--(b4) (1.5,5.5)--(c) (d1)--(d2)--(d3) (e1)--(e2);
    \draw[red] (a1)--(b1)--(-1.5,1.5) (-3.5,3.5)--(d1)--(e1)--(f1) (a2)--(b2)--(-0.5,2.5) (-2.5,4.5)--(d2)--(e2) (-1.5,5.5)--(d3) (a3)--(b3)--(c) (a4)--(b4);
    \node at (1.5,3.5) {\Huge$\udots$};
    \node at (-1.5,3.5) {\Huge$\ddots$};
    \node at (0,5) {\Huge$\cdots$};
    \node at ($(a1)!0.6!(a2)$) {$d_{N}$};
    \node at ($(a3)!0.6!(a4)$) {$d_{N-2k+4}$};
    \node at ($(a4)!0.6!(a5)$) {$d_{N-2k+2}$};
    \node at ($(b1)!0.6!(b2)$) {$d_{N}$};
    \node at ($(b3)!0.6!(b4)$) {$d_{N-2k+4}$};
    \node at ($(d1)!0.6!(d2)$) {$d_{N}$};
    \node at ($(d2)!0.6!(d3)$) {$d_{N-2}$};
    \node at ($(e1)!0.6!(e2)$) {$d_{N}$};
    \node at ($(a1)!0.4!(b1)$) {$D_{N-2k+2}$};
    \node at ($(d1)!0.4!(e1)$) {$D_{N-2}$};
    \node at ($(e1)!0.4!(f1)$) {$D_{N}$};
    \node at ($(a2)!0.4!(b2)$) {$D_{N-2k+2}$};
    \node at ($(d2)!0.4!(e2)$) {$D_{N-2}$};
    \node at ($(a3)!0.4!(b3)$) {$D_{N-2k+2}$};
    \node at ($(b3)!0.4!(c)$) {$D_{N-2k+4}$};
    \node at ($(a4)!0.4!(b4)$) {$D_{N-2k+2}$};
    \end{tikzpicture}}
    \label{eq:Spkfull}
\end{equation}
The moduli space $\mathcal{M}$ of $Sp(k)-[D_N]$ consists of $k+1$ branches: its Coulomb branch, $k-1$ mixed branches, and its Higgs branch:

\begin{equation}
    \begin{split}
        \mathcal{M}(Sp(k)-[D_N])\quad=\quad&\mathcal{C}(Sp(k)-[D_N])\cup\\
        &\bigcup_{l=1}^{k-1}\mathcal{C}(Sp(k-l)-[D_{N-2l}])\times\mathcal{H}(Sp(l)-[D_N])\\
        &\cup\mathcal{H}(Sp(k)-[D_N])\;.
    \end{split}
\end{equation}

We turn to electric theories that have multiple magnetic quivers associated to them in the next section.

\section{Bad theories, multiple cones -- multiple bases}
\label{Bad}
One can ask the question what happens when the classical Higgs branch is a union of several cones, these are examples of \emph{bad} theories. In the following we consider only theories with complete Higgsing. Take $SU(2)$ with 2 Flavours. The classical Higgs branch is the union of two $a_1$ with trivial intersection. The Hasse diagram is
\begin{equation}
    \HH=\quad\begin{tikzpicture}
        \node[hasse] (1) at (0,0) {};
        \node[hasse] (2) at (-1,2) {};
        \node[hasse] (3) at (1,2) {};
        
        \draw (2)--(1)--(3);
        
        \node at (-1,1) {$a_1$};
        \node at (1,1) {$a_1$};
    \end{tikzpicture}\quad\begin{tikzpicture}
    \node at (-1,0) {for};
        \node[gauge] (1) at (0,0) {};
        \node[flavour] (4) at (0,1) {};
        \draw(1)--(4);
        
        \node at (0.7,0) {SU(2)};
        \node at (0.5,1) {2};
    \end{tikzpicture}\,.
\end{equation}
The inversion of the Hasse diagram is
\begin{equation}
    \mathfrak{I}(\HH)=\quad\begin{tikzpicture}
        \node[hasse] (1) at (0,0) {};
        \node[hasse] (2) at (-1,-2) {};
        \node[hasse] (3) at (1,-2) {};
        
        \draw[red] (2)--(1)--(3);
        
        \node at (-1,-1) {$A_1$};
        \node at (1,-1) {$A_1$};
    \end{tikzpicture}\quad=\HC\quad\begin{tikzpicture}
    \node at (-1,0) {for};
        \node[gauge] (1) at (0,0) {};
        \node[flavour] (4) at (0,1) {};
        \draw(1)--(4);
        
        \node at (0.7,0) {SU(2)};
        \node at (0.5,1) {2};
    \end{tikzpicture}\,.
\end{equation}
Applying the notion of inversion to the Hasse diagram of the classical Higgs branch suggests that there are two separate leaves which are most singular in the Coulomb branch, i.e. the Hasse diagram of the Coulomb branch has two lowest points; two bases. This agrees with the results of \cite{Assel:2018exy} based on the abelianisation approach of \cite{Bullimore:2015lsa} already applied to study bad $U(n)$ theories in \cite{Assel:2017jgo}. We can now obtain the Hasse diagram of the full moduli space by applying the previous procedure and obtain
\begin{equation}
    \begin{tikzpicture}
        \node[hasse] (1) at (0,0) {};
        \node[hasse] (2) at (-1,-2) {};
        \node[hasse] (3) at (1,-2) {};
        \node[hasse] (4) at (-2,0) {};
        \node[hasse] (5) at (2,0) {};
        
        \draw[red] (2)--(1)--(3);
        \draw[blue] (4)--(2) (3)--(5);
        
        \node at (-1,-1) {$A_1$};
        \node at (1,-1) {$A_1$};
        \node at (-2,-1) {$a_1$};
        \node at (2,-1) {$a_1$};
    \end{tikzpicture}\quad\begin{tikzpicture}
    \node at (-1,0) {for};
        \node[gauge] (1) at (0,0) {};
        \node[flavour] (4) at (0,1) {};
        \draw(1)--(4);
        
        \node at (0.7,0) {SU(2)};
        \node at (0.5,1) {2};
    \end{tikzpicture}\,,
\end{equation}
which fits Figure 2 in \cite{Assel:2018exy}. The two cones in the classical Higgs branch get separated along the quantum Coulomb branch. An effect also observed in 4 dimensions \cite{Argyres:1996eh}. If we focus on the local structure around one singularity it looks like the moduli space of $(1)-[2]$ which has the Hasse diagram
\begin{equation}
    \begin{tikzpicture}
        \node[hasse] (1) at (0,0) {};
        \node[hasse] (2) at (-1,-2) {};
        \node[hasse] (4) at (-2,0) {};
        
        \draw[blue] (2)--(1);
        \draw[red] (4)--(2);
        
        \node at (0,-1) {$a_1$};
        \node at (-2,-1) {$A_1$};
    \end{tikzpicture}\quad\begin{tikzpicture}
    \node at (-1,0) {for};
        \node[gauge] (1) at (0,0) {};
        \node[flavour] (4) at (0,1) {};
        \draw(1)--(4);
        
        \node at (0.5,0) {1};
        \node at (0.5,1) {2};
    \end{tikzpicture}\,,
\end{equation}
which is consistent with the literature.

\subsection{Sp(k)-[D${}_{2k}$]}
The example of $SU(2)=Sp(1)$ with two fundamental flavours is part of the family
\begin{equation}
    \begin{tikzpicture}
        \node[gauge] (1) at (0,0) {};
        \node[flavour] (2) at (0,1) {};
        \draw (1)--(2);
        \node at (0.7,0) {$Sp(k)$};
        \node at (0.8,1) {$D_{2k}$};
    \end{tikzpicture}\,,
\end{equation}
whose Higgs branch is studied extensively in \cite{Ferlito:2016grh}. The corresponding classical Higgs branch Hasse diagram and its inversion are
\begin{equation}
    \HH=\qquad\begin{tikzpicture}
        \node[hasse] (1) at (0,0) {};
        \node[hasse] (2) at (0,2) {};
        \node[hasse] (3) at (0,4) {};
        \node[hasse] (4) at (0,6) {};
        \node[hasse] (5) at (0,8) {};
        \node[hasse] (6) at (-1,10) {};
        \node[hasse] (7) at (1,10) {};
        \draw (1)--(2) (3)--(4)--(5)--(6) (5)--(7);
        \node at (0,3) {$\vdots$};
        \node at (0.5,1) {$d_{2k}$};
        \node at (0.5,5) {$d_{6}$};
        \node at (0.5,7) {$d_{4}$};
        \node at (-1,9) {$a_{1}$};
        \node at (1,9) {$a_{1}$};
    \end{tikzpicture}\;;\qquad \mathfrak{I}(\HH)=\HC=\qquad\begin{tikzpicture}
    \begin{scope}[rotate=180]
        \node[hasse] (1) at (0,0) {};
        \node[hasse] (2) at (0,2) {};
        \node[hasse] (3) at (0,4) {};
        \node[hasse] (4) at (0,6) {};
        \node[hasse] (5) at (0,8) {};
        \node[hasse] (6) at (-1,10) {};
        \node[hasse] (7) at (1,10) {};
        \draw[red] (1)--(2) (3)--(4)--(5)--(6) (5)--(7);
        \node at (0,3) {$\vdots$};
        \node at (0.5,1) {$D_{2k}$};
        \node at (0.5,5) {$D_{6}$};
        \node at (0.5,7) {$D_{4}$};
        \node at (-1,9) {$A_{1}$};
        \node at (1,9) {$A_{1}$};
    \end{scope}
    \end{tikzpicture}\,.
\end{equation}
The dimensions and partial order of the Coulomb branch Hasse diagram obtained from inversion matches the structure described in equation (3.13) of \cite{Assel:2018exy}. Using the algorithm to obtain the Hasse diagram of the entire Moduli space, we see that for this entire one-parameter family, the picture of the Higgs branch, which classically intersects along a subvariety, becomes much more complicated in the quantum theory.\\

For $k=2$ the Hasse diagram of the entire moduli space is
\begin{equation}
    \begin{tikzpicture}
    \node[hasse] (1) at (-1,0) {};
    \node[hasse] (2) at (1,0) {};
    \node[hasse] (3) at (0,1) {};
    \node[hasse] (4) at (-0.5,2.5) {};
    \node[hasse] (5) at (0.5,2.5) {};
    \node[hasse] (6) at (-2,1) {};
    \node[hasse] (7) at (-3,2.5) {};
    \node[hasse] (8) at (2,1) {};
    \node[hasse] (9) at (3,2.5) {};
    \draw[red] (1)--(3)--(2) (3)--(4) (6)--(5)--(8);
    \draw[blue] (1)--(6)--(7) (2)--(8)--(9) (3)--(5);
    \node at ($(1)!0.5!(3)$) {$A_1$};
    \node at ($(2)!0.5!(3)$) {$A_1$};
    \node at ($(3)!0.4!(4)$) {$D_4$};
    \node at ($(3)!0.5!(5)$) {$d_4$};
    \node at ($(6)!0.4!(5)$) {$A_1$};
    \node at ($(8)!0.5!(5)$) {$A_1$};
    \node at ($(1)!0.5!(6)$) {$d_4$};
    \node at ($(2)!0.5!(8)$) {$d_4$};
    \node at ($(6)!0.5!(7)$) {$a_1$};
    \node at ($(8)!0.5!(9)$) {$a_1$};
    \end{tikzpicture}\quad\begin{tikzpicture}
    \node at (-1,0) {for};
        \node[gauge] (1) at (0,0) {};
        \node[flavour] (2) at (0,1) {};
        \draw (1)--(2);
        \node at (0,-0.5) {$Sp(2)$};
        \node at (0,1.5) {$D_4$};
    \end{tikzpicture}
\end{equation}
One can identify the Hasse diagram of the entire moduli space of $SU(2)-[D_2]$ as a subdiagram. The moduli space of $Sp(2)-[D_4]$ consists of 4 branches: Two Higgs branches, which emanate from different bases; one Coulomb branch which has two bases; and one mixed branch, which is a product of the Coulomb branch of $SU(2)-[D_2]$ (two bases) and the Higgs branch of $SU(2)-[D_4]$. If we focus on the Hasse diagram emanating from just one base, we obtain the Hasse diagram:

\begin{equation}
    \begin{tikzpicture}
    \node[hasse] (1) at (0,0) {};
    \node[hasse] (2) at (1,1) {};
    \node[hasse] (3) at (2,2) {};
    \node[hasse] (4) at (-1,1) {};
    \node[hasse] (5) at (0,2) {};
    \node[hasse] (6) at (-2,2) {};
    \draw[blue] (1)--(2)--(3) (4)--(5);
    \draw[red] (1)--(4)--(6) (2)--(5);
    \node at ($(1)!0.5!(2)$) {$d_4$};
    \node at ($(2)!0.5!(3)$) {$a_1$};
    \node at ($(4)!0.5!(5)$) {$d_4$};
    \node at ($(1)!0.5!(4)$) {$A_1$};
    \node at ($(4)!0.5!(6)$) {$D_4$};
    \node at ($(2)!0.5!(5)$) {$A_1$};
    \end{tikzpicture}
    \begin{tikzpicture}
    \node at (-1,0.5) {for};
    \node[gauge,label=below:{$U(1)$}] (1) at (0,0) {};
    \node[gauge,label=below:{$Sp(1)$}] (2) at (1,0) {};
    \node[flavour,label=right:{$D_4$}] (3) at (1,1) {};
    \draw (1)--(2)--(3);
    \end{tikzpicture}\;,
\end{equation}
which is easily checked.\\

For $k=3$ the Hasse diagram of the entire moduli space is:

\begin{equation}
    \begin{tikzpicture}
        \node[hasse] (1) at (-1,0) {};
        \node[hasse] (2) at (1,0) {};
        \node[hasse] (3) at (-0.5,2) {};
        \node[hasse] (4) at (-1,4) {};
        \node[hasse] (5) at (-1.5,6) {};
        \draw[red] (1)--(3)--(4)--(5) (2)--(3);
        \node[hasse] (6) at (-2,2) {};
        \node[hasse] (7) at (-3,4) {};
        \node[hasse] (8) at (-4,6) {};
        \node[hasse] (9) at (2,2) {};
        \node[hasse] (10) at (3,4) {};
        \node[hasse] (11) at (4,6) {};
        \draw[blue] (1)--(6)--(7)--(8) (2)--(9)--(10)--(11);
        \node[hasse] (12) at (0.5,4) {};
        \node[hasse] (13) at (1,6) {};
        \draw[blue] (3)--(12)--(13);
        \node[hasse] (14) at (-0.5,6) {};
        \draw[blue] (4)--(14);
        \draw[red] (6)--(12)--(9) (12)--(14) (7)--(13)--(10);
        
        \node at ($(1)!0.5!(3)$) {$A_1$};
        \node at ($(2)!0.5!(3)$) {$A_1$};
        \node at ($(3)!0.7!(4)$) {$D_4$};
        \node at ($(4)!0.7!(5)$) {$D_6$};
        \node at ($(1)!0.5!(6)$) {$d_6$};
        \node at ($(6)!0.5!(7)$) {$d_4$};
        \node at ($(7)!0.5!(8)$) {$a_1$};
        \node at ($(2)!0.5!(9)$) {$d_6$};
        \node at ($(9)!0.5!(10)$) {$d_4$};
        \node at ($(10)!0.5!(11)$) {$a_1$};
        \node at ($(3)!0.5!(12)$) {$d_6$};
        \node at ($(12)!0.5!(13)$) {$d_4$};
        \node at ($(4)!0.4!(14)$) {$d_6$};
        \node at ($(6)!0.35!(12)$) {$A_1$};
        \node at ($(9)!0.5!(12)$) {$A_1$};
        \node at ($(12)!0.5!(14)$) {$D_4$};
        \node at ($(7)!0.2!(13)$) {$A_1$};
        \node at ($(10)!0.5!(13)$) {$A_1$};
    \end{tikzpicture}\quad\begin{tikzpicture}
    \node at (-1,0) {for};
        \node[gauge] (1) at (0,0) {};
        \node[flavour] (2) at (0,1) {};
        \draw (1)--(2);
        \node at (0,-0.5) {$Sp(3)$};
        \node at (0,1.5) {$D_6$};
    \end{tikzpicture}\,,
    \label{eq:badSp3}
\end{equation}
which is straight forward to generalise to any $k$ and agrees with the analysis in \cite{Assel:2018exy}. In this case we are not able to identify a theory, which has as it's Hasse diagram the Hasse diagram emanating from a single base of \eqref{eq:badSp3}. One can identify both the Hasse diagram for $Sp(2)-[D_4]$ and $Sp(1)-[D_2]$ as subdiagrams. As can be seen from taking the closure of all maximal leaves in \eqref{eq:badSp3}, the moduli space of $Sp(3)-[D_6]$ consists of 5 branches: Two Higgs branches, which emanate from the two different bases; a Coulomb branch, which has two bases; a mixed branch, which is the product of the Coulomb branch of $Sp(2)-[D_4]$ (two bases) and the Higgs branch of $Sp(1)-[D_6]$; and another mixed branch, which is the product of the Coulomb branch of $Sp(1)-[D_2]$ (two bases) and the Higgs branch of $Sp(2)-[D_6]$.

\subsection{SU(k)-[2k-2]}
The SU(2) example with 2 flavours is also part of the family
\begin{equation}
    \begin{tikzpicture}
        \node[gauge] (1) at (0,0) {};
        \node[flavour] (2) at (0,1) {};
        \draw (1)--(2);
        \node at (0.8,0) {$SU(N)$};
        \node at (0.8,1) {$2N-2$};
    \end{tikzpicture}\,.
\end{equation}
The Hasse diagram of the classical Higgs branch and its inversion which is expected to be the Hasse diagram of the Coulomb branch are
\begin{equation}
    \HH=\qquad\begin{tikzpicture}
        \node[hasse] (1) at (0,0) {};
        \node[hasse] (2) at (0,2) {};
        \node[hasse] (3) at (0,4) {};
        \node[hasse] (4) at (0,6) {};
        \node[hasse] (5) at (0,8) {};
        \node[hasse] (6) at (-1,10) {};
        \node[hasse] (7) at (1,10) {};
        \draw (1)--(2) (3)--(4)--(5)--(6) (5)--(7);
        \node at (0,3) {$\vdots$};
        \node at (0.6,1) {$a_{2N-3}$};
        \node at (0.5,5) {$a_{5}$};
        \node at (0.5,7) {$a_{3}$};
        \node at (-1,9) {$a_{1}$};
        \node at (1,9) {$a_{1}$};
    \end{tikzpicture}\;;\qquad \mathfrak{I}(\HH)=\HC=\qquad\begin{tikzpicture}
    \begin{scope}[rotate=180]
        \node[hasse] (1) at (0,0) {};
        \node[hasse] (2) at (0,2) {};
        \node[hasse] (3) at (0,4) {};
        \node[hasse] (4) at (0,6) {};
        \node[hasse] (5) at (0,8) {};
        \node[hasse] (6) at (-1,10) {};
        \node[hasse] (7) at (1,10) {};
        \draw[red] (1)--(2) (3)--(4)--(5)--(6) (5)--(7);
        \node at (0,3) {$\vdots$};
        \node at (0.6,1) {$A_{2N-3}$};
        \node at (0.5,5) {$A_{5}$};
        \node at (0.5,7) {$A_{3}$};
        \node at (-1,9) {$A_{1}$};
        \node at (1,9) {$A_{1}$};
    \end{scope}
    \end{tikzpicture}\,.
\end{equation}
Again it is straight forward to produce the Hasse diagram of the entire moduli space. The Hasse diagram for the entire moduli space for $SU(3)$ with 4 fundamental hypermultiplets is:

\begin{equation}
    \begin{tikzpicture}
    \node[hasse] (1) at (-1,0) {};
    \node[hasse] (2) at (1,0) {};
    \node[hasse] (3) at (0,1) {};
    \node[hasse] (4) at (-0.5,2.5) {};
    \node[hasse] (5) at (0.5,2.5) {};
    \node[hasse] (6) at (-2,1) {};
    \node[hasse] (7) at (-3,2.5) {};
    \node[hasse] (8) at (2,1) {};
    \node[hasse] (9) at (3,2.5) {};
    \draw[red] (1)--(3)--(2) (3)--(4) (6)--(5)--(8);
    \draw[blue] (1)--(6)--(7) (2)--(8)--(9) (3)--(5);
    \node at ($(1)!0.5!(3)$) {$A_1$};
    \node at ($(2)!0.5!(3)$) {$A_1$};
    \node at ($(3)!0.4!(4)$) {$A_3$};
    \node at ($(3)!0.5!(5)$) {$a_3$};
    \node at ($(6)!0.4!(5)$) {$A_1$};
    \node at ($(8)!0.5!(5)$) {$A_1$};
    \node at ($(1)!0.5!(6)$) {$a_3$};
    \node at ($(2)!0.5!(8)$) {$a_3$};
    \node at ($(6)!0.5!(7)$) {$a_1$};
    \node at ($(8)!0.5!(9)$) {$a_1$};
    \end{tikzpicture}\quad\begin{tikzpicture}
    \node at (-1,0) {for};
        \node[gauge] (1) at (0,0) {};
        \node[flavour] (2) at (0,1) {};
        \draw (1)--(2);
        \node at (0,-0.5) {SU(3)};
        \node at (0,1.5) {4};
    \end{tikzpicture}
    \label{SU(3)4}
\end{equation}
The moduli space of $SU(3)-[4]$ consists of four branches: two Higgs branches, which emanate from the different bases; its Coulomb branch, which has two bases; and a mixed branch, which is the product of the Coulomb branch of $SU(2)-[2]$ (two bases) and the Higgs branch of $(1)-[4]$.\\

We propose that for certain theories the procedure of inversion employed in Section \ref{Moduli} and \ref{Bad} can be applied to analyse the Coulomb branch and full moduli space even if the theory is bad. It should be noted, that the notion of inversion here is applied in a more intricate way:
\begin{enumerate}
    \item Start with the Hasse diagram of the classical Higgs branch (computed for example from quiver subtraction on magnetic quivers).
    \item Use inversion to obtain the Hasse diagram of the Coulomb branch $\HC$, draw it in red.
    \item Draw in blue: For every node in $\HC$ add the inversion of the associated sub-diagram going from this node to the top. For the lowest node this gives $\HH$ which may be split into disconnected diagrams.
    \item Draw in red: For every node in $\HH$ add the inversion of the associated sub-diagram going to the top and link it up with the diagrams added in 2. such that Coulomb and Higgs (red and blue) directions commute.
\end{enumerate}
One obtains the Hasse diagram of the full moduli space, which does not include the Hasse diagram of the classical Higgs branch as a subdiagram. Rather there is a smaller Higgs branch emanating from every base of the Coulomb branch.

In the next section we focus on the Hasse diagram of the entire moduli space for non-invertible Higgs branch Hasse diagrams.

\section{The Hasse diagram of the full moduli space -- non-invertible}
\label{noninv}
After a discussion of the Hasse diagram for the entire moduli space, generated by using inversion, one can ask if it is possible to generate the Hasse diagram for the entire moduli space for a theory where the Hasse diagram of the Higgs or Coulomb branch are not invertible. If a brane construction is known, one can map out the moduli space identifying all leaves and elementary slices. A demonstration of the brane construction for $O(3)-[D_4]$ is given in the Appendix \ref{app:O}. We provide the moduli space Hasse diagram for several $O(k)-[D_N]$ theories, which are obtainable e.g. from brane constructions similar to the one given in the Appendix \ref{app:O}. The knowledge obtained about non-invertible theories from brane constructions will be useful for theories where no brane construction is known, as in Section \ref{G2}.\\

$O(1)-[C_2]:$
\begin{equation}
\begin{tikzpicture}
    \node[hasse] (1) at (0,0) {};
    \node[hasse] (2) at (1,1) {};
    \draw[blue] (1)--(2);
    \node at ($(1)!0.5!(2)$) {$c_2$};
\end{tikzpicture}\quad\begin{tikzpicture}
    \node at (-1,0) {for};
        \node[gauge] (1) at (0,0) {};
        \node[flavour] (4) at (0,1) {};
        \draw(1)--(4);
        
        \node at (0.7,0) {$O(1)$};
        \node at (0.7,1) {$C_2$};
    \end{tikzpicture}\;.
\end{equation}
This theory has no Coulomb branch, since it is rank 0. The Higgs branch is the minimal nilpotent orbit of $C_2$ and makes up the entire moduli space.\\

$O(2)-[C_3]:$
\begin{equation}
    \begin{tikzpicture}
    \node[hasse] (1) at (0,0) {};
    \node[hasse] (2) at (1,1) {};
    \node[hasse] (3) at (2,2) {};
    \node[hasse] (4) at (-1,1) {};
    \draw[blue] (1)--(2)--(3);
    \draw[red] (1)--(4);
    \node at ($(1)!0.5!(2)$) {$c_3$};
    \node at ($(2)!0.5!(3)$) {$c_2$};
    \node at ($(1)!0.5!(4)$) {$D_5$};
    \end{tikzpicture}\quad\begin{tikzpicture}
    \node at (-1,0) {for};
        \node[gauge] (1) at (0,0) {};
        \node[flavour] (4) at (0,1) {};
        \draw(1)--(4);
        
        \node at (0.7,0) {$O(2)$};
        \node at (0.7,1) {$C_3$};
    \end{tikzpicture}\;.
\end{equation}
This theory has a Coulomb branch which is the Kleinian singularity $D_5$ \cite{Cabrera:2017njm}, which is also the Coulomb branch of $Sp(1)-[D_5]$, see \eqref{eq:Spkfull}. The Higgs branch consists of of 3 leaves with two elementary slices $c_3$ and $c_2$, it is closure of the the next-to-minimal orbit of $\mathfrak{s}\mathfrak{p}(3)$, which is different from the Higgs branch of $Sp(1)-[D_5]$. The entire moduli space of $O(2)-[C_3]$ consists of a union of its Coulomb branch and its Higgs branch, which intersect at the origin. One can see that the $O(1)-[C_2]$ theory can be obtained from partially Higgsing along the Higgs branch.\\

$O(3)-[C_4]:$ 
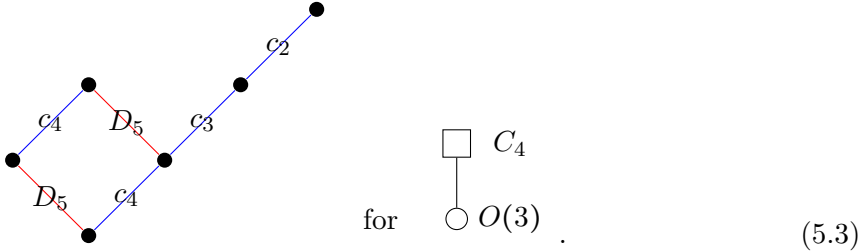
\begin{equation}
    \begin{tikzpicture}
    \node[hasse] (1) at (0,0) {};
    \node[hasse] (2) at (1,1) {};
    \node[hasse] (3) at (2,2) {};
    \node[hasse] (4) at (-1,1) {};
    \draw[blue] (1)--(2)--(3);
    \draw[red] (1)--(4);
    \node at ($(1)!0.5!(2)$) {$c_3$};
    \node at ($(2)!0.5!(3)$) {$c_2$};
    \node at ($(1)!0.5!(4)$) {$D_5$};
    \node[hasse] (a1) at (-1,-1) {};
    \node[hasse] (a2) at (-2,0) {};
    \draw[blue] (a2)--(4) (a1)--(1);
    \draw[red] (a1)--(a2);
    \node at ($(a1)!0.5!(1)$) {$c_4$};
    \node at ($(a1)!0.5!(a2)$) {$D_5$};
    \node at ($(a2)!0.5!(4)$) {$c_4$};
    \end{tikzpicture}\quad\begin{tikzpicture}
    \node at (-1,0) {for};
        \node[gauge] (1) at (0,0) {};
        \node[flavour] (4) at (0,1) {};
        \draw(1)--(4);
        
        \node at (0.7,0) {$O(3)$};
        \node at (0.7,1) {$C_4$};
    \end{tikzpicture}\;.
\end{equation}
The Coulomb branch of this theory is the same as for the theory before. The Higgs branch of this theory consists of 4 leaves, with elementary slices $c_4$, $c_3$ and $c_2$. There is a crucial qualitative difference between the moduli space of $O(2)-[C_3]$ and this theory! The moduli space is a union of two cones, but it is not the union of the Coulomb branch and the Higgs branch intersecting at the origin. It is the union of:
\begin{enumerate}[label=\alph*)]
    \item The so called \emph{enhanced Coulomb branch}: $\mathcal{C}_{\mathrm{enhanced}}$\\
    which is a direct product of the Higgs branch of $O(1)-[C_4]$, $c_4$, and the Coulomb branch of $O(2)-[C_3]$, $D_5$. The enhanced Coulomb branch is a type of mixed branch.
    (the name was originally introduced for $4d$ $\mathcal{N}=2$ theories \cite{Argyres:2016xmc} \footnote{J.F.G. thanks Mario Martone for an explanation of the enhanced Coulomb Branch.})
    \item The Higgs branch: $\mathcal{H}$ 
\end{enumerate}
The two branches intersect not just at the origin but along the $c_4$ elementary slice to the origin:

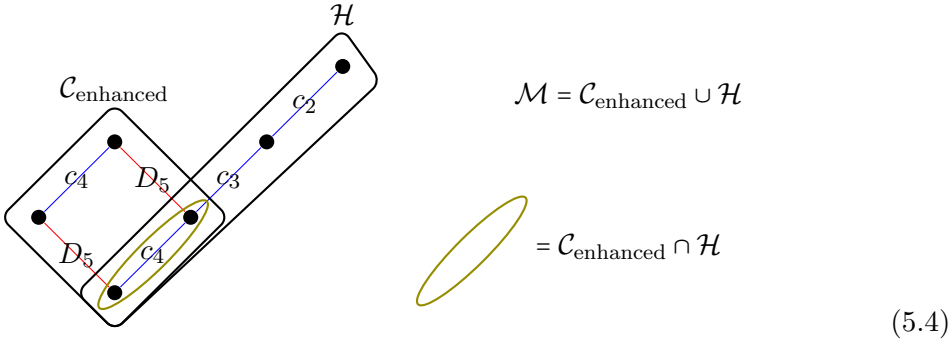
\begin{equation}
    \begin{tikzpicture}
    \node[hasse] (1) at (0,0) {};
    \node[hasse] (2) at (1,1) {};
    \node[hasse] (3) at (2,2) {};
    \node[hasse] (4) at (-1,1) {};
    \draw[blue] (1)--(2)--(3);
    \draw[red] (1)--(4);
    \node at ($(1)!0.5!(2)$) {$c_3$};
    \node at ($(2)!0.5!(3)$) {$c_2$};
    \node at ($(1)!0.5!(4)$) {$D_5$};
    \node[hasse] (a1) at (-1,-1) {};
    \node[hasse] (a2) at (-2,0) {};
    \draw[blue] (a2)--(4) (a1)--(1);
    \draw[red] (a1)--(a2);
    \node at ($(a1)!0.5!(1)$) {$c_4$};
    \node at ($(a1)!0.5!(a2)$) {$D_5$};
    \node at ($(a2)!0.5!(4)$) {$c_4$};
    \draw[thick, rounded corners] (-1,-1.5) -- (-2.5, 0) -- (-1, 1.5) -- (0.5, 0) --  cycle;
    \draw[thick, rounded corners] (-1,-1.5) -- (-1.5, -1) -- (2,2.5) -- (2.5,1.8) --  cycle;
    \node at (-1,1.7) {$\mathcal{C}_{\mathrm{enhanced}}$};
    \node at (2,2.7) {$\mathcal{H}$};
    \draw[olive, thick, rotate=45] (-0.7,0) ellipse (1.0cm and 0.2cm);
    \end{tikzpicture}
\quad\begin{tikzpicture}
    \node at (-1,-2) {};
    \node at (1,1) {$\mathcal{M}=\mathcal{C}_{\mathrm{enhanced}}\cup\mathcal{H}$};
    \draw[olive, thick, rotate=45] (-1.5,0) ellipse (1.0cm and 0.2cm);
    \node at (1,-1) {$=\mathcal{C}_{\mathrm{enhanced}}\cap\mathcal{H}$};
    \end{tikzpicture}
\end{equation}
The $O(2)-[C_3]$ theory is reached from partially Higgsing along the Higgs branch, Higgsing further along the Higgs branch an $O(1)-[C_3]$ is reached. An $O(1)-[C_4]$ theory is reached from partially Higgsing along the Coulomb branch.\\

$O(4)-[C_5]:$
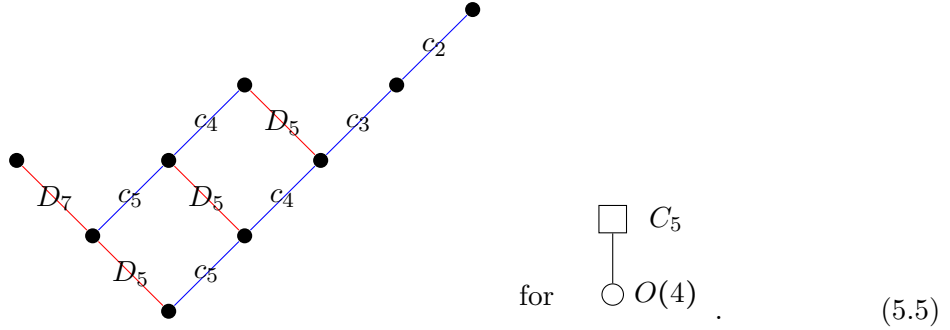
\begin{equation}
    \begin{tikzpicture}
    \node[hasse] (1) at (0,0) {};
    \node[hasse] (2) at (1,1) {};
    \node[hasse] (3) at (2,2) {};
    \node[hasse] (4) at (-1,1) {};
    \draw[blue] (1)--(2)--(3);
    \draw[red] (1)--(4);
    \node at ($(1)!0.5!(2)$) {$c_3$};
    \node at ($(2)!0.5!(3)$) {$c_2$};
    \node at ($(1)!0.5!(4)$) {$D_5$};
    \node[hasse] (a1) at (-1,-1) {};
    \node[hasse] (a2) at (-2,0) {};
    \draw[blue] (a2)--(4) (a1)--(1);
    \draw[red] (a1)--(a2);
    \node at ($(a1)!0.5!(1)$) {$c_4$};
    \node at ($(a1)!0.5!(a2)$) {$D_5$};
    \node at ($(a2)!0.5!(4)$) {$c_4$};
    \node[hasse] (b1) at (-2,-2) {};
    \node[hasse] (b2) at (-3,-1) {};
    \node[hasse] (b3) at (-4,0) {};
    \draw[blue] (b1)--(a1) (b2)--(a2);
    \draw[red] (b1)--(b2)--(b3);
    \node at ($(b1)!0.5!(a1)$) {$c_5$};
    \node at ($(b2)!0.5!(a2)$) {$c_5$};
    \node at ($(b1)!0.5!(b2)$) {$D_5$};
    \node at ($(b2)!0.5!(b3)$) {$D_7$};
    \end{tikzpicture}\quad\begin{tikzpicture}
    \node at (-1,0) {for};
        \node[gauge] (1) at (0,0) {};
        \node[flavour] (4) at (0,1) {};
        \draw(1)--(4);
        
        \node at (0.7,0) {$O(4)$};
        \node at (0.7,1) {$C_5$};
    \end{tikzpicture}\;.
\end{equation}
This theory has an entire moduli space which consists of: a Coulomb branch (not enhanced) which is the Coulomb branch of $Sp(2)-[D_7]$, see \eqref{eq:Spkfull}; the Higgs branch which is the (next-to)${}^3$-minimal nilpotent orbit closure of $\mathfrak{s}\mathfrak{p}(5)$; and also a mixed branch which is a product of the Higgs branch of $Sp(2)-[C_5]$, the closure of the next-to-minimal nilpotent orbit of $\mathfrak{s}\mathfrak{p}(5)$, and the Coulomb branch of $O(2)-[C_3]$, $D_5$.\\

$O(5)-[C_6]:$
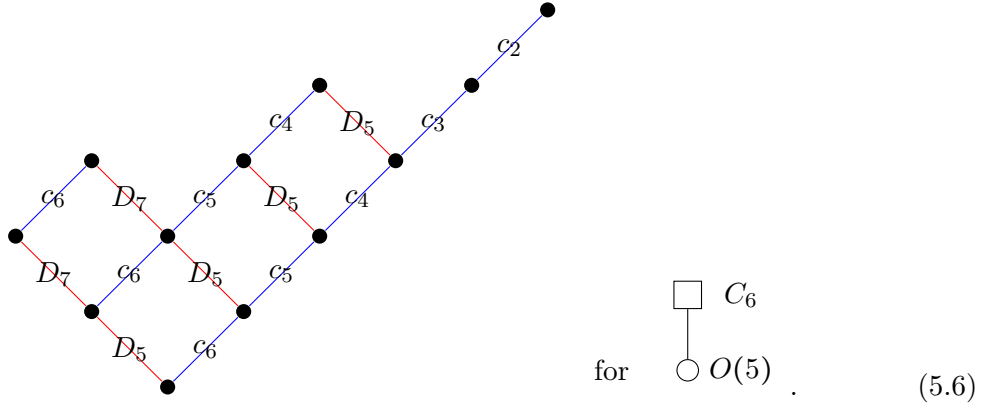
\begin{equation}
    \begin{tikzpicture}
    \node[hasse] (1) at (0,0) {};
    \node[hasse] (2) at (1,1) {};
    \node[hasse] (3) at (2,2) {};
    \node[hasse] (4) at (-1,1) {};
    \draw[blue] (1)--(2)--(3);
    \draw[red] (1)--(4);
    \node at ($(1)!0.5!(2)$) {$c_3$};
    \node at ($(2)!0.5!(3)$) {$c_2$};
    \node at ($(1)!0.5!(4)$) {$D_5$};
    \node[hasse] (a1) at (-1,-1) {};
    \node[hasse] (a2) at (-2,0) {};
    \draw[blue] (a2)--(4) (a1)--(1);
    \draw[red] (a1)--(a2);
    \node at ($(a1)!0.5!(1)$) {$c_4$};
    \node at ($(a1)!0.5!(a2)$) {$D_5$};
    \node at ($(a2)!0.5!(4)$) {$c_4$};
    \node[hasse] (b1) at (-2,-2) {};
    \node[hasse] (b2) at (-3,-1) {};
    \node[hasse] (b3) at (-4,0) {};
    \draw[blue] (b1)--(a1) (b2)--(a2);
    \draw[red] (b1)--(b2)--(b3);
    \node at ($(b1)!0.5!(a1)$) {$c_5$};
    \node at ($(b2)!0.5!(a2)$) {$c_5$};
    \node at ($(b1)!0.5!(b2)$) {$D_5$};
    \node at ($(b2)!0.5!(b3)$) {$D_7$};
    \node[hasse] (c1) at (-3,-3) {};
    \node[hasse] (c2) at (-4,-2) {};
    \node[hasse] (c3) at (-5,-1) {};
    \draw[blue] (c1)--(b1) (c2)--(b2) (c3)--(b3);
    \draw[red] (c1)--(c2)--(c3);
    \node at ($(c1)!0.5!(b1)$) {$c_6$};
    \node at ($(c2)!0.5!(b2)$) {$c_6$};
    \node at ($(c3)!0.5!(b3)$) {$c_6$};
    \node at ($(c1)!0.5!(c2)$) {$D_5$};
    \node at ($(c2)!0.5!(c3)$) {$D_7$};
    \end{tikzpicture}\quad\begin{tikzpicture}
    \node at (-1,0) {for};
        \node[gauge] (1) at (0,0) {};
        \node[flavour] (4) at (0,1) {};
        \draw(1)--(4);
        
        \node at (0.7,0) {$O(5)$};
        \node at (0.7,1) {$C_6$};
    \end{tikzpicture}\;.
\end{equation}
This theory has a moduli space which consists of: an enhanced Coulomb branch which is a product of the Higgs branch of $O(1)-[C_6]$, $c_6$, with the Coulomb branch of $O(4)-[C_5]$; a mixed branch which is the product of the Higgs branch of $O(3)-[C_6]$, the next-to-next-to-minimal nilpotent orbit closure of $\mathfrak{s}\mathfrak{p}(6)$, and the Coulomb branch of $O(2)-[C_3]$, a $D_5$ singularity; and a Higgs branch, which is the (next-to)${}^4$-minimal orbit closure of $\mathfrak{s}\mathfrak{p}(6)$.\\

\paragraph{Enhanced Coulomb branch} In general the Coulomb branch of $O(k)$ is enhanced if $1<k=2r+1$ is odd. In this case enhancement of the Coulomb branch stems from the $O(1)$ gauge theory with matter coupled to it, which remains on a general point on the Coulomb branch, where the gauge group is broken to $U(1)^r\times O(1)$. The matter coupled to the $O(1)$ is responsible for the enhancement of the Coulomb branch. If $k=2r$ is even, then there is no enhancement of the Coulomb branch. This is in agreement with the analysis for $4d$ $\mathcal{N}=2$ $SO$ theories in \cite{Argyres:2016xmc}.

\subsection{O(2r)-[C${}_N$], $N\geq2r$}
\label{sec:Oeven}
The general Hasse diagram is
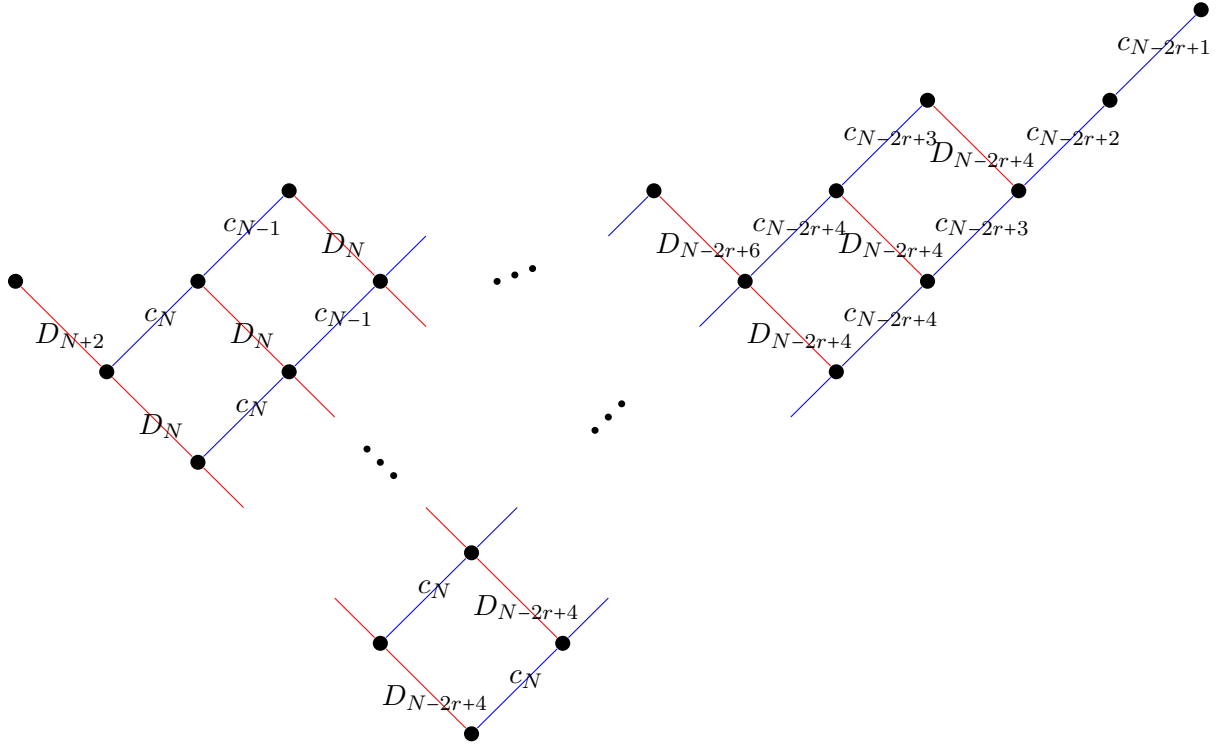
\begin{equation}
    \makebox[\textwidth][c]{\begin{tikzpicture}[scale=1.2]
    \node[hasse] (a1) at (0,0) {};
    \node[hasse] (a2) at (1,1) {};
    \node[hasse] (a3) at (4,4) {};
    \node[hasse] (a4) at (5,5) {};
    \node[hasse] (a5) at (6,6) {};
    \node[hasse] (a6) at (7,7) {};
    \node[hasse] (a7) at (8,8) {};
    \node[hasse] (b1) at (-1,1) {};
    \node[hasse] (b2) at (0,2) {};
    \node[hasse] (b3) at (3,5) {};
    \node[hasse] (b4) at (4,6) {};
    \node[hasse] (b5) at (5,7) {};
    \node[hasse] (c) at (2,6) {};
    \node[hasse] (d1) at (-3,3) {};
    \node[hasse] (d2) at (-2,4) {};
    \node[hasse] (d3) at (-1,5) {};
    \node[hasse] (e1) at (-4,4) {};
    \node[hasse] (e2) at (-3,5) {};
    \node[hasse] (e3) at (-2,6) {};
    \node[hasse] (f1) at (-5,5) {};
    \draw[blue] (a1)--(a2)--(1.5,1.5) (3.5,3.5)--(a3)--(a4)--(a5)--(a6)--(a7) (b1)--(b2)--(0.5,2.5) (2.5,4.5)--(b3)--(b4)--(b5) (1.5,5.5)--(c) (d1)--(d2)--(d3)--(-0.5,5.5) (e1)--(e2)--(e3);
    \draw[red] (a1)--(b1)--(-1.5,1.5) (-2.5,2.5)--(d1)--(e1)--(f1) (a2)--(b2)--(-0.5,2.5) (-1.5,3.5)--(d2)--(e2) (-0.5,4.5)--(d3)--(e3) (a3)--(b3)--(c) (a4)--(b4) (a5)--(b5);
    \node at (1.5,3.5) {\Huge$\udots$};
    \node at (-1,3) {\Huge$\ddots$};
    \node[rotate=20] at (0.5,5) {\Huge$\cdots$};
    \node at ($(a1)!0.6!(a2)$) {$c_{N}$};
    \node at ($(a3)!0.6!(a4)$) {$c_{N-2r+4}$};
    \node at ($(a4)!0.6!(a5)$) {$c_{N-2r+3}$};
    \node at ($(a5)!0.6!(a6)$) {$c_{N-2r+2}$};
    \node at ($(a6)!0.6!(a7)$) {$c_{N-2r+1}$};
    \node at ($(b1)!0.6!(b2)$) {$c_{N}$};
    \node at ($(b3)!0.6!(b4)$) {$c_{N-2r+4}$};
    \node at ($(b4)!0.6!(b5)$) {$c_{N-2r+3}$};
    \node at ($(d1)!0.6!(d2)$) {$c_{N}$};
    \node at ($(d2)!0.6!(d3)$) {$c_{N-1}$};
    \node at ($(e1)!0.6!(e2)$) {$c_{N}$};
    \node at ($(e2)!0.6!(e3)$) {$c_{N-1}$};
    \node at ($(a1)!0.4!(b1)$) {$D_{N-2r+4}$};
    \node at ($(a2)!0.4!(b2)$) {$D_{N-2r+4}$};
    \node at ($(a3)!0.4!(b3)$) {$D_{N-2r+4}$};
    \node at ($(a4)!0.4!(b4)$) {$D_{N-2r+4}$};
    \node at ($(a5)!0.4!(b5)$) {$D_{N-2r+4}$};
    \node at ($(b3)!0.4!(c)$) {$D_{N-2r+6}$};
    \node at ($(d1)!0.4!(e1)$) {$D_{N}$};
    \node at ($(d2)!0.4!(e2)$) {$D_{N}$};
    \node at ($(d3)!0.4!(e3)$) {$D_{N}$};
    \node at ($(e1)!0.4!(f1)$) {$D_{N+2}$};
    \end{tikzpicture}}
\end{equation}
The moduli space $\mathcal{M}$ of $O(2r)-[C_N]$ consists of $r+1$ branches: its Coulomb branch, $r-1$ mixed branches, and its Higgs branch:

\begin{equation}
    \begin{split}
        \mathcal{M}(O(2r)-[C_N])\quad=\quad&\mathcal{C}(O(2r)-[C_N])\cup\\
        &\bigcup_{l=1}^{r-1}\mathcal{C}(O(2r-2l)-[C_{N-2l}])\times\mathcal{H}(O(2l)-[C_N])\\
        &\cup\mathcal{H}(O(2r)-[C_N])\;.
    \end{split}
\end{equation}
Note that $\mathcal{C}(O(2r)-[C_N])=\mathcal{C}(Sp(r)-[D_{N+2}])$ \cite{Cremonesi:2014uva,Bourget:2020xdz}.

\subsection{O(2r+1)-[C${}_N$], $N\geq2r+1$}
\label{sec:Oodd}
The general Hasse diagram is
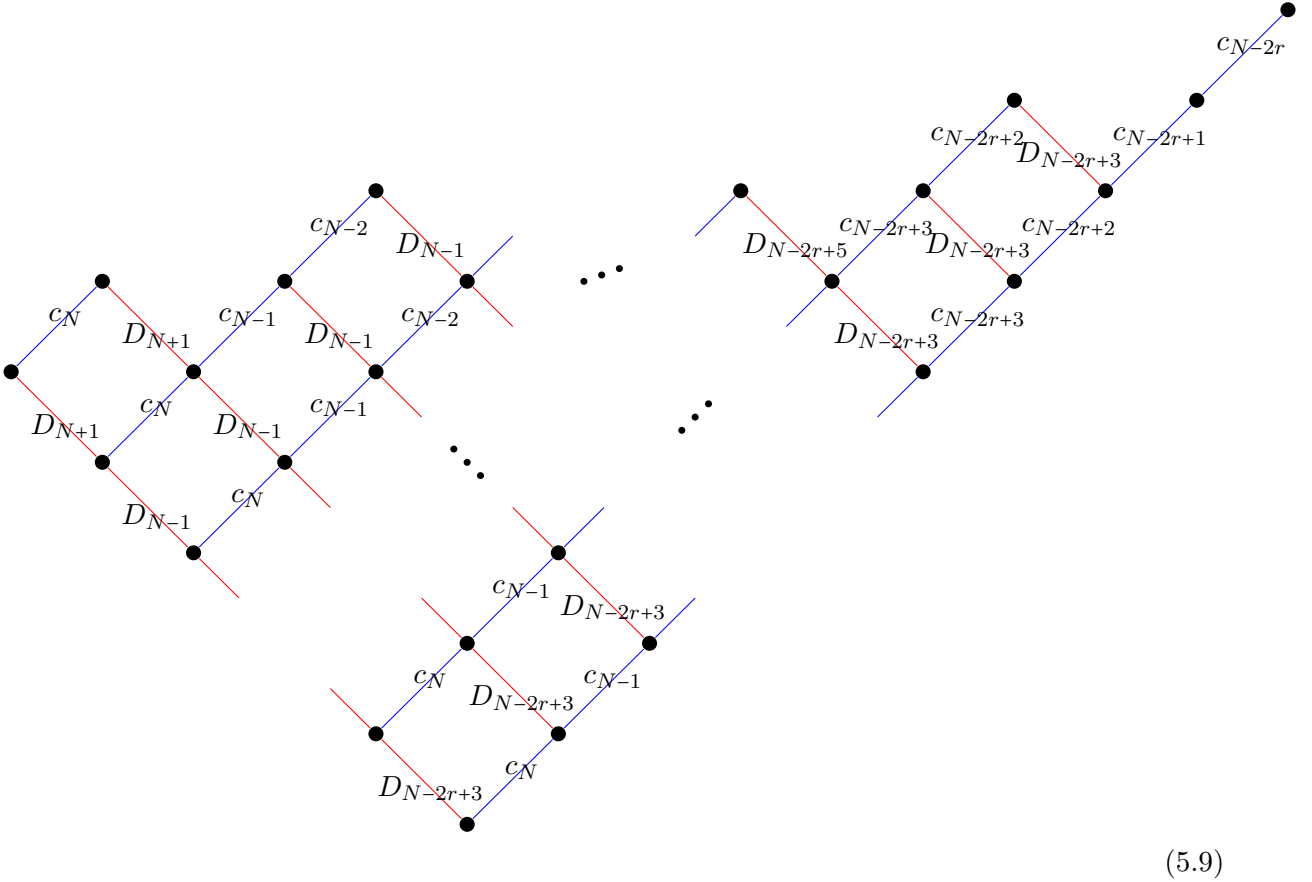
\begin{equation}
    \makebox[\textwidth][c]{\begin{tikzpicture}[scale=1.2]
    \node[hasse] (a0) at (-1,-1) {};
    \node[hasse] (b0) at (-2,0) {};
    \node[hasse] (d0) at (-4,2) {};
    \node[hasse] (e0) at (-5,3) {};
    \node[hasse] (f0) at (-6,4) {};
    \node[hasse] (a1) at (0,0) {};
    \node[hasse] (a2) at (1,1) {};
    \node[hasse] (a3) at (4,4) {};
    \node[hasse] (a4) at (5,5) {};
    \node[hasse] (a5) at (6,6) {};
    \node[hasse] (a6) at (7,7) {};
    \node[hasse] (a7) at (8,8) {};
    \node[hasse] (b1) at (-1,1) {};
    \node[hasse] (b2) at (0,2) {};
    \node[hasse] (b3) at (3,5) {};
    \node[hasse] (b4) at (4,6) {};
    \node[hasse] (b5) at (5,7) {};
    \node[hasse] (c) at (2,6) {};
    \node[hasse] (d1) at (-3,3) {};
    \node[hasse] (d2) at (-2,4) {};
    \node[hasse] (d3) at (-1,5) {};
    \node[hasse] (e1) at (-4,4) {};
    \node[hasse] (e2) at (-3,5) {};
    \node[hasse] (e3) at (-2,6) {};
    \node[hasse] (f1) at (-5,5) {};
    \draw[blue] (a0)--(a1)--(a2)--(1.5,1.5) (3.5,3.5)--(a3)--(a4)--(a5)--(a6)--(a7) (b0)--(b1)--(b2)--(0.5,2.5) (2.5,4.5)--(b3)--(b4)--(b5) (1.5,5.5)--(c) (d0)--(d1)--(d2)--(d3)--(-0.5,5.5) (e0)--(e1)--(e2)--(e3) (f0)--(f1);
    \draw[red] (a0)--(b0)--(-2.5,0.5) (-3.5,1.5)--(d0)--(e0)--(f0) (a1)--(b1)--(-1.5,1.5) (-2.5,2.5)--(d1)--(e1)--(f1) (a2)--(b2)--(-0.5,2.5) (-1.5,3.5)--(d2)--(e2) (-0.5,4.5)--(d3)--(e3) (a3)--(b3)--(c) (a4)--(b4) (a5)--(b5);
    \node at (1.5,3.5) {\Huge$\udots$};
    \node at (-1,3) {\Huge$\ddots$};
    \node[rotate=20] at (0.5,5) {\Huge$\cdots$};
    \node at ($(a0)!0.6!(a1)$) {$c_{N}$};
    \node at ($(a1)!0.6!(a2)$) {$c_{N-1}$};
    \node at ($(a3)!0.6!(a4)$) {$c_{N-2r+3}$};
    \node at ($(a4)!0.6!(a5)$) {$c_{N-2r+2}$};
    \node at ($(a5)!0.6!(a6)$) {$c_{N-2r+1}$};
    \node at ($(a6)!0.6!(a7)$) {$c_{N-2r}$};
    \node at ($(b0)!0.6!(b1)$) {$c_{N}$};
    \node at ($(b1)!0.6!(b2)$) {$c_{N-1}$};
    \node at ($(b3)!0.6!(b4)$) {$c_{N-2r+3}$};
    \node at ($(b4)!0.6!(b5)$) {$c_{N-2r+2}$};
    \node at ($(d0)!0.6!(d1)$) {$c_{N}$};
    \node at ($(d1)!0.6!(d2)$) {$c_{N-1}$};
    \node at ($(d2)!0.6!(d3)$) {$c_{N-2}$};
    \node at ($(e0)!0.6!(e1)$) {$c_{N}$};
    \node at ($(e1)!0.6!(e2)$) {$c_{N-1}$};
    \node at ($(e2)!0.6!(e3)$) {$c_{N-2}$};
    \node at ($(f0)!0.6!(f1)$) {$c_{N}$};
    \node at ($(a0)!0.4!(b0)$) {$D_{N-2r+3}$};
    \node at ($(a1)!0.4!(b1)$) {$D_{N-2r+3}$};
    \node at ($(a2)!0.4!(b2)$) {$D_{N-2r+3}$};
    \node at ($(a3)!0.4!(b3)$) {$D_{N-2r+3}$};
    \node at ($(a4)!0.4!(b4)$) {$D_{N-2r+3}$};
    \node at ($(a5)!0.4!(b5)$) {$D_{N-2r+3}$};
    \node at ($(b3)!0.4!(c)$) {$D_{N-2r+5}$};
    \node at ($(d0)!0.4!(e0)$) {$D_{N-1}$};
    \node at ($(d1)!0.4!(e1)$) {$D_{N-1}$};
    \node at ($(d2)!0.4!(e2)$) {$D_{N-1}$};
    \node at ($(d3)!0.4!(e3)$) {$D_{N-1}$};
    \node at ($(e0)!0.4!(f0)$) {$D_{N+1}$};
    \node at ($(e1)!0.4!(f1)$) {$D_{N+1}$};
    \end{tikzpicture}}
\end{equation}
The moduli space $\mathcal{M}$ of $O(2r+1)-[C_N]$ consists of $r+1$ branches: its enhanced Coulomb branch which is a type of mixed branch (l=0), $r-1$ additional mixed branches, and its Higgs branch:

\begin{equation}
    \begin{split}
        \mathcal{M}(O(2r+1)-[C_N])\quad=\quad&\bigcup_{l=0}^{r-1}\mathcal{C}(O(2r-2l+1)-[C_{N-2l}])\times\mathcal{H}(O(2l+1)-[C_N])\\
        &\cup\mathcal{H}(O(2r+1)-[C_N])\;.
    \end{split}
\end{equation}
Note that $\mathcal{C}(O(2r+1)-[C_N])=\mathcal{C}(O(2r)-[C_{N-1}])=\mathcal{C}(Sp(r)-[D_{N+1}])$ \cite{Cremonesi:2014uva,Bourget:2020xdz}.

\section{More exotic theories}
\label{exotic}

\subsection{G${}_2$-[C${}_{N}$], $N\geq3$}
\label{G2}
In section \ref{noninv} the moduli space of the entire Higgs branch of $O(k)$ theories with fundamental matter is studied, without using inversion of the Higgs branch Hasse diagram, but by studying a brane system. In this section we obtain the Hasse diagram of $G_2$ with fundamental hypermultiplets by using inversion of its Higgs branch Hasse diagram, even though it is not invertible. While this sounds contradictory, the lessons of \ref{noninv} are useful in the following study. They turn out to have remarkable implications on the moduli space of this class of theories. The Higgs branch Hasse diagram of $G_2$ with $N\geq3$ fundamental flavours is \cite{Bourget:2019aer}

\begin{equation}
    \begin{tikzpicture}
    \node[hasse] (1) at (0,0) {};
    \node[hasse] (2) at (0,1) {};
    \node[hasse] (3) at (0,2) {};
    \node[hasse] (4) at (0,3) {};
    \draw[blue] (1)--(2)--(3)--(4);
    \node at (0.4,0.5) {$c_N$};
    \node at (0.6,1.5) {$a_{2N-3}$};
    \node at (0.6,2.5) {$d_{2N-4}$};
    \end{tikzpicture}\quad\begin{tikzpicture}
    \node at (-1,0) {for};
        \node[gauge] (1) at (0,0) {};
        \node[flavour] (4) at (0,1) {};
        \draw(1)--(4);
        
        \node at (0.7,0) {$G_2$};
        \node at (0.8,1) {$C_N$};
    \end{tikzpicture}\;.
    \label{G2Higgs}
\end{equation}
This Hasse diagram is not invertible because of the presence of the $c_N$ line. However there is only a single $c_N$ line at the bottom of the Hasse diagram which suggests, that there is a $O(1)-[N]$ theory living on the Coulomb branch of $G_2$ with N flavours. This suggests that the Coulomb branch is enhanced by a $c_N$ line. This reasoning is in agreement with the $4d$ $\mathcal{N}=2$ analysis in \cite{Argyres:2016xmc}. Before proceeding we observe that the top of the Hasse diagram for the case of $N=3$ is $d_2$ which implies that the Higgs branch of this theory is a union of two cones. In turn, by inversion, this implies that the Coulomb branch has a multiple base. This requires special attention and is treated in section \ref{G2bad}. We proceed with $N\geq4$.

\subsubsection{G${}_2$-[C${}_N$], $N\geq4$}
\label{G2N>3}
Ignoring the $c_N$ and performing inversion on the rest of the Hasse diagram \eqref{G2Higgs} we obtain the Coulomb branch Hasse diagram
\begin{equation}
    \begin{tikzpicture}
    \node[hasse] (1) at (0,0) {};
    \node[hasse] (2) at (0,1) {};
    \node[hasse] (3) at (0,2) {};
    \draw[red] (1)--(2)--(3);
    \node at (0.6,0.5) {$D_{2N-4}$};
    \node at (0.6,1.5) {$A_{2N-3}$};
    \end{tikzpicture}\;,
    \label{G2Coulomb}
\end{equation}
which is the Coulomb branch Hasse diagram of $SU(3)-[2N-2]$ flavours\footnote{Note however, that while the Hasse diagrams of the Coulomb branches of $G_2-[C_N]$ and $SU(3)-[2N-2]$ are the same, the geometric spaces are different \cite{Cremonesi:2013lqa,Hanany:2016ezz}}. This is expected, as the top part of the Hasse diagram \eqref{G2Higgs} is the Higgs branch Hasse diagram of $SU(3)-[2N-2]$ flavours:

\begin{equation}
    \begin{tikzpicture}
    \node[hasse] (2) at (0,1) {};
    \node[hasse] (3) at (0,2) {};
    \node[hasse] (4) at (0,3) {};
    \draw[blue] (2)--(3)--(4);
    \node at (0.6,1.5) {$a_{2N-3}$};
    \node at (0.6,2.5) {$d_{2N-4}$};
    \end{tikzpicture}\;,
    \label{SU3Higgs}
\end{equation}
and in addition, $SU(3)$ is the commutant of $O(1)$ in $G_2$, which means that the $SU(3)$ theory is obtained from $G_2$ by partially Higgsing along the $c_N$ in the Higgs branch. One can obtain the Hasse diagram of the entire moduli space of $G_2$ with $N\geq4$ flavours by following the algorithm presented in Section \ref{Moduli} and combining \eqref{G2Higgs} and \eqref{G2Coulomb}:

    \begin{equation}
        \begin{tikzpicture}[scale=1.3]
        \node[hasse] (a1) at (0,0) {};
        \node[hasse] (a2) at (1,1) {};
        \node[hasse] (a3) at (2,2) {};
        \node[hasse] (a4) at (3,3) {};
        \node[hasse] (b1) at (-1,1) {};
        \node[hasse] (b2) at (0,2) {};
        \node[hasse] (b3) at (1,3) {};
        \node[hasse] (c1) at (-2,2) {};
        \node[hasse] (c2) at (-1,3) {};
        \draw[blue] (a1)--(a2)--(a3)--(a4) (b1)--(b2)--(b3) (c1)--(c2);
        \draw[red] (a1)--(b1)--(c1) (a2)--(b2)--(c2) (a3)--(b3);
        \node at ($(a1)!0.6!(a2)$) {$c_{N}$};
        \node at ($(a2)!0.6!(a3)$) {$a_{2N-3}$};
        \node at ($(a3)!0.6!(a4)$) {$d_{2N-4}$};
        \node at ($(b1)!0.6!(b2)$) {$c_{N}$};
        \node at ($(b2)!0.6!(b3)$) {$a_{2N-3}$};
        \node at ($(c1)!0.6!(c2)$) {$c_{N}$};
        \node at ($(a1)!0.4!(b1)$) {$D_{2N-4}$};
        \node at ($(a2)!0.4!(b2)$) {$D_{2N-4}$};
        \node at ($(a3)!0.4!(b3)$) {$D_{2N-4}$};
        \node at ($(b1)!0.4!(c1)$) {$A_{2N-3}$};
        \node at ($(b2)!0.4!(c2)$) {$A_{2N-3}$};
        \end{tikzpicture}\quad\begin{tikzpicture}
    \node at (-1,0) {for};
        \node[gauge] (1) at (0,0) {};
        \node[flavour] (4) at (0,1) {};
        \draw(1)--(4);
        
        \node at (0.7,0) {$G_2$};
        \node at (0.7,1) {$C_N$};
    \end{tikzpicture}\;.
    \end{equation}
    We get the nice result that the moduli space of $G_2$ with $N\geq4$ flavours consists of three branches: An enhanced Coulomb branch, a mixed branch and a Higgs branch. However the geometry of the enhanced Coulomb branch is unclear, as $\mathcal{C}(G_2-[C_N])\neq\mathcal{C}(SU(3)-[2N-2])$ \cite{Cremonesi:2013lqa,Hanany:2016ezz}, and hence it is difficult to tell if the enhanced Coulomb branch is a product of $\mathcal{C}(G_2-[C_N])$ and $\mathcal{H}(O(1)-[C_N])$, or a more complicated space. We therefore only make a claim about the Hasse diagram of the full moduli space and leave further investigation for future work.

\subsubsection{G${}_2$-[C${}_3$]}
\label{G2bad}

    The Hasse diagram of the classical Higgs branch of $G_2$ with 3 flavours is.
    
    
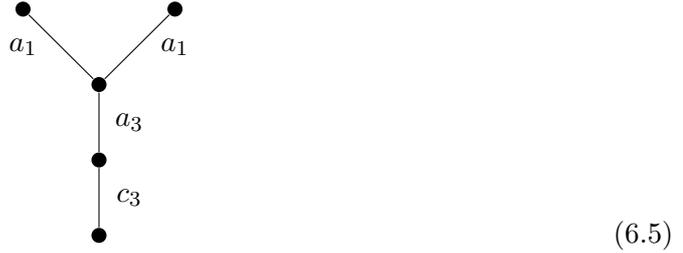
\begin{equation}
        \begin{tikzpicture}
        \node[hasse] (1) at (0,0) {};
        \node[hasse] (2) at (0,1) {};
        \node[hasse] (3) at (0,2) {};
        \node[hasse] (4) at (-1,3) {};
        \node[hasse] (5) at (1,3) {};
        \draw (1)--(2)--(3)--(4) (3)--(5);
        \node at (0.4,0.5) {$c_3$};
        \node at (0.4,1.5) {$a_3$};
        \node at (1,2.5) {$a_1$};
        \node at (-1,2.5) {$a_1$};
        \end{tikzpicture}
        \label{eq:G23}
    \end{equation}
    The case $N=3$ is special, since the top of the Higgs branch Hasse diagram is $d_2=a_1\cup a_1$, which is an indication that theory theory is bad. Nevertheless, following the approach of Section \ref{Bad} by using inversion, and the approach of Section \ref{G2N>3} by ignoring the single $c_3$ line, we obtain the Coulomb branch Hasse diagram:
    
    
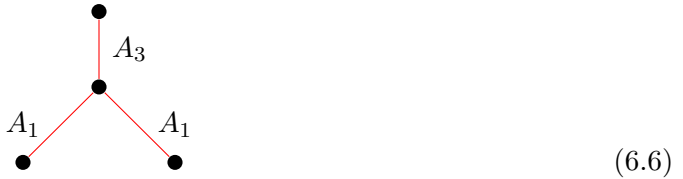
\begin{equation}
        \begin{tikzpicture}
        \node[hasse] (1) at (-1,0) {};
        \node[hasse] (2) at (1,0) {};
        \node[hasse] (3) at (0,1) {};
        \node[hasse] (4) at (0,2) {};
        \draw[red] (1)--(3)--(2) (3)--(4);
        \node at (-1,0.5) {$A_1$};
        \node at (1,0.5) {$A_1$};
        \node at (0.4,1.5) {$A_3$};
        \end{tikzpicture}
    \end{equation}
    Which is the Coulomb branch Hasse diagram of $SU(3)-[4]$ \eqref{SU(3)4}. The $c_3$ line in \eqref{eq:G23} is again an indication of an enhanced Coulomb branch, the Hasse diagram of the entire moduli space is proposed to be
    
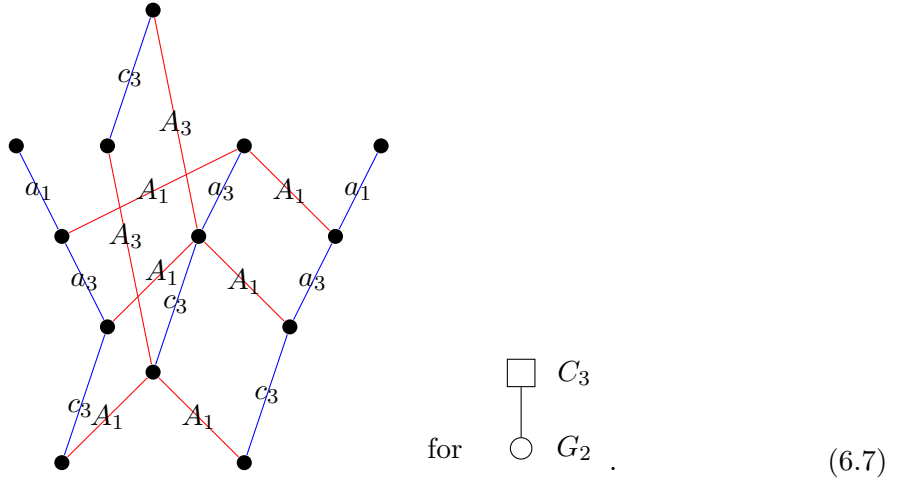
\begin{equation}
        \begin{tikzpicture}[scale=1.2]
        \node[hasse] (1) at (-1,0) {};
        \node[hasse] (2) at (1,0) {};
        \node[hasse] (3) at (0,1) {};
        \node[hasse] (4) at (-0.5,3.5) {};
        \draw[red] (1)--(3)--(2) (3)--(4);
        \node[hasse] (a1) at (-0.5,1.5) {};
        \node[hasse] (a2) at (1.5,1.5) {};
        \node[hasse] (a3) at (0.5,2.5) {};
        \node[hasse] (a4) at (0,5) {};
        \draw[blue] (1)--(a1) (2)--(a2) (3)--(a3) (4)--(a4);
        \draw[red] (a1)--(a3)--(a4) (a2)--(a3);
        \node[hasse] (b1) at (-1,2.5) {};
        \node[hasse] (b2) at (2,2.5) {};
        \node[hasse] (b3) at (1,3.5) {};
        \draw[blue] (a1)--(b1) (a2)--(b2) (a3)--(b3);
        \draw[red] (b1)--(b3)--(b2);
        \node[hasse] (c1) at (-1.5,3.5) {};
        \node[hasse] (c2) at (2.5,3.5) {};
        \draw[blue] (b1)--(c1) (b2)--(c2);
        \node at ($(1)!0.5!(3)$) {$A_1$};
        \node at ($(2)!0.5!(3)$) {$A_1$};
        \node at ($(a1)!0.6!(a3)$) {$A_1$};
        \node at ($(a2)!0.5!(a3)$) {$A_1$};
        \node at ($(b1)!0.5!(b3)$) {$A_1$};
        \node at ($(b2)!0.5!(b3)$) {$A_1$};
        \node at ($(3)!0.6!(4)$) {$A_3$};
        \node at ($(a3)!0.5!(a4)$) {$A_3$};
        \node at ($(1)!0.4!(a1)$) {$c_3$};
        \node at ($(2)!0.5!(a2)$) {$c_3$};
        \node at ($(3)!0.5!(a3)$) {$c_3$};
        \node at ($(4)!0.5!(a4)$) {$c_3$};
        \node at ($(a1)!0.5!(b1)$) {$a_3$};
        \node at ($(a2)!0.5!(b2)$) {$a_3$};
        \node at ($(a3)!0.5!(b3)$) {$a_3$};
        \node at ($(b1)!0.5!(c1)$) {$a_1$};
        \node at ($(b2)!0.5!(c2)$) {$a_1$};
        \end{tikzpicture}\quad\begin{tikzpicture}
    \node at (-1,0) {for};
        \node[gauge] (1) at (0,0) {};
        \node[flavour] (4) at (0,1) {};
        \draw(1)--(4);
        
        \node at (0.7,0) {$G_2$};
        \node at (0.7,1) {$C_3$};
    \end{tikzpicture}\;.
    \end{equation}
    We can identify the Hasse diagram of $SU(3)-[4]$ \eqref{SU(3)4} as a subdiagram. The moduli space of $G_2-[C_3]$ is conjectured to consist of 4 cones: An enhanced Coulomb branch, which is the product of the Coulomb branch of $G_2-[C_3]$ (with two bases) with the Higgs branch of $O(1)-[C_3]$; two Higgs branches which emanate from different bases; and a mixed branch, which is the product of the Coulomb branch (with two bases) of $SU(2)-[2]$ and the Higgs branch of $SO(3)-[C_3]$.
    
\subsection{A $C_N$ flavour theory}
    We may use the lessons learned from the $G_2$ case to study more exotic theories, which do not have an electric quiver description. For example, consider the dimensional reduction of the $4d$ $\mathcal{N}=2$ SCFT with $C_5$ flavour symmetry given in Table 1 of \cite{Argyres:2016xmc}, which we can call $X_5$. We know that the global symmetry is $Sp(5)$, hence (assuming there exists a unitary magnetic quiver) the magnetic quiver describing the Higgs branch has to be minimally unbalanced C-type. Minimally unbalanced quivers are classified in \cite{Cabrera:2018uvz}. We are also provided with the dimension of the Higgs branch which is $dim_{\mathbb{H}}(\mathcal{H})=16$, the only option for the magnetic quiver is:

\begin{equation}
\begin{tikzpicture}
    \node[gauge,label=below:{1}] (1) at (0,0) {};
    \node[gauge,label=below:{2}] (2) at (1,0) {};
    \node[gauge,label=below:{3}] (3) at (2,0) {};
    \node[gauge,label=below:{4}] (4) at (3,0) {};
    \node[gauge,label=below:{5}] (5) at (4,0) {};
    \node[gauge,label=below:{2}] (6) at (5,0) {};
    \draw (1)--(2)--(3)--(4) (5)--(6) (4.8,-0.2)--(5.2,-0.2)--(5.2,0.2)--(4.8,0.2)--(4.8,-0.2) (3.6,0.2)--(3.4,0)--(3.6,-0.2);
    \draw[transform canvas={yshift=-1.5pt}] (4)--(5);
    \draw[transform canvas={yshift=1.5pt}] (4)--(5);
\end{tikzpicture}
\label{eq:C5E6quiv}
\end{equation}
where the gauge node with a box around it is the particular choice of \emph{ungauging scheme}, the topic of investigation in \cite{Hanany:2020jzl}. The  Coulomb branch Hasse diagram of \eqref{eq:C5E6quiv} is \cite[Conclusion]{Bourget:2019aer}:
\begin{equation}
    \HC\eqref{eq:C5E6quiv}=
    \begin{tikzpicture}
    \node[hasse] (1) at (0,0) {};
    \node[hasse] (2) at (0,1) {};
    \node[hasse] (3) at (0,2) {};
    \draw[red] (1)--(2)--(3);
    \node at (0.4,0.5) {$c_5$};
    \node at (0.4,1.5) {$e_6$};
    \end{tikzpicture}
\end{equation}
As \eqref{eq:C5E6quiv} is the magnetic quiver of $X_5$, the Higgs branch Hasse diagram of $X_5$ is
\begin{equation}
    \HH(X)=
    \begin{tikzpicture}
    \node[hasse] (1) at (0,0) {};
    \node[hasse] (2) at (0,1) {};
    \node[hasse] (3) at (0,2) {};
    \draw[blue] (1)--(2)--(3);
    \node at (0.4,0.5) {$c_5$};
    \node at (0.4,1.5) {$e_6$};
    \end{tikzpicture}\;. 
\end{equation}
The $c_5$ line is an indication of the existence of an enhanced Coulomb branch, which is indeed what \cite{Argyres:2016xmc} find. Furthermore the 4d theory is a SCFT of rank 1 and hence has only a one-dimensional Coulomb branch. Using inversion in a similar manner as in Section \ref{G2}, by ignoring the $c_5$ line, we obtain that the Coulomb branch is $E_6$. The Hasse diagram of the entire moduli space is thus:

\begin{equation}
        \begin{tikzpicture}[scale=1.3]
        \node[hasse] (a1) at (0,0) {};
        \node[hasse] (a2) at (1,1) {};
        \node[hasse] (a3) at (2,2) {};
        \node[hasse] (b1) at (-1,1) {};
        \node[hasse] (b2) at (0,2) {};
        \draw[blue] (a1)--(a2)--(a3) (b1)--(b2);
        \draw[red] (a1)--(b1) (a2)--(b2);
        \node at ($(a1)!0.6!(a2)$) {$c_{5}$};
        \node at ($(a2)!0.6!(a3)$) {$e_6$};
        \node at ($(b1)!0.6!(b2)$) {$c_{5}$};
        \node at ($(a1)!0.4!(b1)$) {$E_6$};
        \node at ($(a2)!0.4!(b2)$) {$E_6$};
        \end{tikzpicture}\;.
\end{equation}
And we see that there is a partial Higgsing to the $E_6$ theory mentioned in Table \ref{tab:ADEguys}. This agrees with the analysis in $4$d \cite{Apruzzi:2020pmv}. A study of the entire moduli space of the dimensional reductions of $4d$ $\mathcal{N}=2$ SCFTs would be an interesting undertaking and could provide inspiration for the study of the moduli space of the $4d$ theories.\\

The magnetic quiver \eqref{eq:C5E6quiv} is part of the family:

\begin{equation}
\begin{tikzpicture}
    \node[gauge,label=below:{1}] (1) at (0,0) {};
    \node[gauge,label=below:{2}] (2) at (1,0) {};
    \node (3) at (2,0) {$\cdots$};
    \node[gauge,label=below:{N-1}] (4) at (3,0) {};
    \node[gauge,label=below:{N}] (5) at (4,0) {};
    \node[gauge,label=below:{2}] (6) at (5,0) {};
    \draw (1)--(2)--(3)--(4) (5)--(6) (4.8,-0.2)--(5.2,-0.2)--(5.2,0.2)--(4.8,0.2)--(4.8,-0.2) (3.6,0.2)--(3.4,0)--(3.6,-0.2);
    \draw[transform canvas={yshift=-1.5pt}] (4)--(5);
    \draw[transform canvas={yshift=1.5pt}] (4)--(5);
\end{tikzpicture}
\label{eq:CNE6quiv}
\end{equation}
The Coulomb branch Hasse diagram of \eqref{eq:CNE6quiv} for $N\geq5$ is \cite[Conclusion]{Bourget:2019aer}:
\begin{equation}
    \begin{tikzpicture}
    \node[hasse] (1) at (0,0) {};
    \node[hasse] (2) at (0,1) {};
    \node[hasse] (3) at (0,2) {};
    \node[hasse] (4) at (0,3) {};
    \node[hasse] (5) at (0,4) {};
    \node[hasse] (6) at (0,5) {};
    \draw[red] (1)--(2)--(3)--(0,2.3) (0,2.7)--(4)--(5)--(6);
    \node at (0,2.5) {$\vdots$};
    \node at (0.5,0.5) {$c_N$};
    \node at (0.5,1.5) {$c_{N-1}$};
    \node at (0.5,3.5) {$c_5$};
    \node at (0.5,4.5) {$e_6$};
    \end{tikzpicture}
\end{equation}
If \eqref{eq:CNE6quiv} is the magnetic quiver of a 3d $\mathcal{N}=4$ theory $X_N$, then this theory has a Higgs branch Hasse diagram:
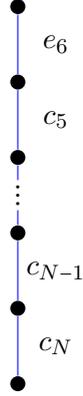
\begin{equation}
    \begin{tikzpicture}
    \node[hasse] (1) at (0,0) {};
    \node[hasse] (2) at (0,1) {};
    \node[hasse] (3) at (0,2) {};
    \node[hasse] (4) at (0,3) {};
    \node[hasse] (5) at (0,4) {};
    \node[hasse] (6) at (0,5) {};
    \draw[blue] (1)--(2)--(3)--(0,2.3) (0,2.7)--(4)--(5)--(6);
    \node at (0,2.5) {$\vdots$};
    \node at (0.5,0.5) {$c_N$};
    \node at (0.5,1.5) {$c_{N-1}$};
    \node at (0.5,3.5) {$c_5$};
    \node at (0.5,4.5) {$e_6$};
    \end{tikzpicture}
    \label{eq:XnHiggs}
\end{equation}
Since the top of the Higgs branch Hasse digram \eqref{eq:XnHiggs} is an $e_6$ transition, we expect from inversion, that the bottom of the Coulomb branch Hasse diagram of $X_N$ will be $E_6$, as for $X_5$. Furthermore we see that the $c$-part of the Hasse diagram is the Higgs branch Hasse diagram of $O(N-4)-[C_N]$, see Sections \ref{sec:Oeven} and \ref{sec:Oodd}. This indicates, that the $O(N-4)-[C_N]$ theory is reached from $X_n$ through a partial Higgsing along the Coulomb branch of $X_N$, moving along the $E_6$ line in the Coulomb branch. Hence we can obtain the Hasse diagram of the entire moduli space of the $X_N$ theory by modifying the Hasse diagram of $O(N-4)-[C_N]$. We obtain:

\paragraph{N even:}
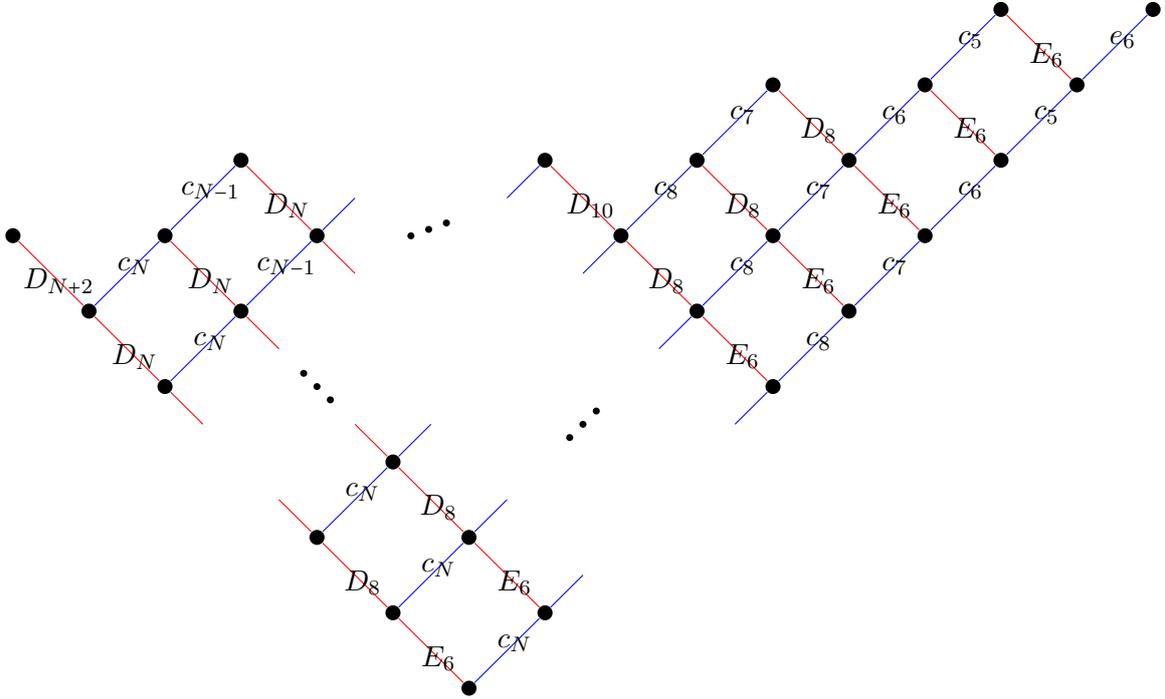
\begin{equation}
    \makebox[\textwidth][c]{\begin{tikzpicture}[scale=1]
    \node[hasse] (a1) at (0,0) {};
    \node[hasse] (a2) at (1,1) {};
    \node[hasse] (a3) at (4,4) {};
    \node[hasse] (a4) at (5,5) {};
    \node[hasse] (a5) at (6,6) {};
    \node[hasse] (a6) at (7,7) {};
    \node[hasse] (a7) at (8,8) {};
    \node[hasse] (b1) at (-1,1) {};
    \node[hasse] (b2) at (0,2) {};
    \node[hasse] (b3) at (3,5) {};
    \node[hasse] (b4) at (4,6) {};
    \node[hasse] (b5) at (5,7) {};
    \node[hasse] (c) at (2,6) {};
    \node[hasse] (d1) at (-3,3) {};
    \node[hasse] (d2) at (-2,4) {};
    \node[hasse] (d3) at (-1,5) {};
    \node[hasse] (e1) at (-4,4) {};
    \node[hasse] (e2) at (-3,5) {};
    \node[hasse] (e3) at (-2,6) {};
    \node[hasse] (f1) at (-5,5) {};
    \draw[blue] (a1)--(a2)--(1.5,1.5) (3.5,3.5)--(a3)--(a4)--(a5)--(a6)--(a7) (b1)--(b2)--(0.5,2.5) (2.5,4.5)--(b3)--(b4)--(b5) (1.5,5.5)--(c) (d1)--(d2)--(d3)--(-0.5,5.5) (e1)--(e2)--(e3);
    \draw[red] (a1)--(b1)--(-1.5,1.5) (-2.5,2.5)--(d1)--(e1)--(f1) (a2)--(b2)--(-0.5,2.5) (-1.5,3.5)--(d2)--(e2) (-0.5,4.5)--(d3)--(e3) (a3)--(b3)--(c) (a4)--(b4) (a5)--(b5);
    \node at (2.5,2.5) {\Huge$\udots$};
    \node at (-1,3) {\Huge$\ddots$};
    \node[rotate=20] at (0.5,5) {\Huge$\cdots$};
    \node at ($(a1)!0.6!(a2)$) {$c_{N}$};
    \node at ($(a3)!0.6!(a4)$) {$c_{8}$};
    \node at ($(a4)!0.6!(a5)$) {$c_{7}$};
    \node at ($(a5)!0.6!(a6)$) {$c_{6}$};
    \node at ($(a6)!0.6!(a7)$) {$c_{5}$};
    \node at ($(b1)!0.6!(b2)$) {$c_{N}$};
    \node at ($(b3)!0.6!(b4)$) {$c_{8}$};
    \node at ($(b4)!0.6!(b5)$) {$c_{7}$};
    \node at ($(d1)!0.6!(d2)$) {$c_{N}$};
    \node at ($(d2)!0.6!(d3)$) {$c_{N-1}$};
    \node at ($(e1)!0.6!(e2)$) {$c_{N}$};
    \node at ($(e2)!0.6!(e3)$) {$c_{N-1}$};
    \node at ($(a1)!0.4!(b1)$) {$D_{8}$};
    \node at ($(a2)!0.4!(b2)$) {$D_{8}$};
    \node at ($(a3)!0.4!(b3)$) {$D_{8}$};
    \node at ($(a4)!0.4!(b4)$) {$D_{8}$};
    \node at ($(a5)!0.4!(b5)$) {$D_{8}$};
    \node at ($(b3)!0.4!(c)$) {$D_{10}$};
    \node at ($(d1)!0.4!(e1)$) {$D_{N}$};
    \node at ($(d2)!0.4!(e2)$) {$D_{N}$};
    \node at ($(d3)!0.4!(e3)$) {$D_{N}$};
    \node at ($(e1)!0.4!(f1)$) {$D_{N+2}$};
    \node[hasse] (1) at (1,-1) {};
    \node[hasse] (2) at (2,0) {};
    \node[hasse] (3) at (5,3) {};
    \node[hasse] (4) at (6,4) {};
    \node[hasse] (5) at (7,5) {};
    \node[hasse] (6) at (8,6) {};
    \node[hasse] (7) at (9,7) {};
    \node[hasse] (8) at (10,8) {};
    \draw[red] (1)--(a1) (2)--(a2) (3)--(a3) (4)--(a4) (5)--(a5) (6)--(a6) (7)--(a7);
    \draw[blue] (1)--(2)--(2.5,0.5) (4.5,2.5)--(3)--(4)--(5)--(6)--(7)--(8);
    \node at ($(1)!0.4!(a1)$) {$E_6$};
    \node at ($(2)!0.4!(a2)$) {$E_6$};
    \node at ($(3)!0.4!(a3)$) {$E_6$};
    \node at ($(4)!0.4!(a4)$) {$E_6$};
    \node at ($(5)!0.4!(a5)$) {$E_6$};
    \node at ($(6)!0.4!(a6)$) {$E_6$};
    \node at ($(7)!0.4!(a7)$) {$E_6$};
    \node at ($(1)!0.6!(2)$) {$c_{N}$};
    \node at ($(3)!0.6!(4)$) {$c_{8}$};
    \node at ($(4)!0.6!(5)$) {$c_{7}$};
    \node at ($(5)!0.6!(6)$) {$c_{6}$};
    \node at ($(6)!0.6!(7)$) {$c_{5}$};
    \node at ($(7)!0.6!(8)$) {$e_6$};
    \end{tikzpicture}}
\end{equation}
The moduli space $\mathcal{M}$ of $X_N$, for N even, consists of $\frac{N}{2}$ branches: its Coulomb branch, $\frac{N}{2}-2$ mixed branches, and its Higgs branch:

\begin{equation}
    \begin{split}
        \mathcal{M}(X_N)\quad=\quad&\mathcal{C}(X_N)\cup\\
        &\bigcup_{l=1}^{\frac{N}{2}-2}\mathcal{C}(X_{N-2l})\times\mathcal{H}(O(2l)-[C_N])\\
        &\cup\mathcal{H}(X_N)\;.
    \end{split}
\end{equation}

\paragraph{N odd:}
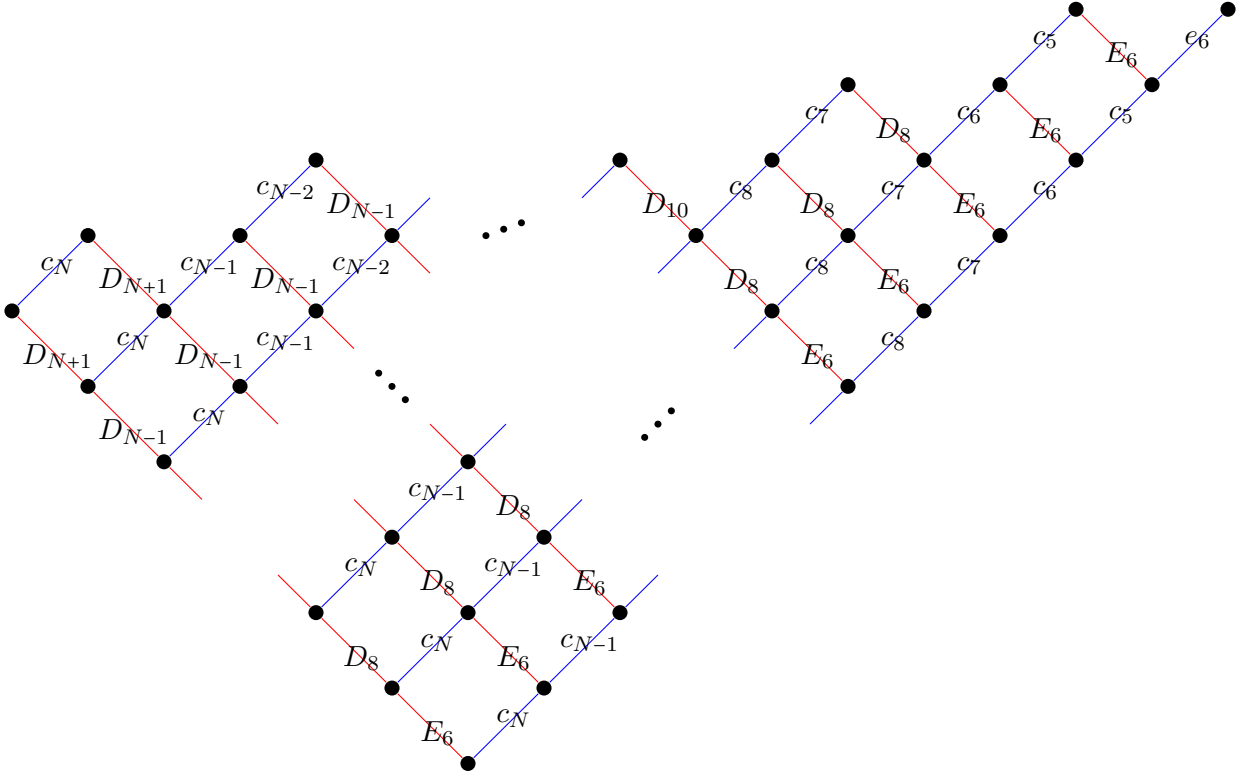
\begin{equation}
    \makebox[\textwidth][c]{\begin{tikzpicture}[scale=1]
    \node[hasse] (a0) at (-1,-1) {};
    \node[hasse] (b0) at (-2,0) {};
    \node[hasse] (d0) at (-4,2) {};
    \node[hasse] (e0) at (-5,3) {};
    \node[hasse] (f0) at (-6,4) {};
    \node[hasse] (a1) at (0,0) {};
    \node[hasse] (a2) at (1,1) {};
    \node[hasse] (a3) at (4,4) {};
    \node[hasse] (a4) at (5,5) {};
    \node[hasse] (a5) at (6,6) {};
    \node[hasse] (a6) at (7,7) {};
    \node[hasse] (a7) at (8,8) {};
    \node[hasse] (b1) at (-1,1) {};
    \node[hasse] (b2) at (0,2) {};
    \node[hasse] (b3) at (3,5) {};
    \node[hasse] (b4) at (4,6) {};
    \node[hasse] (b5) at (5,7) {};
    \node[hasse] (c) at (2,6) {};
    \node[hasse] (d1) at (-3,3) {};
    \node[hasse] (d2) at (-2,4) {};
    \node[hasse] (d3) at (-1,5) {};
    \node[hasse] (e1) at (-4,4) {};
    \node[hasse] (e2) at (-3,5) {};
    \node[hasse] (e3) at (-2,6) {};
    \node[hasse] (f1) at (-5,5) {};
    \draw[blue] (a0)--(a1)--(a2)--(1.5,1.5) (3.5,3.5)--(a3)--(a4)--(a5)--(a6)--(a7) (b0)--(b1)--(b2)--(0.5,2.5) (2.5,4.5)--(b3)--(b4)--(b5) (1.5,5.5)--(c) (d0)--(d1)--(d2)--(d3)--(-0.5,5.5) (e0)--(e1)--(e2)--(e3) (f0)--(f1);
    \draw[red] (a0)--(b0)--(-2.5,0.5) (-3.5,1.5)--(d0)--(e0)--(f0) (a1)--(b1)--(-1.5,1.5) (-2.5,2.5)--(d1)--(e1)--(f1) (a2)--(b2)--(-0.5,2.5) (-1.5,3.5)--(d2)--(e2) (-0.5,4.5)--(d3)--(e3) (a3)--(b3)--(c) (a4)--(b4) (a5)--(b5);
    \node at (2.5,2.5) {\Huge$\udots$};
    \node at (-1,3) {\Huge$\ddots$};
    \node[rotate=20] at (0.5,5) {\Huge$\cdots$};
    \node at ($(a0)!0.6!(a1)$) {$c_{N}$};
    \node at ($(a1)!0.6!(a2)$) {$c_{N-1}$};
    \node at ($(a3)!0.6!(a4)$) {$c_{8}$};
    \node at ($(a4)!0.6!(a5)$) {$c_{7}$};
    \node at ($(a5)!0.6!(a6)$) {$c_{6}$};
    \node at ($(a6)!0.6!(a7)$) {$c_{5}$};
    \node at ($(b0)!0.6!(b1)$) {$c_{N}$};
    \node at ($(b1)!0.6!(b2)$) {$c_{N-1}$};
    \node at ($(b3)!0.6!(b4)$) {$c_{8}$};
    \node at ($(b4)!0.6!(b5)$) {$c_{7}$};
    \node at ($(d0)!0.6!(d1)$) {$c_{N}$};
    \node at ($(d1)!0.6!(d2)$) {$c_{N-1}$};
    \node at ($(d2)!0.6!(d3)$) {$c_{N-2}$};
    \node at ($(e0)!0.6!(e1)$) {$c_{N}$};
    \node at ($(e1)!0.6!(e2)$) {$c_{N-1}$};
    \node at ($(e2)!0.6!(e3)$) {$c_{N-2}$};
    \node at ($(f0)!0.6!(f1)$) {$c_{N}$};
    \node at ($(a0)!0.4!(b0)$) {$D_{8}$};
    \node at ($(a1)!0.4!(b1)$) {$D_{8}$};
    \node at ($(a2)!0.4!(b2)$) {$D_{8}$};
    \node at ($(a3)!0.4!(b3)$) {$D_{8}$};
    \node at ($(a4)!0.4!(b4)$) {$D_{8}$};
    \node at ($(a5)!0.4!(b5)$) {$D_{8}$};
    \node at ($(b3)!0.4!(c)$) {$D_{10}$};
    \node at ($(d0)!0.4!(e0)$) {$D_{N-1}$};
    \node at ($(d1)!0.4!(e1)$) {$D_{N-1}$};
    \node at ($(d2)!0.4!(e2)$) {$D_{N-1}$};
    \node at ($(d3)!0.4!(e3)$) {$D_{N-1}$};
    \node at ($(e0)!0.4!(f0)$) {$D_{N+1}$};
    \node at ($(e1)!0.4!(f1)$) {$D_{N+1}$};
    \node[hasse] (0) at (0,-2) {};
    \node[hasse] (1) at (1,-1) {};
    \node[hasse] (2) at (2,0) {};
    \node[hasse] (3) at (5,3) {};
    \node[hasse] (4) at (6,4) {};
    \node[hasse] (5) at (7,5) {};
    \node[hasse] (6) at (8,6) {};
    \node[hasse] (7) at (9,7) {};
    \node[hasse] (8) at (10,8) {};
    \draw[red] (0)--(a0) (1)--(a1) (2)--(a2) (3)--(a3) (4)--(a4) (5)--(a5) (6)--(a6) (7)--(a7);
    \draw[blue] (0)--(1)--(2)--(2.5,0.5) (4.5,2.5)--(3)--(4)--(5)--(6)--(7)--(8);
    \node at ($(0)!0.4!(a0)$) {$E_6$};
    \node at ($(1)!0.4!(a1)$) {$E_6$};
    \node at ($(2)!0.4!(a2)$) {$E_6$};
    \node at ($(3)!0.4!(a3)$) {$E_6$};
    \node at ($(4)!0.4!(a4)$) {$E_6$};
    \node at ($(5)!0.4!(a5)$) {$E_6$};
    \node at ($(6)!0.4!(a6)$) {$E_6$};
    \node at ($(7)!0.4!(a7)$) {$E_6$};
    \node at ($(0)!0.6!(1)$) {$c_{N}$};
    \node at ($(1)!0.6!(2)$) {$c_{N-1}$};
    \node at ($(3)!0.6!(4)$) {$c_{8}$};
    \node at ($(4)!0.6!(5)$) {$c_{7}$};
    \node at ($(5)!0.6!(6)$) {$c_{6}$};
    \node at ($(6)!0.6!(7)$) {$c_{5}$};
    \node at ($(7)!0.6!(8)$) {$e_6$};
    \end{tikzpicture}}
\end{equation}
The moduli space $\mathcal{M}$ of $X_N$, for N odd, consists of $\frac{N-1}{2}$ branches: its enhanced Coulomb branch (l=0), $\frac{N-1}{2}-2$ further mixed branches, and its Higgs branch:

\begin{equation}
    \begin{split}
        \mathcal{M}(X_N)\quad=\quad&\bigcup_{l=0}^{\frac{N-1}{2}-2}\mathcal{C}(X_{N-2l})\times\mathcal{H}(O(2l+1)-[C_N])\\
        &\cup\mathcal{H}(X_N)\;.
    \end{split}
\end{equation}
Note that $\mathcal{C}(X_{2r+1})=\mathcal{C}(X_{2r})$.

\section{Conclusion and Outlook}
\label{Outlook}
This paper argues that for certain $3d$ $\mathcal{N}=4$ quiver gauge theories the Hasse diagram of the Coulomb branch and the classical Higgs branch are related by an operation called inversion. The Hasse diagram of the full moduli space of a good or ugly theory is obtained by combining the Hasse diagrams of the Coulomb and Higgs branch at the origin and then applying inversion to the other nodes, where all transverse slices in Coulomb directions are coloured red and all transverse slices in Higgs directions are coloured blue. 3d Mirror symmetry expresses itself in the Hasse diagram by colouring the red lines blue and the blue lines red.\\

An example of how to use the information of the classical Higgs branch and the method of inversion to obtain information about the Coulomb branch of a bad theory is given. If the proposed conjecture holds then there exists a simple method to obtain the singularity structure of the Coulomb branch of many bad theories. As one can construct theories which have a Higgs branch consisting of any number of cones \cite{Ncones}, which are not all identical, this opens up interesting questions about the singularity structure of their Coulomb branches.\\

When a Hasse diagram consists of transitions which are not Kleinian singularities or closures of minimal nilpotent orbits of ADE then one cannot apply inversion in a straight forward manner. However, we are able to obtain the Hasse diagram of the entire moduli space for $O(k)$ theories from a brane construction. For $O(2r+1)$ theories we observe the existence of an enhanced Coulomb branch. For $G_2$ a combination of inversion and knowledge of the enhanced Coulomb branch is used to obtain the Hasse diagram of the entire moduli space, although the Hasse diagram, strictly speaking, is not invertible.\\

\paragraph{Outlook}

As shown in \cite{Bourget:2019rtl} the classical Higgs branch coordinate ring may contain nilpotent operators and the Higgs branch is therefore not a variety but rather a non-reduced scheme. How to include this information remains to be explored. The case of incomplete Higgsing is omitted and deserves attention.\\

Another interesting question to ask is what happens to the Hasse diagram, when (only some) mass/FI parameters are turned on, (partially) lifting one branch and resolving the other. In this case we expect only a subdiagram of the Hasse diagram of the full moduli space to remain, where the bottom node does not correspond to the origin, but rather to a resolution/deformation of a singularity. A systematic study of this remains to be done.\\

Hasse diagrams may be used to study the symplectic duality proposed in \cite{Webster1407}. This duality was explored as a duality between the Coulomb and Higgs branch of a $3d$ $\mathcal{N}=4$ quiver gauge theory in several works. See for example \cite{Nakajima:2015txa,Nakajima2015,Braverman:2016wma,Webster2016arXiv161106541W} from a mathematical perspective and \cite{Bullimore:2016nji,Balasubramanian:2018pbp} from a more physical perspective. Studying the relationship between inversion and symplectic duality may lead to a better understanding of symplectic singularities, and even moduli spaces which are not symplectic singularities.\\

One can ask what part of this analysis survives in higher dimensions. Clearly the Coulomb branch in dimensions larger that 3 is no longer hyper-K\"ahler, hence the analysis is quite different. However one can hope to find a Hasse diagram of the entire moduli space, where the Coulomb part does not represent the stratification of a symplectic singularity, but rather a different geometric structure or simply the partial ordering given by Higgsing the theory.\\

Coulomb branch moduli were used to identify the leaves of the Higgs branch of 5d $\mathcal{N}=1$ theories living on $(p,q)-5$brane webs in \cite{Bourget:2019aer}. However 5d gauge theories should be viewed as flowing from a UV fixed point, which is a 5d $\mathcal{N}=1$ SCFT. A characterisation of the full moduli space of a 5d $\mathcal{N}=1$ SCFT and the Higgs branch as a function of the various (global and local) deformations of the brane web remains an interesting open problem.\\

In six dimensions there is no notion of a Coulomb branch, but there is a Tensor branch, and a Hasse diagram describing the moduli space of a SCFT in terms of Higgs and Tensor branch directions could lead to a better understanding of those theories.

\section*{Acknowledgements}

We would like to thank Antoine Bourget, Rudolph Kalveks, Mario Martone, Dominik Miketa, Jamie Rogers, Travis Schedler, Marcus Sperling, Ben Webster and Zhenghao Zhong for helpful discussions. This work was supported by the STFC Consolidated Grant  ST/P000762/1.

\appendix

\section{Quiver Subtraction}
\label{app:subtract}
In this Appendix we review the operation called \emph{quiver subtraction} developed in \cite{Cabrera:2018ann, Bourget:2019aer} for unitary quivers and give some additional insight for the non-unitary case, which was discussed in the realm of nilpotent orbits in \cite{Hanany:2019tji}. We focus only on quiver subtraction of an elementary slice, when the slice is a Kleinian singularity or the closure of a minimal nilpotent orbit\footnote{This is shown to be enough for affine grassmanians of simply laced groups in \cite[Theorem A]{2003math......5095M}, where it is also noted, that the affine grassmannians of non-simply laced groups involve other elementary slices. Different types of elementary slices also appear in the realm of exceptional nilpotent orbits \cite{2015arXiv150205770F}}. We slightly abuse notation, by referring to both the quiver and its $3d$ $\mathcal{N}=4$ Coulomb branch as a 'slice' to be subtracted.\\

Note that the subtraction presented here is of the form $\mathsf{Q}-\mathsf{D}=\mathsf{Q}'$, where $\mathsf{D}$ is a elementary slice, as presented in \cite{Bourget:2019aer}. The algorithm for subtraction is different from the algorithm for $\mathsf{Q}-\mathsf{Q}'=\mathsf{D}$ in \cite{Cabrera:2018ann}.

\paragraph{Unitary quivers} For unitary quivers the algorithm presented in A.2 of \cite{Bourget:2019aer} is as follows:\\
Note: An important step in order for quiver subtraction to give the entire Hasse diagram of the Coulomb branch of a unitary quiver $\mathsf{Q}$, is to work with flavourless/unframed quivers\footnote{A \emph{framed} unitary quiver, i.e.\ a unitary quiver with flavour nodes and gauge nodes, can be turned into an \emph{unframed} unitary quiver, i.e.\ a unitary quiver with only gauge nodes, by collecting all flavours into a single U(1) node and adding enough hypers that the remaining quiver keeps its balance. To mathematicians this is known as the Crawley-Boevey trick referring to an observation in \cite{Crawley-Boevey2001}.}. Otherwise elementary slices may be missed.\\

\underline{Quiver Subtraction of an Elementary Slice:}$\qquad\mathsf{Q}-\mathsf{D}=\mathsf{Q}'$
\begin{enumerate}
    \item Align quivers $\mathsf{Q}$ and $\mathsf{D}$ (this may be done in different ways, not necessarily related by action of the automorphism group of the quiver $\mathsf{Q}$).
    \item Subtract the ranks of $\mathsf{D}$ from the ranks of $\mathsf{Q}$ to obtain a quiver $\mathsf{S}$.
    \item Restore the balance of the nodes: Add a $\mathrm{U}(1)$ node\footnote{See the Caution just below.} to $\mathsf{S}$ and connect it, with possible edge multiplicity, to the remaining non-zero nodes of $\mathsf{S}$ to create a new quiver $\mathsf{Q}'$  such that the balance of the nodes in $\mathsf{Q}'$ and $\mathsf{Q}$ match.
\end{enumerate}
Caution: there is an intricacy when subtracting the same slice twice. If the same quiver can be subtracted twice in a row, there are two possibilities:
\begin{enumerate}
    \item It is subtracted on all the same nodes as before: In this case a problem arises, when following step 3. of the algorithm. The procedure has to be modified. Details will be presented in an upcoming work.
    \item At least one node is different: In this case there is no problem.
\end{enumerate}
For U(k) with N fundamental hypermultiplets, we obtain the Coulomb branch Hasse diagram
\begin{equation}
		\begin{tikzpicture}[scale=0.8]
		\node[hasse] (1) at (0,0) {};
		\node[hasse] (2) at (0,2) {};
		\node[hasse] (3) at (0,4) {};
		\node[hasse] (4) at (0,6) {};
		\node[hasse] (5) at (0,8) {};
		\node[hasse] (6) at (0,10) {};
		\draw[red] (1)--(2)--(3) (4)--(5)--(6);
		\node at (0,5) {$\vdots$};
		\node at (1,1) {$A_{N-2k+1}$};
		\node at (1,3) {$A_{N-2k+3}$};
		\node at (0.7,7) {$A_{N-2}$};
		\node at (0.5,9) {$A_{N}$};
		\end{tikzpicture}
    \begin{tikzpicture}
    	\node at (-2,-2) {};
    	\node at (-1,0) {for};
        \node[gauge] (1) at (0,0) {};
        \node[flavour] (2) at (0,1) {};
        \draw (1)--(2);
        \node at (0.7,0) {U(k)};
        \node at (0.5,1) {N};
    \end{tikzpicture}
    \begin{tikzpicture}
    	\node at (-2,-2) {};
    	\node at (-1,0) {=};
        \node[gauge] (1) at (0,0) {};
        \node[gauge] (2) at (0,1) {};
        \draw[transform canvas={xshift=-1.5pt}] (1)--(2);
        \draw[transform canvas={xshift=1.5pt}] (1)--(2);
        \node at (0.3,0.5) {N};
        \node at (0.7,0) {U(k)};
        \node at (0.7,1) {U(1)};
    \end{tikzpicture}
\end{equation}

\paragraph{Non-Unitary quivers, simple gauge node, fundamental matter}
In the cases of quivers where not all gauge nodes are unitary, there is a complication, coming from the fact, that the fundamental representations of different groups may be complex, real or pseudo-real; opposed to the unitary case, where only complex fundamental or bifundamental representations appear. We propose, that when performing quiver subtraction $\mathsf{Q}-\mathsf{D}$ the kind of matter representation in $\mathsf{Q}$ and $\mathsf{D}$ should agree. We give examples for quivers that have a single gauge node and enough matter for the theory to be good.\\

For SU(k) with N fundamental hypermultiplets, we obtain the Coulomb branch Hasse diagram:

\begin{equation}
		\begin{tikzpicture}[scale=0.8]
		\node[hasse] (1) at (0,0) {};
		\node[hasse] (2) at (0,2) {};
		\node[hasse] (3) at (0,4) {};
		\node[hasse] (4) at (0,6) {};
		\node[hasse] (5) at (0,8) {};
		\node[hasse] (6) at (0,10) {};
		\draw[red] (1)--(2)--(3) (4)--(5)--(6);
		\node at (0,5) {$\vdots$};
		\node at (1,1) {$D_{N-2k+4}$};
		\node at (1,3) {$A_{N-2k+5}$};
		\node at (0.7,7) {$A_{N-2}$};
		\node at (0.5,9) {$A_{N}$};
		\end{tikzpicture}
    \begin{tikzpicture}
    	\node at (-2,-2) {};
    	\node at (-1,0) {for};
        \node[gauge] (1) at (0,0) {};
        \node[flavour] (2) at (0,1) {};
        \draw (1)--(2);
        \node at (0.8,0) {SU(k)};
        \node at (0.5,1) {N};
    \end{tikzpicture}
\end{equation}

For Sp(k) with 2N fundamental half-hypermultiplets, we obtain the Coulomb branch Hasse diagram:

\begin{equation}
		\begin{tikzpicture}[scale=0.8]
		\node[hasse] (1) at (0,0) {};
		\node[hasse] (2) at (0,2) {};
		\node[hasse] (3) at (0,4) {};
		\node[hasse] (4) at (0,6) {};
		\node[hasse] (5) at (0,8) {};
		\node[hasse] (6) at (0,10) {};
		\draw[red] (1)--(2)--(3) (4)--(5)--(6);
		\node at (0,5) {$\vdots$};
		\node at (1,1) {$D_{N-2k+2}$};
		\node at (0.8,3) {$D_{N-2k}$};
		\node at (0.8,7) {$D_{N-2}$};
		\node at (0.5,9) {$D_{N}$};
		\end{tikzpicture}
    \begin{tikzpicture}
    	\node at (-2,-2) {};
    	\node at (-1,0) {for};
        \node[gauge] (1) at (0,0) {};
        \node[flavour] (2) at (0,1) {};
        \draw (1)--(2);
        \node at (0.8,0) {Sp(k)};
        \node at (0.7,1) {$D_N$};
    \end{tikzpicture}
\end{equation}
Where

\begin{equation}
\begin{tikzpicture}
\node at (-1,0.5) {$A_n=$};
        \node[gauge] (1) at (0,0) {};
        \node[flavour] (2) at (0,1) {};
        \draw (1)--(2);
        \node at (0.7,0) {U(1)};
        \node at (0.5,1) {n};
\end{tikzpicture}
\begin{tikzpicture}
\node at (-2,0.5) {and};
\node at (-1,0.5) {$D_n=$};
        \node[gauge] (1) at (0,0) {};
        \node[flavour] (2) at (0,1) {};
        \draw (1)--(2);
        \node at (1.3,0) {SU(2)=Sp(1)};
        \node at (0.5,1) {$D_n$};
\end{tikzpicture}\;.
\end{equation}

The fundamental representation of U(n) and SU(n) (for $n>2$ in SU(n)) is complex. While the fundamental representation of Sp(n) is pseudo-real. Hence for U(n) or SU(n) gauge nodes one has to subtract a U(1) piece with the same number of flavours and for the SU(2) and Sp(n) case one has to subtract SU(2)=Sp(1) pieces with the same number of flavours.\\

Rules for quiver subtraction of framed orthosymplectic quivers are given for `Slodowy intersections' in \cite{Hanany:2019tji}. Unframed orthosymplectic quivers are studied in \cite{Bourget:2020xdz,Bourget:2020gzi}. A complete set of rules for quiver subtraction that works for any type of quiver is still being developed.

\subsection{Quiver Addition is not unique}
\label{app:add}
If two symplectic singularities have the same Hasse diagram, it does not imply that they are the same. This is related to the fact that \emph{quiver addition} is not unique. We may define quiver addition as the inverse of quiver subtraction, in the sense, that $\mathsf{Q}=\mathsf{Q}'+\mathsf{D}$ if $\mathsf{Q}'=\mathsf{Q}-\mathsf{D}$. Note that the order is crucial, just as with subtraction. Let us take the example \eqref{eq:3leafexample}\footnote{here we use the unframed version, ungauging the top U(1) node in $\mathsf{Q}_1$ of \eqref{eq:add} yields the framed quiver from \eqref{eq:3leafexample}, the quivers are equivalent.} and compare it to a second quiver
\begin{equation}
    \begin{tikzpicture}
        \node at (-2,0) {$\mathsf{Q}_1=$};
        \node[gauge] (1) at (0,0) {};
        \node[gauge] (2) at (-1,0) {};
        \node[gauge] (3) at (1,0) {};
        \node[gauge] (4) at (0,1) {};
        \draw(2)--(1)--(3);
        \draw[transform canvas={xshift=-1.5pt}] (1)--(4);
        \draw[transform canvas={xshift=1.5pt}] (1)--(4);
        
        \node at (-1,-0.5) {1};
        \node at (0,-0.5) {2};
        \node at (1,-0.5) {1};
        \node at (0.5,1) {1};
    \end{tikzpicture}\,,\qquad\begin{tikzpicture}
        \node at (-2,0) {$\mathsf{Q}_2=$};
        \node[gauge] (1) at (0,0) {};
        \node[gauge] (2) at (-1,0) {};
        \node[gauge] (3) at (1,0) {};
        \node[gauge] (4) at (-1,1) {};
        \node[gauge] (5) at (1,1) {};
        \draw (4)--(2)--(1)--(3)--(5);
        \draw[transform canvas={yshift=-1.5pt}] (5)--(4);
        \draw[transform canvas={yshift=1.5pt}] (5)--(4);
        
        \node at (-1,-0.5) {1};
        \node at (0,-0.5) {1};
        \node at (1,-0.5) {1};
        \node at (-1.5,1) {1};
        \node at (1.5,1) {1};
    \end{tikzpicture}\,;\qquad\begin{tikzpicture}
        \node at (-1,0) {$\HC=$};
        \node[hasse] (1) at (0,-0.5) {};
        \node[hasse] (2) at (0,0.5) {};
        \node[hasse] (3) at (0,1.5) {};
        \draw[red] (1)--(2)--(3);
        \node at (0.3,0) {$a_3$};
        \node at (0.3,1) {$a_1$};
    \end{tikzpicture}\,,\qquad\begin{tikzpicture}
        \node at (-1,0) {$\HH=$};
        \node[hasse] (1) at (0,-0.5) {};
        \node[hasse] (2) at (0,0.5) {};
        \node[hasse] (3) at (0,1.5) {};
        \draw[blue] (1)--(2)--(3);
        \node at (0.3,0) {$A_1$};
        \node at (0.3,1) {$A_3$};
    \end{tikzpicture}\;.
    \label{eq:add}
\end{equation}
Both quivers can be written as
\begin{equation}
    \mathsf{Q}_i=a_3+a_1\qquad i\in\{1,2\}\,.
\end{equation}
It is easy to check, that $\HC(\mathsf{Q}_1)=\HC(\mathsf{Q}_2)$ and $\HH(\mathsf{Q}_1)=\HH(\mathsf{Q}_2)$ are the same, in fact the Hasse diagram of their entire moduli space is identical. However, a Hilbert series calculation yields that the two quivers have different Coulomb branches and Higgs branches.\\

Note that for both $\mathsf{Q}_1$ and $\mathsf{Q}_2$ the Hasse diagram of their respective Higgs and Coulomb branch are related by inversion, $\HH(\mathsf{Q}_1)=\HH(\mathsf{Q}_2)=\mathcal{I}(\HC(\mathsf{Q}_1))=\mathcal{I}(\HC(\mathsf{Q}_2))$.\\

A systematic study of how quivers can be added in inequivalent ways remains to be done.

\section{Computing transverse slices, $\mathsf{T}(\mathcal{L}_1,\mathcal{L}_2)$, in the full moduli space -- from brane constructions.}
\label{app:O}
Brane constructions play an invaluable role in analysing supersymmetric gauge theories and this appendix is dedicated to yet one further example of the power of branes. If a good or ugly $3d$ $\mathcal{N}=4$ theory admits a brane construction, then the leaves in the moduli space of the theory correspond to different phases in the brane system. Examples of this were already given in Figures \ref{fig:nminA3branes} and \ref{fig:nminA3dualbranes}. In Figure \ref{fig:O(3)-C4} we give another example.

Let us consider a concrete theory realised on a brane system. At every phase $P$ of the brane system, corresponding to a leaf $\mathcal{L}$ in the moduli space, one can compute both an electric quiver $\mathsf{Q}_e(P)$ and a magnetic quiver $\mathsf{Q}_m(P)$ for the moduli which were turned on. Computing the Coulomb branch of the quiver $\mathsf{Q}_e(P)$, provides one with the Coulomb part of the closure of the leaf $\mathcal{L}$ and computing the Coulomb branch of $\mathsf{Q}_m(P)$ provides one with the Higgs part of the closure of the leaf $\mathcal{L}$, combining this one obtains the closure of the leaf $\mathcal{L}$, i.e.\ the transverse slice $\mathsf{T}(\{0\},\mathcal{L})$:

\begin{equation}
    \bar{\mathcal{L}}=\mathsf{T}(\{0\},\mathcal{L})=\mathcal{C}(\mathsf{Q}_e(P))\times \mathcal{C}(\mathsf{Q}_m(P))
\end{equation}

If one compares the moduli which were turned on additionally between two phases $P_1$ and $P_2$, where the corresponding leaves obey $\mathcal{L}_1\subset\bar{\mathcal{L}}_2$, one can read an electric quiver $\mathsf{Q}_e(P_2-P_1)$ and a magnetic quiver $\mathsf{Q}_m(P_2-P_1)$. Computing the Coulomb branch of the quiver $\mathsf{Q}_e(P_2-P_1)$, provides one with the Coulomb part of the transverse slice $\mathsf{T}(\mathcal{L}_1,\mathcal{L}_2)$ and computing the Coulomb branch of $\mathsf{Q}_m(P)$ provides one with the Higgs part of the transverse slice $\mathsf{T}(\mathcal{L}_1,\mathcal{L}_2)$, combining this one obtains the full transverse slice $\mathsf{T}(\mathcal{L}_1,\mathcal{L}_2)$:

\begin{equation}
    \mathsf{T}(\mathcal{L}_1,\mathcal{L}_2)=\mathcal{C}(\mathsf{Q}_e(P_2-P_1))\times \mathcal{C}(\mathsf{Q}_m(P_2-P_1))\;.
\end{equation}

This is an extension of the previously developed ideas of realising Kraft-Procesi transitions in brane systems in \cite{Cabrera:2016vvv,Cabrera:2017njm} to the entire moduli space.\\

\subsection{O(3)-[C${}_4$]}

As an example in Figure \ref{fig:O(3)-C4} we provide the various phases of the brane set up for $O(3)-[C_4]$ using an $O5^+$ plane. As an illustration we compute both the enhanced Coulomb branch of this theory, which is the closure of the leaf $\mathcal{L}_{c)}$, i.e.\ the transverse slice between the leaves $\mathcal{L}_{a)}$ (the origin) and $\mathcal{L}_{c)}$, corresponding to a) and c) in Figure \ref{fig:O(3)-C4}; and also transverse slice between the leaves $\mathcal{L}_{d)}$ and $\mathcal{L}_{e)}$, corresponding to d) and e) in Figure \ref{fig:O(3)-C4}.

\paragraph{The slice between a) and c) of Figure \ref{fig:O(3)-C4}.} In order to read off the electric and magnetic quivers correctly, one has to focus on the concrete moduli one is trying to describe. Since a) corresponds to the origin of the moduli space, we have to take into account all moduli in the phase c). The brane diagram is:

\begin{equation}
    \begin{tikzpicture}[scale=0.4]
        \draw[red] (0,-4)--(0,4) (13,-4)--(13,4);
            \draw[blue] (1,-3)--(4,3) (2,-3)--(5,3) (3,-3)--(6,3) (4,-3)--(7,3) (6,-3)--(9,3) (7,-3)--(10,3) (8,-3)--(11,3) (9,-3)--(12,3);
            \draw[green] (5,-3)--(8,3);
            \draw[dashed] (-0.5,0)--(13.5,0);
            \draw (0,-0.1)--(2.45,-0.1) (10.45,-0.1)--(13,-0.1);
            \draw[transform canvas={xshift=-0.8*0.5cm,yshift=-0.8*1cm}] (2.5,0)--(3.5,0) (9.5,0)--(10.5,0);
            \draw[transform canvas={xshift=-0.8*0.7cm,yshift=-0.8*1.4cm}] (3.5,0)--(4.5,0) (8.5,0)--(9.5,0);
            \draw[transform canvas={xshift=-0.8*0.5cm,yshift=-0.8*1cm}] (4.5,0)--(5.5,0) (7.5,0)--(8.5,0);
            \draw[transform canvas={xshift=-0.8*0.7cm,yshift=-0.8*1.4cm}] (5.5,0)--(7.5,0);
            \draw (0,-3.5)--(13,-3.5);
            \draw (0,3.5)--(13,3.5);
    \end{tikzpicture}\;.
    \label{eq:c)}
\end{equation}
In order to obtain the electric quiver, we momentarily ignore\footnote{Instead of `ignoring' the Higgs moduli one could also send them to infinity in this case.} the Higgs branch moduli in \eqref{eq:c)} and consider the brane system:

\begin{equation}
    \begin{tikzpicture}[scale=0.4]
        \draw[red] (0,-4)--(0,4) (13,-4)--(13,4);
            \draw[blue] (1,-3)--(4,3) (2,-3)--(5,3) (3,-3)--(6,3) (4,-3)--(7,3) (6,-3)--(9,3) (7,-3)--(10,3) (8,-3)--(11,3) (9,-3)--(12,3);
            \draw[green] (5,-3)--(8,3);
            \draw[dashed] (-0.5,0)--(13.5,0);
            \draw (0,-0.1)--(2.45,-0.1) (10.45,-0.1)--(13,-0.1);
            \draw (0,-3.5)--(13,-3.5);
            \draw (0,3.5)--(13,3.5);
    \end{tikzpicture}\;.
\end{equation}
Performing a Hanany-Witten transition we obtain a brane diagram for which an electric quiver is easily read:

\begin{equation}
    \begin{tikzpicture}[scale=0.4]
        \draw[red] (0,-4)--(0,4) (13,-4)--(13,4);
            \draw[blue] (1-5,-3)--(4-5,3) (2,-3)--(5,3) (3,-3)--(6,3) (4,-3)--(7,3) (6,-3)--(9,3) (7,-3)--(10,3) (8,-3)--(11,3) (9+5,-3)--(12+5,3);
            \draw[green] (5,-3)--(8,3);
            \draw[dashed] (-0.5,0)--(13.5,0);
            \draw (0,-3.5)--(13,-3.5);
            \draw (0,3.5)--(13,3.5);
    \end{tikzpicture}
    \quad\textnormal{electric quiver:}\quad\mathsf{Q}_e(P_{c)})=\quad
    \begin{tikzpicture}
    \node[gauge,label=right:{$O(2)$}] (1) at (0,0) {};
    \node[flavour,label=right:{$C_3$}] (2) at (0,1) {};
    \draw (1)--(2);
    \end{tikzpicture}\;.
\end{equation}
The Coulomb branch of $\mathsf{Q}_e(P_{c)})$ is the $D_5$ singularity.\\
In order to obtain the magnetic quiver, we momentarily ignore the Coulomb branch moduli in \eqref{eq:c)} and consider the brane system:

\begin{equation}
    \begin{tikzpicture}[scale=0.4]
        \draw[red] (0,-4)--(0,4) (13,-4)--(13,4);
            \draw[blue] (1,-3)--(4,3) (2,-3)--(5,3) (3,-3)--(6,3) (4,-3)--(7,3) (6,-3)--(9,3) (7,-3)--(10,3) (8,-3)--(11,3) (9,-3)--(12,3);
            \draw[green] (5,-3)--(8,3);
            \draw[dashed] (-0.5,0)--(13.5,0);
            \draw (0,-0.1)--(2.45,-0.1) (10.45,-0.1)--(13,-0.1);
            \draw[transform canvas={xshift=-0.8*0.5cm,yshift=-0.8*1cm}] (2.5,0)--(3.5,0) (9.5,0)--(10.5,0);
            \draw[transform canvas={xshift=-0.8*0.7cm,yshift=-0.8*1.4cm}] (3.5,0)--(4.5,0) (8.5,0)--(9.5,0);
            \draw[transform canvas={xshift=-0.8*0.5cm,yshift=-0.8*1cm}] (4.5,0)--(5.5,0) (7.5,0)--(8.5,0);
            \draw[transform canvas={xshift=-0.8*0.7cm,yshift=-0.8*1.4cm}] (5.5,0)--(7.5,0);
    \end{tikzpicture}\;.
\end{equation}
Performing a Hanany-Witten transition we obtain a brane diagram for which a magnetic quiver is easily read:

\begin{equation}
    \begin{tikzpicture}[scale=0.4]
            \draw[red] (3,-4)--(3,4) (10,-4)--(10,4);
            \draw[blue] (1,-3)--(4,3) (2,-3)--(5,3) (3,-3)--(6,3) (4,-3)--(7,3) (6,-3)--(9,3) (7,-3)--(10,3) (8,-3)--(11,3) (9,-3)--(12,3);
            \draw[green] (5,-3)--(8,3);
            \draw[dashed] (-0.5,0)--(13.5,0);
            \draw[transform canvas={xshift=-0.8*0.5cm,yshift=-0.8*1cm}] (2.5,0)--(3.5,0) (9.5,0)--(10.5,0);
            \draw[transform canvas={xshift=-0.8*0.7cm,yshift=-0.8*1.4cm}] (3.5,0)--(4.5,0) (8.5,0)--(9.5,0);
            \draw[transform canvas={xshift=-0.8*0.5cm,yshift=-0.8*1cm}] (4.5,0)--(5.5,0) (7.5,0)--(8.5,0);
            \draw[transform canvas={xshift=-0.8*0.7cm,yshift=-0.8*1.4cm}] (5.5,0)--(7.5,0);
    \end{tikzpicture}
    \quad\textnormal{magnetic quiver:}\quad\mathsf{Q}_m(P_{c)})=\quad
    \begin{tikzpicture}
    \node[gauge,label=below:{$1$}] (1) at (0,0) {};
    \node[gauge,label=below:{$1$}] (2) at (1,0) {};
    \node[gauge,label=below:{$1$}] (3) at (2,0) {};
    \node[gauge,label=below:{$1$}] (4) at (3,0) {};
    \node[flavour,label=left:{$1$}] (0) at (0,1) {};
    \draw (0)--(1)--(2)--(3) (2.6,0.2)--(2.4,0)--(2.6,-0.2);
    \draw[transform canvas={yshift=-1.5pt}] (3)--(4);
    \draw[transform canvas={yshift=1.5pt}] (3)--(4);
    \end{tikzpicture}\;.
\end{equation}
The Coulomb branch of $\mathsf{Q}_m(P_{c)})$ is the minimal nilpotent orbit closure $c_4$. Hence the Hasse diagram between a) and c) is:

\begin{equation}
    \begin{tikzpicture}
    \node[hasse] (1) at (0,0) {};
    \node[hasse] (2) at (-1,1) {};
    \node[hasse] (3) at (1,1) {};
    \node[hasse] (4) at (0,2) {};
    \draw[blue] (1)--(3) (2)--(4);
    \draw[red] (1)--(2) (3)--(4);
    \node at ($(1)!0.5!(3)$) {$c_4$};
    \node at ($(2)!0.5!(4)$) {$c_4$};
    \node at ($(1)!0.5!(2)$) {$D_5$};
    \node at ($(3)!0.5!(4)$) {$D_5$};
    \end{tikzpicture}\;,
\end{equation}
and we have computed the enhanced Coulomb branch of $O(3)-[C_4]$.

\paragraph{The slice between d) and e) of Figure \ref{fig:O(3)-C4}.} This case is more involved than the previous one, as one has to be extremely careful in identifying the relevant moduli. Let us compare the phases $P_{d)}$ and $P_{e)}$ of the brane system:

\begin{equation}
    \begin{tikzpicture}[scale=0.4]
            \node at (6.5,-5) {d)};
            \node at (21.5,-5) {e)};
            \draw[red] (0,-4)--(0,4) (13,-4)--(13,4);
            \draw[blue] (1,-3)--(4,3) (2,-3)--(5,3) (3,-3)--(6,3) (4,-3)--(7,3) (6,-3)--(9,3) (7,-3)--(10,3) (8,-3)--(11,3) (9,-3)--(12,3);
            \draw[green] (5,-3)--(8,3);
            \draw[dashed] (-0.5,0)--(13.5,0);
            \draw (0,-0.1)--(2.45,-0.1) (10.45,-0.1)--(13,-0.1);
            \draw[transform canvas={xshift=-0.8*0.5cm,yshift=-0.8*1cm}] (2.5,0)--(3.5,0) (9.5,0)--(10.5,0);
            \draw[transform canvas={xshift=-0.8*0.7cm,yshift=-0.8*1.4cm}] (3.5,0)--(4.5,0) (8.5,0)--(9.5,0);
            \draw[transform canvas={xshift=-0.8*0.5cm,yshift=-0.8*1cm}] (4.5,0)--(5.5,0) (7.5,0)--(8.5,0);
            \draw[transform canvas={xshift=-0.8*0.7cm,yshift=-0.8*1.4cm}] (5.5,0)--(7.5,0);
            \draw (0,-0.2)--(13,-0.2);
            \draw (0,0.2)--(13,0.2);
            \begin{scope}[shift={(15,0)}]
            \draw[red] (0,-4)--(0,4) (13,-4)--(13,4);
            \draw[blue] (1,-3)--(4,3) (2,-3)--(5,3) (3,-3)--(6,3) (4,-3)--(7,3) (6,-3)--(9,3) (7,-3)--(10,3) (8,-3)--(11,3) (9,-3)--(12,3);
            \draw[green] (5,-3)--(8,3);
            \draw[dashed] (-0.5,0)--(13.5,0);
            \draw (0,-0.1)--(2.45,-0.1) (10.45,-0.1)--(13,-0.1);
            \draw[transform canvas={xshift=-0.8*0.5cm,yshift=-0.8*1cm}] (2.5,0)--(3.5,0) (9.5,0)--(10.5,0);
            \draw[transform canvas={xshift=-0.8*0.7cm,yshift=-0.8*1.4cm}] (3.5,0)--(4.5,0) (8.5,0)--(9.5,0);
            \draw[transform canvas={xshift=-0.8*0.5cm,yshift=-0.8*1cm}] (4.5,0)--(5.5,0) (7.5,0)--(8.5,0);
            \draw[transform canvas={xshift=-0.8*0.7cm,yshift=-0.8*1.4cm}] (5.5,0)--(7.5,0);
            \draw (0,-0.2)--(3.4,-0.2) (9.4,-0.2)--(13,-0.2);
            \draw[transform canvas={xshift=-0.8*0.6cm,yshift=-0.8*1.2cm}] (3.5,0)--(4.5,0) (8.5,0)--(9.5,0);
            \draw[transform canvas={xshift=-0.8*0.4cm,yshift=-0.8*0.8cm}] (4.5,0)--(5.5,0) (7.5,0)--(8.5,0);
            \draw[transform canvas={xshift=-0.8*0.6cm,yshift=-0.8*1.2cm}] (5.5,0)--(7.5,0);
            \draw (0,0.2)--(13,0.2);
            \end{scope}
    \end{tikzpicture}\;.
\end{equation}
The first order of business is to forget exactly those moduli we see turned on in e) which were already turned on in d) in order identify the extra moduli which were tuned in order to move from the phase $P_{d)}$ to the phase $P_{e)}$. This means, we ignore those Higgs branch moduli in d) which are turned on in e). We see the relevant part of the brane system:

\begin{equation}
    \begin{tikzpicture}[scale=0.4]
            \draw[red] (0,-4)--(0,4) (13,-4)--(13,4);
            \draw[blue] (1,-3)--(4,3) (2,-3)--(5,3) (3,-3)--(6,3) (4,-3)--(7,3) (6,-3)--(9,3) (7,-3)--(10,3) (8,-3)--(11,3) (9,-3)--(12,3);
            \draw[green] (5,-3)--(8,3);
            \draw[dashed] (-0.5,0)--(13.5,0);
            \draw (0,-0.1)--(2.45,-0.1) (10.45,-0.1)--(13,-0.1);
            \draw (0,-0.2)--(3.4,-0.2) (9.4,-0.2)--(13,-0.2);
            \draw[transform canvas={xshift=-0.8*0.6cm,yshift=-0.8*1.2cm}] (3.5,0)--(4.5,0) (8.5,0)--(9.5,0);
            \draw[transform canvas={xshift=-0.8*0.4cm,yshift=-0.8*0.8cm}] (4.5,0)--(5.5,0) (7.5,0)--(8.5,0);
            \draw[transform canvas={xshift=-0.8*0.6cm,yshift=-0.8*1.2cm}] (5.5,0)--(7.5,0);
            \draw (0,0.2)--(13,0.2);
    \end{tikzpicture}\;.
\end{equation}
Now we also have to ignore the moduli, which we are still able to turn on but haven't, i.e.\ the brane resting at the origin, and we obtain the brane system:

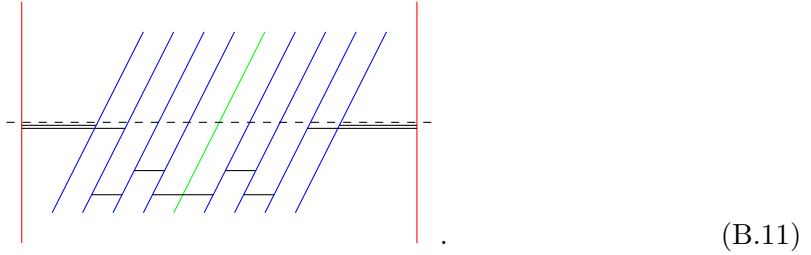
\begin{equation}
    \begin{tikzpicture}[scale=0.4]
            \draw[red] (0,-4)--(0,4) (13,-4)--(13,4);
            \draw[blue] (1,-3)--(4,3) (2,-3)--(5,3) (3,-3)--(6,3) (4,-3)--(7,3) (6,-3)--(9,3) (7,-3)--(10,3) (8,-3)--(11,3) (9,-3)--(12,3);
            \draw[green] (5,-3)--(8,3);
            \draw[dashed] (-0.5,0)--(13.5,0);
            \draw (0,-0.1)--(2.45,-0.1) (10.45,-0.1)--(13,-0.1);
            \draw (0,-0.2)--(3.4,-0.2) (9.4,-0.2)--(13,-0.2);
            \draw[transform canvas={xshift=-0.8*0.6cm,yshift=-0.8*1.2cm}] (3.5,0)--(4.5,0) (8.5,0)--(9.5,0);
            \draw[transform canvas={xshift=-0.8*0.4cm,yshift=-0.8*0.8cm}] (4.5,0)--(5.5,0) (7.5,0)--(8.5,0);
            \draw[transform canvas={xshift=-0.8*0.6cm,yshift=-0.8*1.2cm}] (5.5,0)--(7.5,0);
    \end{tikzpicture}\;.
\end{equation}
There is no electric quiver $\mathsf{Q}_e(P_{e)}-P_{d)})$ to read from this brane system, hence the Coulomb part of the transverse slice $\mathsf{T}(\mathcal{L}_{d)},\mathcal{L}_{e)})$ is trivial. We can perform Hanany-Witten transitions in order to obtain a brane system from which a magnetic quiver can be read straight forwardly:

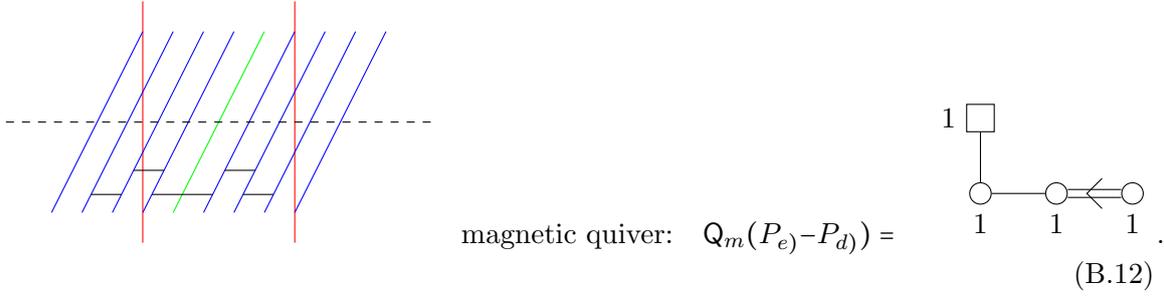
\begin{equation}
    \begin{tikzpicture}[scale=0.4]
            \draw[red] (4,-4)--(4,4) (9,-4)--(9,4);
            \draw[blue] (1,-3)--(4,3) (2,-3)--(5,3) (3,-3)--(6,3) (4,-3)--(7,3) (6,-3)--(9,3) (7,-3)--(10,3) (8,-3)--(11,3) (9,-3)--(12,3);
            \draw[green] (5,-3)--(8,3);
            \draw[dashed] (-0.5,0)--(13.5,0);
            \draw[transform canvas={xshift=-0.8*0.6cm,yshift=-0.8*1.2cm}] (3.5,0)--(4.5,0) (8.5,0)--(9.5,0);
            \draw[transform canvas={xshift=-0.8*0.4cm,yshift=-0.8*0.8cm}] (4.5,0)--(5.5,0) (7.5,0)--(8.5,0);
            \draw[transform canvas={xshift=-0.8*0.6cm,yshift=-0.8*1.2cm}] (5.5,0)--(7.5,0);
    \end{tikzpicture}
    \quad\textnormal{magnetic quiver:}\quad\mathsf{Q}_m(P_{e)}-P_{d)})=\quad
    \begin{tikzpicture}
    \node[gauge,label=below:{$1$}] (2) at (1,0) {};
    \node[gauge,label=below:{$1$}] (3) at (2,0) {};
    \node[gauge,label=below:{$1$}] (4) at (3,0) {};
    \node[flavour,label=left:{$1$}] (0) at (1,1) {};
    \draw (0)--(2)--(3) (2.6,0.2)--(2.4,0)--(2.6,-0.2);
    \draw[transform canvas={yshift=-1.5pt}] (3)--(4);
    \draw[transform canvas={yshift=1.5pt}] (3)--(4);
    \end{tikzpicture}\;.
\end{equation}
Hence the transverse slice $\mathsf{T}(\mathcal{L}_{d)},\mathcal{L}_{e)})$ consists only of a Higgs part, which is $c_3$. The relevant Hasse diagram is:

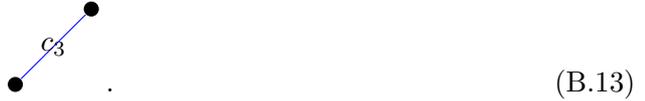
\begin{equation}
    \begin{tikzpicture}
    \node[hasse] (1) at (0,0) {};
    \node[hasse] (2) at (1,1) {};
    \draw[blue] (1)--(2);
    \node at ($(1)!0.5!(2)$) {$c_3$};
\end{tikzpicture}\;.
\end{equation}

We propose that this procedure can be used, whenever a theory admits a brane construction for which the rules of reading magnetic quivers are understood, in order to compute any transverse slice in the moduli space of the theory.

\begin{landscape}
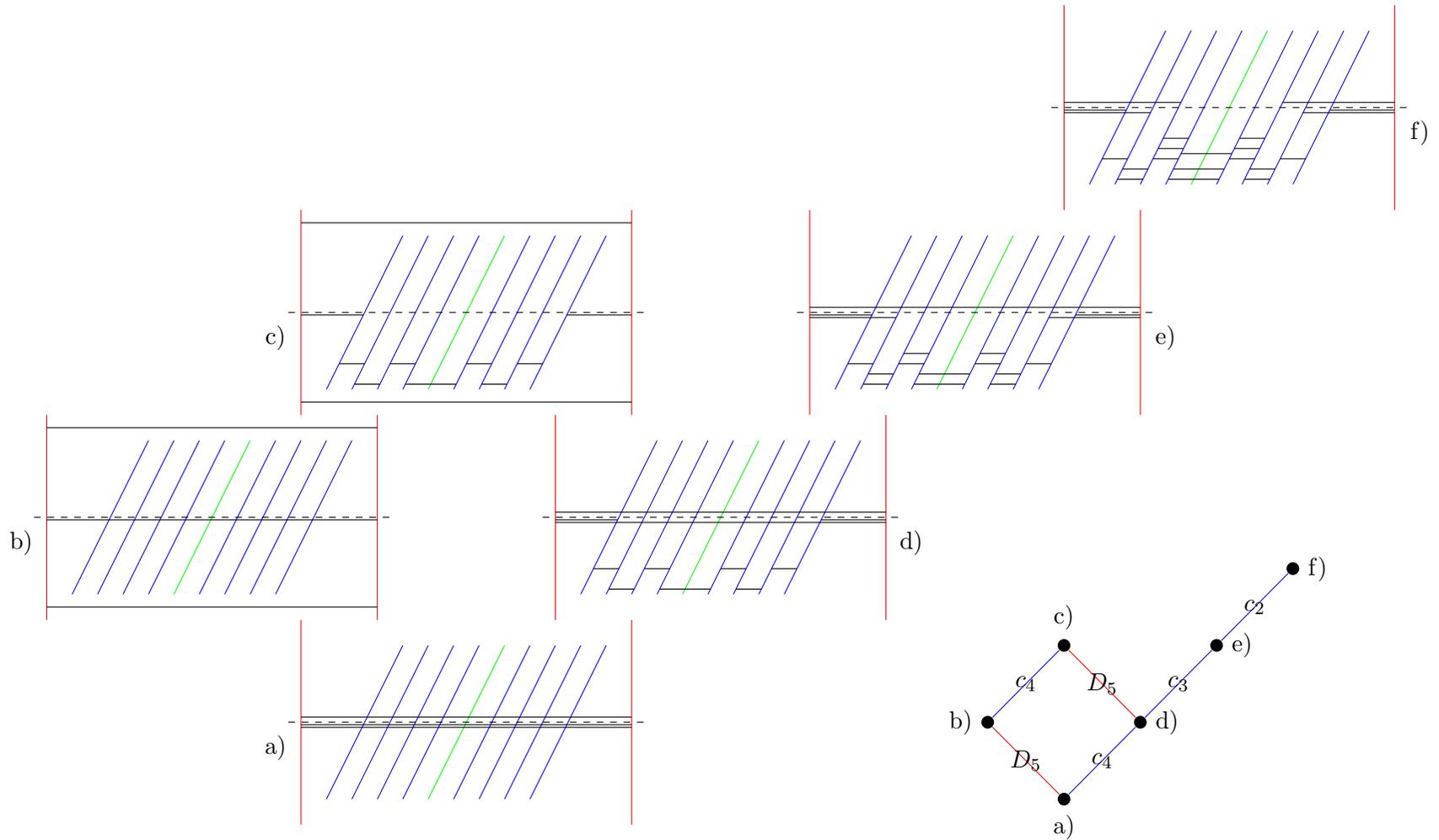
\begin{figure}
    \centering
        \makebox[\textwidth][c]{\begin{tikzpicture}[scale=0.4]
            \node at (-1,-1) {a)};
            \node at (-11,7) {b)};
            \node at (-1,15) {c)};
            \node at (24,7) {d)};
            \node at (34,15) {e)};
            \node at (44,23) {f)};
            \draw[red] (0,-4)--(0,4) (13,-4)--(13,4);
            \draw[blue] (1,-3)--(4,3) (2,-3)--(5,3) (3,-3)--(6,3) (4,-3)--(7,3) (6,-3)--(9,3) (7,-3)--(10,3) (8,-3)--(11,3) (9,-3)--(12,3);
            \draw[green] (5,-3)--(8,3);
            \draw[dashed] (-0.5,0)--(13.5,0);
            \draw (0,-0.1)--(13,-0.1);
            \draw (0,-0.2)--(13,-0.2);
            \draw (0,0.2)--(13,0.2);
            \begin{scope}[shift={(-10,8)}]
            \draw[red] (0,-4)--(0,4) (13,-4)--(13,4);
            \draw[blue] (1,-3)--(4,3) (2,-3)--(5,3) (3,-3)--(6,3) (4,-3)--(7,3) (6,-3)--(9,3) (7,-3)--(10,3) (8,-3)--(11,3) (9,-3)--(12,3);
            \draw[green] (5,-3)--(8,3);
            \draw[dashed] (-0.5,0)--(13.5,0);
            \draw (0,-0.1)--(13,-0.1);
            \draw (0,-3.5)--(13,-3.5);
            \draw (0,3.5)--(13,3.5);
            \end{scope}
            \begin{scope}[shift={(0,16)}]
            \draw[red] (0,-4)--(0,4) (13,-4)--(13,4);
            \draw[blue] (1,-3)--(4,3) (2,-3)--(5,3) (3,-3)--(6,3) (4,-3)--(7,3) (6,-3)--(9,3) (7,-3)--(10,3) (8,-3)--(11,3) (9,-3)--(12,3);
            \draw[green] (5,-3)--(8,3);
            \draw[dashed] (-0.5,0)--(13.5,0);
            \draw (0,-0.1)--(2.45,-0.1) (10.45,-0.1)--(13,-0.1);
            \draw[transform canvas={xshift=-0.8*0.5cm,yshift=-0.8*1cm}] (2.5,0)--(3.5,0) (9.5,0)--(10.5,0);
            \draw[transform canvas={xshift=-0.8*0.7cm,yshift=-0.8*1.4cm}] (3.5,0)--(4.5,0) (8.5,0)--(9.5,0);
            \draw[transform canvas={xshift=-0.8*0.5cm,yshift=-0.8*1cm}] (4.5,0)--(5.5,0) (7.5,0)--(8.5,0);
            \draw[transform canvas={xshift=-0.8*0.7cm,yshift=-0.8*1.4cm}] (5.5,0)--(7.5,0);
            \draw (0,-3.5)--(13,-3.5);
            \draw (0,3.5)--(13,3.5);
            \end{scope}
            \begin{scope}[shift={(10,8)}]
            \draw[red] (0,-4)--(0,4) (13,-4)--(13,4);
            \draw[blue] (1,-3)--(4,3) (2,-3)--(5,3) (3,-3)--(6,3) (4,-3)--(7,3) (6,-3)--(9,3) (7,-3)--(10,3) (8,-3)--(11,3) (9,-3)--(12,3);
            \draw[green] (5,-3)--(8,3);
            \draw[dashed] (-0.5,0)--(13.5,0);
            \draw (0,-0.1)--(2.45,-0.1) (10.45,-0.1)--(13,-0.1);
            \draw[transform canvas={xshift=-0.8*0.5cm,yshift=-0.8*1cm}] (2.5,0)--(3.5,0) (9.5,0)--(10.5,0);
            \draw[transform canvas={xshift=-0.8*0.7cm,yshift=-0.8*1.4cm}] (3.5,0)--(4.5,0) (8.5,0)--(9.5,0);
            \draw[transform canvas={xshift=-0.8*0.5cm,yshift=-0.8*1cm}] (4.5,0)--(5.5,0) (7.5,0)--(8.5,0);
            \draw[transform canvas={xshift=-0.8*0.7cm,yshift=-0.8*1.4cm}] (5.5,0)--(7.5,0);
            \draw (0,-0.2)--(13,-0.2);
            \draw (0,0.2)--(13,0.2);
            \end{scope}
            \begin{scope}[shift={(20,16)}]
            \draw[red] (0,-4)--(0,4) (13,-4)--(13,4);
            \draw[blue] (1,-3)--(4,3) (2,-3)--(5,3) (3,-3)--(6,3) (4,-3)--(7,3) (6,-3)--(9,3) (7,-3)--(10,3) (8,-3)--(11,3) (9,-3)--(12,3);
            \draw[green] (5,-3)--(8,3);
            \draw[dashed] (-0.5,0)--(13.5,0);
            \draw (0,-0.1)--(2.45,-0.1) (10.45,-0.1)--(13,-0.1);
            \draw[transform canvas={xshift=-0.8*0.5cm,yshift=-0.8*1cm}] (2.5,0)--(3.5,0) (9.5,0)--(10.5,0);
            \draw[transform canvas={xshift=-0.8*0.7cm,yshift=-0.8*1.4cm}] (3.5,0)--(4.5,0) (8.5,0)--(9.5,0);
            \draw[transform canvas={xshift=-0.8*0.5cm,yshift=-0.8*1cm}] (4.5,0)--(5.5,0) (7.5,0)--(8.5,0);
            \draw[transform canvas={xshift=-0.8*0.7cm,yshift=-0.8*1.4cm}] (5.5,0)--(7.5,0);
            \draw (0,-0.2)--(3.4,-0.2) (9.4,-0.2)--(13,-0.2);
            \draw[transform canvas={xshift=-0.8*0.6cm,yshift=-0.8*1.2cm}] (3.5,0)--(4.5,0) (8.5,0)--(9.5,0);
            \draw[transform canvas={xshift=-0.8*0.4cm,yshift=-0.8*0.8cm}] (4.5,0)--(5.5,0) (7.5,0)--(8.5,0);
            \draw[transform canvas={xshift=-0.8*0.6cm,yshift=-0.8*1.2cm}] (5.5,0)--(7.5,0);
            \draw (0,0.2)--(13,0.2);
            \end{scope}
            \begin{scope}[shift={(30,24)}]
            \draw[red] (0,-4)--(0,4) (13,-4)--(13,4);
            \draw[blue] (1,-3)--(4,3) (2,-3)--(5,3) (3,-3)--(6,3) (4,-3)--(7,3) (6,-3)--(9,3) (7,-3)--(10,3) (8,-3)--(11,3) (9,-3)--(12,3);
            \draw[green] (5,-3)--(8,3);
            \draw[dashed] (-0.5,0)--(13.5,0);
            \draw (0,-0.1)--(2.45,-0.1) (10.45,-0.1)--(13,-0.1);
            \draw[transform canvas={xshift=-0.8*0.5cm,yshift=-0.8*1cm}] (2.5,0)--(3.5,0) (9.5,0)--(10.5,0);
            \draw[transform canvas={xshift=-0.8*0.7cm,yshift=-0.8*1.4cm}] (3.5,0)--(4.5,0) (8.5,0)--(9.5,0);
            \draw[transform canvas={xshift=-0.8*0.5cm,yshift=-0.8*1cm}] (4.5,0)--(5.5,0) (7.5,0)--(8.5,0);
            \draw[transform canvas={xshift=-0.8*0.7cm,yshift=-0.8*1.4cm}] (5.5,0)--(7.5,0);
            \draw (0,-0.2)--(3.4,-0.2) (9.4,-0.2)--(13,-0.2);
            \draw[transform canvas={xshift=-0.8*0.6cm,yshift=-0.8*1.2cm}] (3.5,0)--(4.5,0) (8.5,0)--(9.5,0);
            \draw[transform canvas={xshift=-0.8*0.4cm,yshift=-0.8*0.8cm}] (4.5,0)--(5.5,0) (7.5,0)--(8.5,0);
            \draw[transform canvas={xshift=-0.8*0.6cm,yshift=-0.8*1.2cm}] (5.5,0)--(7.5,0);
            \draw (0,0.2)--(4.6,0.2) (8.6,0.2)--(13,0.2);
            \draw[transform canvas={xshift=-0.8*0.3cm,yshift=-0.8*0.6cm}] (4.5,0)--(5.5,0) (7.5,0)--(8.5,0);
            \draw[transform canvas={xshift=-0.8*0.45cm,yshift=-0.8*0.9cm}] (5.5,0)--(7.5,0);
            \end{scope}
            \begin{scope}[shift={(33,0)}]
                \node[hasse,label=right:{d)}] (1) at (0,0) {};
                \node[hasse,label=right:{e)}] (2) at (3,3) {};
                \node[hasse,label=right:{f)}] (3) at (6,6) {};
                \node[hasse,label=above:{c)}] (4) at (-3,3) {};
                \draw[blue] (1)--(2)--(3);
                \draw[red] (1)--(4);
                \node at ($(1)!0.5!(2)$) {$c_3$};
                \node at ($(2)!0.5!(3)$) {$c_2$};
                \node at ($(1)!0.5!(4)$) {$D_5$};
                \node[hasse,label=below:{a)}] (a1) at (-3,-3) {};
                \node[hasse,label=left:{b)}] (a2) at (-6,0) {};
                \draw[blue] (a2)--(4) (a1)--(1);
                \draw[red] (a1)--(a2);
                \node at ($(a1)!0.5!(1)$) {$c_4$};
                \node at ($(a1)!0.5!(a2)$) {$D_5$};
                \node at ($(a2)!0.5!(4)$) {$c_4$};                
            \end{scope}
        \end{tikzpicture}}
    \caption{Different phases of the brane set up for $O(3)$ with 4 fundamental hypermultiplets, and the Hasse diagram which can be read from the brane systems. The green line represents an $O5^+$ orientifold plane parallel to the D5 branes.}
    \label{fig:O(3)-C4}
\end{figure}
\end{landscape}

\bibliographystyle{JHEP}
\bibliography{bibli.bib}

\end{document}